\documentclass[aps,prd,12pt,a4paper,groupedaddress,preprintnumbers,nofootinbib,floatfix]{revtex4-1}
\pdfoutput=1
\usepackage{epsfig,amsfonts,amsthm}
\usepackage{amsmath,amssymb}
\usepackage{array}
\newcommand{\be}{\begin{equation}}
\newcommand{\ee}{\end{equation}}
\newcommand{\bea}{\begin{eqnarray}}
\newcommand{\eea}{\end{eqnarray}}

\renewcommand{\Re}{\mathrm{Re }}
\renewcommand{\Im}{\mathrm{Im }}
\newcommand{\doublet}[2]{ \left( \begin{array}{c}#1 \\ #2 \end{array}\right) }

\usepackage{color}
\definecolor{Red}{rgb}{1,0,0}
\definecolor{Blue}{rgb}{0,0,1}
\definecolor{Green}{rgb}{0,1,0}

\def\lsim{\mathrel{\rlap{\lower4pt\hbox{\hskip1pt$\sim$}}
    \raise1pt\hbox{$<$}}}         
\def\gsim{\mathrel{\rlap{\lower4pt\hbox{\hskip1pt$\sim$}}
    \raise1pt\hbox{$>$}}}         

\def\beq{\begin{equation}}
\def\eeq{\end{equation}}
\def\bea{\begin{eqnarray}}
\def\eea{\end{eqnarray}}

\def\<{\left\langle}
\def\>{\right\rangle}

\newcommand{\bt}{\begin{tabular}}
\newcommand{\et}{\end{tabular}}

\newcommand{\GeV}{{\ensuremath\rm \,GeV}}

\usepackage{slashed}
\usepackage{amssymb} 
\usepackage{float}

\usepackage[justification=centering]{caption}

\usepackage{tikz}
\usetikzlibrary{decorations.pathmorphing,decorations.markings}

\usepackage{graphicx}

\tikzset{
photon/.style={decorate, decoration={snake,amplitude=2pt, segment length=5pt}, draw=black},
particle/.style={draw=black, postaction={decorate}, decoration={markings,mark=at position .5 with {\arrow[draw=black]{>}}}},
antiparticle/.style={draw=black, postaction={decorate}, decoration={markings,mark=at position .5 with {\arrow[draw=black]{>}}}},
gluon/.style={decorate, draw=black, decoration={coil,amplitude=4pt, segment length=5pt}}
goldstone/.style={draw=green,postaction={decorate},decoration={markings,mark=at position .5 with {\arrow[draw=blue]{>}}}}
}

\begin{document}
\bibliographystyle{OurBibTeX}

\title{
Singlet scalar and 2HDM extensions of the Standard Model: CP-violation and constraints from $(g-2)_\mu$ and $e$EDM.
}


\author{
Venus~Keus}
\email{venus.keus@helsinki.fi}
\affiliation{Department of Physics and Helsinki Institute of Physics,
Gustaf Hallstromin katu 2, FIN-00014 University of Helsinki, Finland}
\affiliation{School of Physics and Astronomy, University of Southampton,
Southampton, SO17 1BJ, United Kingdom}
\author{Niko Koivunen}
\email{niko.koivunen@helsinki.fi}
\author{Kimmo Tuominen}
\email{kimmo.i.tuominen@helsinki.fi}
\affiliation{Department of Physics and Helsinki Institute of Physics,
Gustaf Hallstromin katu 2, FIN-00014 University of Helsinki, Finland}

\begin{abstract}
{We study popular scalar extensions of the Standard Model, namely the  singlet extension, the 2-Higgs doublet model (2HDM) and its extension
by a singlet scalar.
We focus on the contributions of the added scalars to the anomalous magnetic moment of the muon, $(g-2)_\mu$ in the presence of CP-violation, and the electric dipole moment of the electron ($e$EDM) in these models.
In the absence of CP-violation, CP-even and CP-odd scalars contribute with an opposite sign to the anomalous magnetic moment of the muon and as a result these models generally require very light scalars to explain the observed discrepancy in $(g-2)_\mu$.
We study the effect of CP-violation on the anomalous magnetic moment of the muon and its compatibility with the $e$EDM constraints. 
We show that given the current status of the global set of constraints applied on all values of $\cot\beta$, in the CP-violating scalar extensions, there exist no
viable parameter space in agreement with both $a_\mu$ and $e$EDM bounds.
}
\end{abstract}

\maketitle


\section{Introduction}
While the Standard Model (SM) of particle physics agrees very well with data from high energy collider experiments, it still falls short on explaining several observed features of Nature. For example, SM does not
provide sufficient amount of CP-violation to source the Baryon Asymmetry of the Universe (BAU)~\cite{Kuzmin:1985mm}
and the scalar sector of the SM does not provide a first order phase transition~\cite{Kajantie:1996mn}, which would be needed to produce BAU at the electroweak transition. Another example is the need
to understand the origin of neutrino masses and mixing patterns. One possible paradigm to address these issues
is to enlarge the scalar sector of the SM. Many such extensions have been studied in the literature~\cite{Silveira:1985rk,McDonald:1993ex,Cline:2012hg,Alanne:2016wtx, Lindner:2016bgg, Bian:2016zba, Kowalska:2017iqv}.

In addition to providing new sources for CP-violation,
the extra scalars arising from such extensions could
also help to explain the muon anomalous magnetic moment $a_\mu =(g-2)_\mu/2$ which deviates from the SM prediction by
\be
\Delta a_\mu=a_\mu^{exp}-a_\mu^{SM} = (2.87\pm 0.8) \times 10^{-9} ~ (3.6 \sigma)
\label{g-2-excess-SM}
\ee
according to the most recent experiment done at BNL \cite{Bennett:2006fi, Blum:2013xva}.

Finally, extended scalar sectors provide new scalar mass eigenstates which can, for example, provide a Dark Matter (DM) candidate. Typically their stability is guaranteed by an ad-hoc discrete symmetry.
In this paper, we do not consider a DM candidate and therefore, in the
models we study we try to avoid extra symmetries if possible.

We focus on the following well-known scalar extensions of the SM
\begin{itemize}
\item
Real and Complex Singlet extension of the SM (SM+RS, SM+CS)
\item
2-Higgs-doublet model (2HDM)
\item
Complex Singlet extension of the 2HDM (2HDM+CS)
\end{itemize}

For each model, we calculate the contribution of the scalars to $a_\mu$ to see if they can explain the observed discrepancy.
\be
\Delta a_\mu=a_\mu^{exp}-\left(a_\mu^{SM~(without~scalars)}+a_\mu^{scalars}\right) = 0
~ \Rightarrow ~ a_\mu^{scalars}= (2.88\pm 0.8)\times 10^{-9}
\label{g-2-excess}
\ee

We show that in the CP-conserving limit, due to the cancelling effect of the CP-odd and CP-even scalars, one can not
explain the excess in Eq.(\ref{g-2-excess}), unless very light scalars are present and $\tan\beta$ is very large.
However, when CP-violation is introduced, we show that less dramatic values of $\tan\beta$ or scalar masses are required to produce
the observed $a_\mu$.

Having introduced CP-violation, the parameter space of models under
consideration is strongly constrained by the data from ACME collaboration on
electron and neutron Electric Dipole Moment (EDM)~\cite{Baron:2013eja}. The
bounds on electron EDM ($e$EDM) with
\be
d_e < 10.25 \times 10^{-29} ~ \mbox{e cm} = 1.573\times 10^{-15}~ \mbox{GeV}^{-1},
\label{ACME}
\ee
impose the strongest constraints on any Beyond Standard Model (BSM) scenario with CP-violation.
In each of the models under consideration we study if the amount of CP-violation required to explain the $a_\mu$ discrepancy can be accommodated within the limits imposed by Eq.~\eqref{ACME}. As CP-violation is one of the main ingredients of BAU, the identification
of the surviving regions of the parameter space after imposing the $e$EDM
bounds is a necessary prerequisite of BAU studies. The models we have listed above have appeared in the context of electroweak baryogenesis: for a singlet extension of the SM, see e.g. \cite{McDonald:1993ey, Profumo:2007wc,Barger:2008jx,Ahriche:2012ei}, for 2HDM see e.g. \cite{Turok:1990zg, Turok:1991uc, Funakubo:1993jg, Davies:1994id, Cline:1995dg,Laine:2000rm,Fromme:2006cm, Basler:2016obg, Basler:2017uxn} and for a singlet extension of the 2HDM see \cite{Alanne:2016wtx, Bonilla:2014xba, Kakizaki:2016dza}. In this paper, we show, for the first time, how to
implement the constraints from $e$EDMs and from the muon anomalous moment
systematically on these models. In particular we show how this allows to
determine the experimentally favoured patterns of Yukawa interactions in these models.

The paper is organised as follows: For the reader's convenience,
in Section~\ref{section-calculations} we review the computation of $a_\mu$ and
$e$EDM from a generic Lagrangian and show in detail the 1-loop and 2-loop
calculations of
such contributions. In Sections~\ref{section-SM+RS},
~\ref{section-2hdm} and~\ref{section-2hdm+CS} we present the scalar potential, theoretical and experimental constraints and $a_\mu$ and $d_e$ contributions in the SM+RS, SM+CS, 2HDM and 2HDM+CS scenarios, respectively. In Section~ \ref{section-conclusion} we draw our conclusions and present our outlook.

\section{Calculation of $a_\mu$ and $d_e$ contributions}
\label{section-calculations}
By definition, the $a_\mu=(g-2)_\mu/2$ and $e$EDM contributions are
\bea
\mathcal{L}_{a_\mu} &=&
\frac{e}{2 m_\mu} \; a_\mu \; (\bar \mu  \sigma_{\mu\nu} \mu  ) F^{\mu\nu},
\\
\mathcal{L}_{d_e} &=&
-\frac{i}{2} \; d_e \; (\bar e  \sigma_{\mu\nu} \;\gamma_5 e)F^{\mu\nu},
\eea
where $F_{\mu\nu}$ is the electromagnetic field strength and $\sigma_{\mu\nu}=i[\gamma_\mu,\gamma_\nu]/2$. Therefore, the relevant parts of the Lagrangian are
\be
\mathcal{L} \supset
\frac{e\; m_l}{8\pi^2}
\biggl[
c_L  (\bar l \sigma_{\mu\nu} P_L l)F^{\mu\nu}
+
c_R  (\bar l \sigma_{\mu\nu} P_R l)F^{\mu\nu} \biggr] + \mathrm{h.c.},
\ee
where $l$ stands for the relevant lepton ($e$ for $e$EDM calculations and $\mu$ for the muon anomalous magnetic moment).
Expanding the Lagrangian for the explicit forms of the operators, leads to
\bea
\mathcal{L} &\supset &
\frac{e\; m_l}{8\pi^2}(c_L +c^*_R)  ( \bar l \sigma_{\mu\nu} P_L l) F^{\mu\nu}
+
\frac{e\; m_l}{8\pi^2}(c^*_L +c_R)  ( \bar l \sigma_{\mu\nu} P_R l) F^{\mu\nu} \nonumber\\
&=&
\frac{e\; m_l}{8\pi^2}\; \Re(c_L +c^*_R) \; ( \bar l \sigma_{\mu\nu} l) F^{\mu\nu}
-i
\frac{e\; m_l}{8\pi^2}\; \Im(c_L +c^*_R) \; ( \bar l \sigma_{\mu\nu} \gamma_5 l) F^{\mu\nu}
\eea
where $P_L=(1-\gamma_5)/2$ and $P_R=(1+\gamma_5)/2$ are the left and right projection operators.
One, therefore, needs to explicitly calculate
\bea
a_\mu &=& \frac{m_\mu^2}{4\pi^2}\; \Re(c_L +c^*_R).
\label{g-2}
\\
d_e &=& \frac{e\; m_e}{4\pi^2} \; \Im(c_L +c^*_R),
\label{de}
\eea
where $c_L$ and $c_R$ are the Wilson coefficients to be calculated for each loop diagram in Figure~\ref{loopy-fig} separately.

\begin{minipage}{\linewidth}
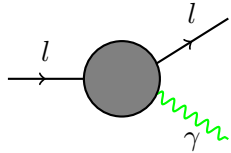
\begin{figure}[H]
\begin{center}
\begin{tikzpicture}[thick,scale=1.0]
\draw[particle] (0,0) -- node[black,above,yshift=0.1cm,xshift=0.0cm] {$l$} (1,0);
\draw[particle] (1.95,0.2) -- node[black,above,yshift=0.1cm,xshift=0.0cm] {$ l$} (2.9,0.8);
\draw[decorate,decoration={snake,amplitude=2pt,segment length=5pt},green] (1.95,-0.2) -- node[black,above,yshift=-0.6cm,xshift=0cm] {$\gamma$} (2.9,-0.8);
\draw[black,fill=gray]  (2,0) node[black,above,sloped,yshift=0.95cm,xshift=1.05cm] {$ $}  arc (360:0:0.5cm) ;
\end{tikzpicture}
\end{center}
\vspace{-5mm}
\caption{The higher order diagrams contributing to muon anomalous magnetic moment ($l=\mu$) and to
$e$EDM ($l=e$).}
\label{loopy-fig}
\end{figure}
\end{minipage}
\vspace{0.2cm}

\subsection{1-loop contributions}
The digram contributing to the $a_\mu$ and $d_e$ at 1-loop is shown in Figure~\ref{1-loop-fig}, where $h_i$ are the neutral scalars in the model with their coupling to electrons and muons represented by $Y^{h_i}_{ee}$ and $Y^{h_i}_{\mu\mu}$, respectively. The charged scalar mediated version of this diagram is sub-dominant and is therefore neglected~\cite{Chun:2015xfx}. The mass of the charged scalar is set to be equal to the mass of the heaviest scalar to comply with the ElectroWeak precision data.
Note also that in the models we study, we only extend the scalar sector of SM and do not add any extra vector or fermion fields, such as right-handed neutrinos.

\begin{minipage}{\linewidth}
\begin{figure}[H]
\begin{center}
\begin{tikzpicture}[thick,scale=1.0]
\fill[black] (1,0) circle (0.06cm);
\fill[black] (2,0) circle (0.06cm);
\fill[black] (3,0) circle (0.06cm);
\draw (0,0) -- node[black,above,yshift=-0.8cm,xshift=0.0cm] {$ $} (4,0);
\draw[particle] (0,0) -- node[black,above,sloped,yshift=-0.3cm,xshift=-0.9cm] {$l$} (1,0);
\draw[particle] (1,0) -- node[black,above,sloped,yshift=-0.0cm,xshift=0.0cm] {$ l$} (2,0);
\draw[particle] (3,0) -- node[black,above,sloped,yshift=-0.3cm,xshift=0.9cm] {$l$} (4,0);
\draw[decorate,decoration={snake,amplitude=2pt,segment length=5pt},green] (2,0) -- node[black,above,yshift=-0.4cm,xshift=0.5cm] {$\gamma$} (3,-1);
\draw[dashed]  (1,0) node[black,above,sloped,yshift=0.95cm,xshift=1.05cm] {$h_i$}  arc (180:0:1cm) ;
\draw[particle] (2,0) -- node[black,above,sloped,yshift=-0.0cm,xshift=0.0cm] {$ l$} (3,0);
\end{tikzpicture}
\end{center}
\vspace{-5mm}
\caption{The 1-loop diagram mediated by neutral scalars $h_i$ contributing to muon anomalous magnetic moment ($l=\mu$) and to
$e$EDM ($l=e$).}
\label{1-loop-fig}
\end{figure}
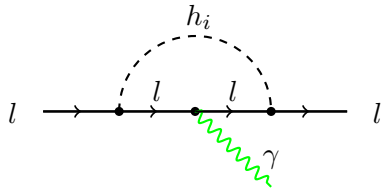
\end{minipage}
\vspace{0.2cm}

The Wilson coefficients are calculated to be
\bea
c_R &=& -
\frac{Y^{h_i}_{ll}}{4 m_l}  
\int_0^1 dx
\int_0^x dy \;
\frac{{Y_{ll}^{h_i}}^* y(y-1) m_l +  \lambda_{ll}^{h_i} (y-1) m_l}{m^2_l [y(y-x)+(1-y)]+m^2_{h_i} y},
\\
c_L &=& -
\frac{{Y^{h_i}_{ll}}^*}{4 m_l}  
\int_0^1 dx
\int_0^x dy \;
\frac{{Y_{ll}^{h_i}} y(y-1) m_l +  {\lambda_{ll}^{h_i}}^* (y-1) m_l}{m^2_l [y(y-x)+(1-y)]+m^2_{h_i} y},
\nonumber
\eea
where $Y^{h_i}_{ll}$ is the scalar $h_i$'s coupling to $ll$ and could in general be complex,
\be 
Y^{h_i}_{ll}=\Re(Y^{h_i}_{ll})+ i \Im(Y^{h_i}_{ll}).
\label{yll-definition}
\ee
The contribution from the 1-loop diagrams to $a_\mu$ and $d_e$ are then caluclated to be
\be
a_\mu^{1-loop} = -\frac{m_\mu^2}{8\pi^2}
\sum^n_{i=1}
\int_0^1 dx \int_0^x dy \frac{ y(y-1) \mid Y^{h_i}_{\mu\mu}\mid^2 +(y-1) \Re((Y^{h_i}_{\mu\mu})^2)}{m_\mu^2[y(y-x)+(1-y) ] +m_{h_i}^2 y},
\label{amu-1loop}
\ee
\be
d_e^{1-loop}=
\frac{e\; m_e}{16\pi^2}
\sum^n_{i=1}
\Im((Y^{h_i}_{ee})^2)
\int_0^1 dx
\int_0^x dy
\frac{(y-1) }{m^2_e [y(y-x)+(1-y)]+m^2_{h_i} y},
\label{de-1loop}
\ee
where $n$ is the number of the scalars mediating the loop in Fig. \ref{1-loop-fig}. Our formulas are in agreement with the known results in \cite{Lindner:2016bgg}, \cite{Chun:2015xfx}-\cite{Broggio:2014mna}.

\subsection{2-loop contributions}
The main 2-loop contributions to $a_\mu$ and $d_e$, shown in Figure~\ref{barzee}, arise from the Bar-Zee diagrams mediated by the scalar states.

\begin{minipage}{\linewidth}
\centering
\begin{minipage}{\linewidth}
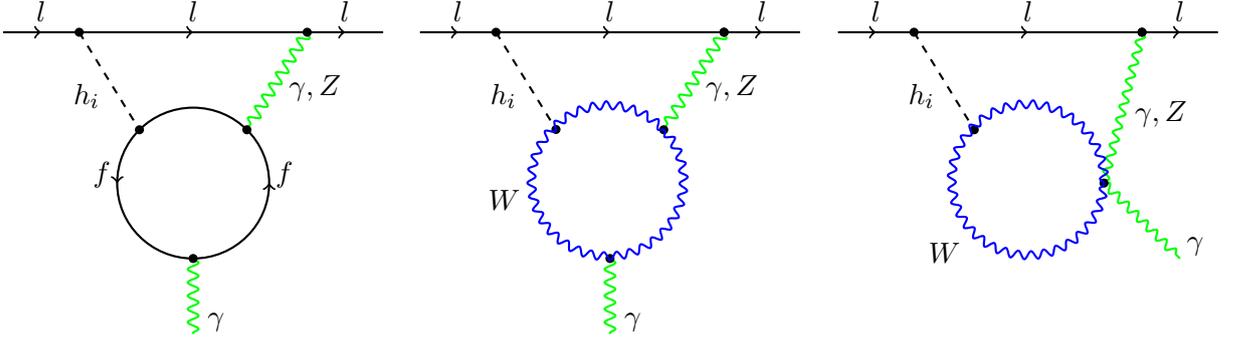
\begin{figure}[H]
\begin{tikzpicture}[thick,scale=1.0]
\draw[particle] (-0.5,0) -- node[black,above,sloped,yshift=-0.0cm,xshift=-0.0cm] {$l$} (0.5,0);
\fill[black] (0.5,0.0) circle (0.06cm);
\draw[particle] (0.5,0) -- node[black,above,sloped,yshift=-0.0cm,xshift=0.0cm] {$l$} (3.5,0);
\draw[particle] (3.5,0) -- node[black,above,sloped,yshift=-0.0cm,xshift=0.0cm] {$l$} (4.5,0);
\draw[dashed] (0.5,0) -- node[black,above,yshift=-0.5cm,xshift=-0.3cm] {$h_i$} (1.293,-1.293);
\fill[black] (1.293,-1.293) circle (0.06cm);
\draw[decorate,decoration={snake,amplitude=2pt,segment length=5pt},green] (3.5,0) -- node[black,above,yshift=-0.4cm,xshift=0.5cm] {$\gamma,Z$} (2.707,-1.293);
\fill[black] (2.707,-1.293) circle (0.06cm);
\draw[decorate,decoration={snake,amplitude=2pt,segment length=5pt},green] (2,-3) -- node[black,above,yshift=-0.6cm,xshift=0.3cm] {$\gamma$} (2,-4);
\fill[black] (2,-3) circle (0.06cm);
\draw[particle]  (2,-1) node[black,above,sloped,yshift=-1.2cm,xshift=-1.2cm] {$f$}  arc (90:270:1cm) ;
\draw[particle]  (2,-3) node[black,above,sloped,yshift=0.8cm,xshift=1.2cm] {$f$}  arc (-90:90:1cm) ;
\fill[black] (3.5,0) circle (0.06cm);
\end{tikzpicture}
\hspace{2mm}
\begin{tikzpicture}[thick,scale=1.0]
\draw[particle] (-0.5,0) -- node[black,above,sloped,yshift=-0.0cm,xshift=-0.0cm] {$l$} (0.5,0);
\fill[black] (0.5,0.0) circle (0.06cm);
\draw[particle] (0.5,0) -- node[black,above,sloped,yshift=-0.0cm,xshift=0.0cm] {$l$} (3.5,0);
\fill[black] (3.5,0) circle (0.06cm);
\draw[particle] (3.5,0) -- node[black,above,sloped,yshift=-0.0cm,xshift=0.0cm] {$l$} (4.5,0);
\draw[dashed] (0.5,0) -- node[black,above,yshift=-0.5cm,xshift=-0.3cm] {$h_i$} (1.293,-1.293);
\fill[black] (1.289,-1.289) circle (0.06cm);
\draw[decorate,decoration={snake,amplitude=2pt,segment length=5pt},green] (3.5,0) -- node[black,above,yshift=-0.4cm,xshift=0.5cm] {$\gamma,Z$} (2.707,-1.293);
\fill[black] (3.5,0) circle (0.06cm);
\fill[black] (2.707,-1.293) circle (0.06cm);
\draw[decorate,decoration={snake,amplitude=2pt,segment length=5pt},green] (2,-3) -- node[black,above,yshift=-0.6cm,xshift=0.3cm] {$\gamma$} (2,-4);
\fill[black] (2,-3) circle (0.06cm);
\draw[decorate,decoration={snake,amplitude=1.5pt,segment length=5pt},blue]  (2.707,-1.293) node[black,above,sloped,yshift=-1.2cm,xshift=-2.1cm] {$W$}  arc (45:405:1cm) ;
\hspace{5.5cm}
\draw[particle] (-0.5,0) -- node[black,above,sloped,yshift=-0.0cm,xshift=-0.0cm] {$l$} (0.5,0);
\fill[black] (0.5,0.0) circle (0.06cm);
\draw[particle] (0.5,0) -- node[black,above,sloped,yshift=-0.0cm,xshift=0.0cm] {$l$} (3.5,0);
\fill[black] (3.5,0) circle (0.06cm);
\draw[particle] (3.5,0) -- node[black,above,sloped,yshift=-0.0cm,xshift=0.0cm] {$l$} (4.5,0);
\draw[dashed] (0.5,0) -- node[black,above,yshift=-0.5cm,xshift=-0.3cm] {$h_i$} (1.293,-1.293);
\fill[black] (1.293,-1.293) circle (0.06cm);
\draw[decorate,decoration={snake,amplitude=1.5pt,segment length=5pt},green] (3.5,0) -- node[black,above,yshift=-0.4cm,xshift=0.5cm] {$\gamma,Z$} (3,-2);
\fill[black] (3.5,0) circle (0.06cm);
\draw[decorate,decoration={snake,amplitude=1.5pt,segment length=5pt},green] (3,-2) -- node[black,above,yshift=-0.6cm,xshift=0.7cm] {$\gamma$} (4,-3);
\fill[black] (3,-2) circle (0.06cm);
\draw[decorate,decoration={snake,amplitude=1.5pt,segment length=5pt},blue]  (3,-2) node[black,above,sloped,yshift=-1.2cm,xshift=-2.1cm] {$W$}  arc (0:360:1cm) ;
\end{tikzpicture}
\vspace{-1mm}
\caption{The Barr--Zee diagrams with largest contributions to muon anomalous magnetic moment $(l=\mu)$ and eEDM $(l=e)$.}
\label{barzee}
\end{figure}
\end{minipage}
\end{minipage}
\vspace{0.5cm}

The diagrams with the $Z$ boson in the loop (instead of $\gamma$) are suppressed by a factor of ($\frac{1}{4}-\sin^2 \theta_W$), which makes their contribution almost two orders of magnitude smaller than diagrams with a photon in the loop. We therefore ignore such diagrams in the calculations that follow. Similarly, contributions from the charged scalars are ignored since
they too are sub-dominant~\cite{Cheung:2001hz}. For our 2-loop calculations, we use the
results of~\cite{Harnik:2012pb}.

The contribution from 2-loop diagrams with heavy fermions $(f=t,b,c,\tau)$\footnote{The subscript $f$ stands for fermion and is not to be confused with the loop function $f(z)$.} and $W$ to the $a_\mu$ are

\begin{footnotesize}
\bea
\label{amu-2loop}
a_{\mu,~f}^{2-loop}&=&
\frac{2}{3}
\left(
\frac{\alpha G_F v^2 m_\mu}{\sqrt{2}\pi^3 m_f}\right) \sum^n_{i=1}
\biggl[
\Re(Y^{h_i}_{\mu\mu}) \Re(Y^{h_i}_{ff}) f(z_{fh_i})
- \Im(Y^{h_i}_{\mu\mu})\Im(Y^{h_i}_{ff}) g(z_{fh_i})
\biggr],
\\
a_{\mu,~W}^{2-loop}&=&
-\left(
\frac{\alpha G_F v m_\mu}{4\sqrt{2}\pi^3}\right) \sum^n_{i=1}
\frac{Y^{h_i}_{WW}}{2m^2_W/v}
\Re(Y^{h_i}_{\mu\mu})
\biggl[
3 f(z_{Wh_i}) +\frac{23}{4} g(z_{Wh_i})+\frac{3}{4} h(z_{Wh_i}) + \frac{f(z_{Wh_i})-g(z_{Wh_i})}{2 z_{Wh_i}}
\biggr], \nonumber
\eea
\end{footnotesize}

\noindent where $z_{AB}= m_{A}^2/m^2_{B}$, $Y_{WW}^{h_i}$ is the scalar $h_i$'s coupling to $WW$. For the SM-Higgs coupling to $WW$, we use the notation $Y_{WW}^{h^{SM}}$ which in the pure SM limit is $2m^2_W/v$.
The $Y^{h_i}_{ff}$ is the scalar $h_i$'s coupling to $ff$ which could in general be complex,
\be 
Y^{h_i}_{ff}=\Re(Y^{h_i}_{ff})+ i \Im(Y^{h_i}_{ff}).
\label{yff-definition}
\ee
and in the pure SM limit is $m_f/v$.

The contribution from 2-loop diagrams to the $d_e$ from heavy fermions, $f$, and $W$ loops are

\begin{footnotesize}
\bea
\label{de-2-loop}
d_{e,~f}^{2-loop}&=&
\frac{e}{3\pi^2}
\left(
\frac{\alpha G_F v^2}{\sqrt{2}\pi m_f}\right) \sum^n_{i=1}
\biggl[
\Im(Y^{h_i}_{ee}) \Re(Y^{h_i}_{ff}) f(z_{fh_i})
+ \Re(Y^{h_i}_{ee})\Im(Y^{h_i}_{ff}) g(z_{fh_i})
\biggr],
\\
d_{e,~W}^{2-loop}&=&
-\frac{e}{8\pi^2}
\left(
\frac{\alpha G_F v}{\sqrt{2}\pi}\right) \sum^n_{i=1}
\frac{Y^{h_i}_{WW}}{2m^2_W/v}
\Im(Y^{h_i}_{ee})
\biggl[
3 f(z_{Wh_i}) +\frac{23}{4} g(z_{Wh_i})+\frac{3}{4} h(z_{Wh_i}) + \frac{f(z_{Wh_i})-g(z_{Wh_i})}{2 z_{Wh_i}}
\biggr].  \nonumber
\eea
\end{footnotesize}
The loop functions $f(z)$, $g(z)$ and $h(z)$ appearing in Eqs.~\eqref{amu-2loop} and~\eqref{de-2-loop} and
are presented in Appendix \ref{loop-functions}.

\section{The real singlet extension (SM+RS)}
\label{section-SM+RS}

The real singlet model is often presented with a $Z_2$ symmetry imposed on the scalar potential in order to stabilise the singlet field as a viable DM candidate \cite{McDonald:1993ex,Burgess:2000yq,Davoudiasl:2004be,Yaguna:2008hd,Lerner:2009xg}.
As mentioned earlier, in this paper we shall not look into DM phenomenology and hence we consider the model in its general form with no extra symmetries.

The most general potential in this case has the following form
\bea
V&=& -\mu_1^2 \Phi^\dagger\Phi -\mu_2^2 S^2 +\lambda_1 (\Phi^\dagger\Phi)^2 +\lambda_2 S^4+\lambda_3 (\Phi^\dagger\Phi) S^2 \nonumber\\
&&+\kappa_1 S + \kappa_2 S (\Phi^\dagger\Phi) +\kappa_3 S^3.
\label{SMrsPot}
\eea
Note that by a translation of $S$, the linear $\kappa_1$ term can be removed.
The fields $\Phi$ and $S$ are, respectively, the SM gauge doublet and singlet with Vacuum
Expectation Values (VEVs) $v$ and $w$. Their field decomposition is as follows,
\be
\Phi= \doublet{$\begin{scriptsize}$ G^+ $\end{scriptsize}$}{\frac{v+\phi_1+i G^0}{\sqrt{2}}} ,\quad
S= \left({\frac{w+\phi_2}{\sqrt{2}}} \right) .
\label{SMrsfields}
\ee

Since $S$ is an $SU(2)$ singlet, it has no direct couplings to the SM gauge bosons or fermions. The field $\Phi$ plays the role of the SM Higgs doublet, therefore, $G^+$ and $G^0$ are the would-be Goldstone bosons which are ``eaten'' by the $W^\pm$ and $Z$ bosons.

The minimum of the potential requires
\bea
\mu^2_1 &=&
\frac{1}{2} \left(2 \lambda_1 v^2+\lambda_3 w^2+\sqrt{2} \kappa_2 w\right),
\\
\mu^2_2 &=&
\frac{1}{4w}
\left(
2 \sqrt{2} \kappa_1+\sqrt{2} \kappa_2 v^2+2\lambda_3 v^2 w+4 \lambda_2 w^3+3 \sqrt{2} \kappa_3 w^2
\right).\nonumber
\eea

The gauge eigenstates $\phi_{1,2}$ are then rotated to the mass eigenstates $h_{1,2}$ with the rotation matrix $R$ defined as
\be
\label{rotation-SMRS}
\phi_i=R_{ij} h_j, \quad
\left(
\begin{array}{c}
\phi_1\\ \phi_2\\
\end{array} \right)
=
\left(
\begin{array}{cc}
\cos\theta & \sin\theta \\
-\sin\theta & \cos\theta \\
\end{array} \right)
\left(
\begin{array}{c}
h_1\\ h_2\\
\end{array} \right),
\ee
where we take $h_1$ to be the observed 125 GeV Higgs boson. The mixing angle $\theta$ is calculated to be
\be
\tan(2\theta)= \frac{4 v w \left(\sqrt{2} \kappa_2+2 \lambda_3 w\right)}{2 \sqrt{2} \kappa_1+v^2 \left(8 \lambda_1 w+\sqrt{2}
\kappa_2\right)-w^2 \left(8 \lambda_2 w+3 \sqrt{2} \kappa_3\right)}.
\ee
The value of $\sin\theta$ is bounded by
experimental~\cite{TheATLASandCMSCollaborations:2015bln} and theoretical~\cite{Falkowski:2015iwa}
constraints 
to be
\be
\mid \sin\theta \mid < 0.33.
\label{sin-theta}
\ee
Throughout this paper we take the conservative limit of $\sin\theta \lesssim 0.3$ into account.

Note that the two neutral scalar mass eigenstates, $h_{1,2}$, are a mixture of
$\phi_{1,2}$ which are CP-even.
Clearly there is no possibility of introducing CP-violation explicitly (through complex parameters in the potential) or spontaneously (through a complex VEV of the doublet and/or singlet).
Hence, CP-violation is introduced
through a higher dimension operator~\cite{Cline:2012hg,Espinosa:2011eu}.
In the absence of a $Z_2$ symmetry, we take this to be the following dimension-5 operator,
\be
\mathcal{L}_{CPV} = \frac{\eta}{\Lambda} \;S \; \bar{Q}_L \; \tilde \Phi \; t_R + {\rm{h.c.}}
\label{higherorder}
\ee
where
\be 
\eta =\Re\eta+i \Im\eta,
\label{eta-definition}
\ee
is the complex CP-violating parameter,
$\Lambda$ is the scale of new physics generating the effective operator, $Q_L$ and $t_R$ are, respectively, the
left-handed doublet and right-handed
quarks of the SM. Note that the
sole source of CP-violation here is the parameter $\eta$, which is only introduced for the top quark couplings.

We use the conventional SM Yukawa couplings as defined by the Lagrangian,
\be
\mathcal{L}_{Yukawa}= Y_{ii}^f \bar f_{L,i} f_{R,i} \phi_1 +{\rm{h.c.}},
\ee
where, as clarified before, $\phi_1$ is the SM Higgs field.

To calculate $a_\mu$ and $d_e$ discussed in Section \ref{section-calculations}, one needs to identify the couplings of the mass eigenstates $h_{1,2}$ to leptons, quarks and the $W$ boson. These are:
\bea
&&Y^{h_i}_{ll}= R_{1i} \left(\frac{m_l}{v} \right) \quad (l=\mu \mbox{~for~} a_\mu, ~~ l=e \mbox{~for~} d_e)
\\
&&Y^{h_i}_{WW}= R_{1i} \left(\frac{2m^2_W}{v}\right), \quad Y^{h_i}_{qq}= R_{1i} \left(\frac{m_q}{v} \right)
\\
&&Y^{h_i}_{tt}= R_{1i} \left(\frac{m_t}{v}\right)+ R_{2i} \left(\frac{v(\Re\eta+ i \Im\eta)}{2\Lambda} \right),
\eea
where $R_{ij}$ are the components of the rotation matrix defined in Eq.~\eqref{rotation-SMRS}.
Note that the only complex coupling is $Y^{h_i}_{tt}=\Re(Y^{h_i}_{tt})+i\Im(Y^{h_i}_{tt})$ with
\be 
\Re(Y^{h_i}_{tt}) =  R_{1i} \left(\frac{m_t}{v}\right)+ R_{2i} \left(\frac{v \Re\eta}{2\Lambda} \right), 
\qquad 
\Im(Y^{h_i}_{tt}) =   R_{2i} \left(\frac{v \Im\eta}{2\Lambda} \right)
\ee
Following from Eq.~\eqref{amu-1loop}-\eqref{de-2-loop}, one can see that since $\Im(Y^{h_i}_{ll})=0$, only the imaginary part of $\eta$ contributes to $d_e$ and only the real part to $a_\mu$,
\be 
d_e \propto \Im\eta, \qquad a_\mu \propto \Re\eta,
\ee
as it will be shown in detail in the next two subsections.
We will therefore quantify our results in terms of the dimensionless quantities
$v \Re(\eta)/(2\Lambda)$ for $a_\mu$ and
$v\Im(\eta)/(2\Lambda)$ for $d_e$.
For the theoretical and experimental constraints, we have adopted the results in \cite{Falkowski:2015iwa}.

\subsection{$a_\mu$ in the SM+RS model}
As shown in detail in Section~\ref{section-calculations}, $a_\mu$ is proportional to the real part of the fermion-scalar couplings. Hence, all 1-loop and 2-loop contributions are non-zero and calculated to be
\begin{footnotesize}
\bea
a^{1-loop}_\mu &=&
-\frac{m_\mu^4}{8\pi^2 v^2}
\sum_{i=1}^2
\int_0^1 dx \int_0^x dy
\frac{ (y+1)(y-1) R_{1i}^2}{m_\mu^2\biggl[y(y-x)+(1-y) \biggr] +m_{h_i}^2 y},
\\
a_{\mu,~t}^{2-loop}&=&
\frac{2}{3}
\left(
\frac{\alpha G_F v m^2_\mu}{\sqrt{2}\pi^3 m_t}\right) \sum^2_{i=1}
\biggl[
R_{1i} \left(
R_{1i} (\frac{m_t}{v})+R_{2i} (\frac{v \Re\eta}{2\Lambda})\right)
f(z_{th_i})
\biggr],\nonumber
\\
a_{\mu,~W}^{2-loop}&=&-
\left(
\frac{\alpha G_F m^2_\mu}{4\sqrt{2}\pi^3}\right)
\sum^2_{i=1}
R_{1i}^2
\biggl[
3 f(z_{Wh_i}) +\frac{23}{4} g(z_{Wh_i})+\frac{3}{4} h(z_{Wh_i}) + \frac{f(z_{Wh_i})-g(z_{Wh_i})}{2 z_{Wh_i}}
\biggr], \nonumber
\\
a_{\mu,~f}^{2-loop}&=&
\frac{2}{3}
\left(
\frac{\alpha G_F  m^2_\mu}{\sqrt{2}\pi^3 }\right) \sum^2_{i=1}
\biggl[
R_{1i}^2
f(z_{fh_i})
\biggr], \quad (f \neq t)
\nonumber
\eea
\end{footnotesize}
where $R_{11}=\cos\theta$ and $R_{12}=\sin\theta$ are the elements of the rotation matrix in Eq.~\eqref{rotation-SMRS}.

\begin{figure}[h!]
\begin{center}
\includegraphics[scale=1]{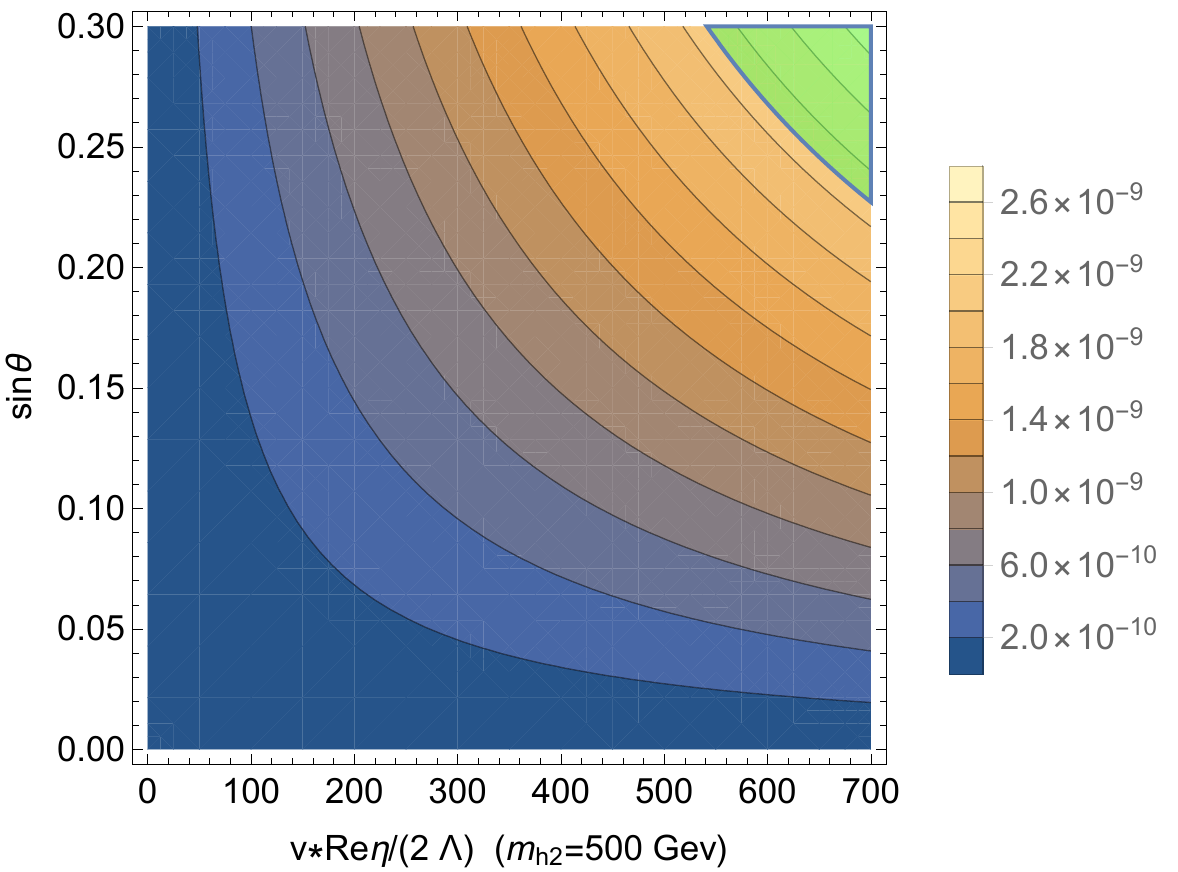}
\caption{The contours showing $a_\mu$ in the SM+RS model. The green region is where the model produces $a_\mu$ within the observed window in Eq.~\eqref{g-2-excess}. At $\eta=0$, where there is no $\mathcal{L}_{CPV}$, the model does not provide large enough contribution to $a_\mu$.}
\label{g-2-SMRS}
\end{center}
\end{figure}

We find that SM+RS model is incapable of explaining the muon anomalous moment
for $m_{h_2}$ of a few hundred GeV, even in the presence of a non-zero
$\mathcal{L}_{CPV}$. In Figure~\ref{g-2-SMRS}, we show contours of $a_\mu$ in the $\sin\theta$-$v\Re(\eta)/(2\Lambda)$-plane for a representative value of $m_{h_2}=500$ GeV. The green region is where the model produces $a_\mu$ within the observed window in Eq.~\eqref{g-2-excess}. Hence, one would need a very large, $\mathcal{O}(10^3)$, non-trivial coupling to top quark in Eq.~\eqref{higherorder}.
Note that the green region shown in the plot is not affected by $e$EDM constraints, which are governed by different couplings. We will discuss these constraints in detail in the next section.

\subsection{$d_e$ in the SM+RS model}
As mentioned before, the only CP-violating coupling is that of the top quark which is introduced in Eq.~\eqref{higherorder} through a dimension-5 operator.
Therefore the only $e$EDM contributions come from the 2-loop Barr-Zee diagrams mediated by the top quark as the 1-loop and $W$-mediated 2-loop diagrams are proportional to the imaginary part of the scalar-electron couplings and are hence zero,
\be
d_e^{1-loop}
\propto  \Im((Y^{hi}_{ee})^2) = 0, \hspace{5mm}
\mbox{and}
\hspace{5mm}
d_{e,~W}^{2-loop}
\propto \Im(Y^{hi}_{ee}) = 0.
\ee

From Eq.~\eqref{de-2-loop}, the 2-loop contributions from the top quark are calculated to be
\be
d_{e,~t}^{2-loop}=\frac{e}{3\pi^2}
\left(
\frac{\alpha G_F v}{\sqrt{2}\pi m_t}\right)
m_e \left(\frac{v \Im\eta}{2\Lambda}\right) \sin\theta \cos\theta
\biggl[-g(z_{th_1})+g(z_{th_2})
\biggr].
\label{de-SMRS}
\ee

\begin{figure}
\begin{center}
\includegraphics[scale=1]{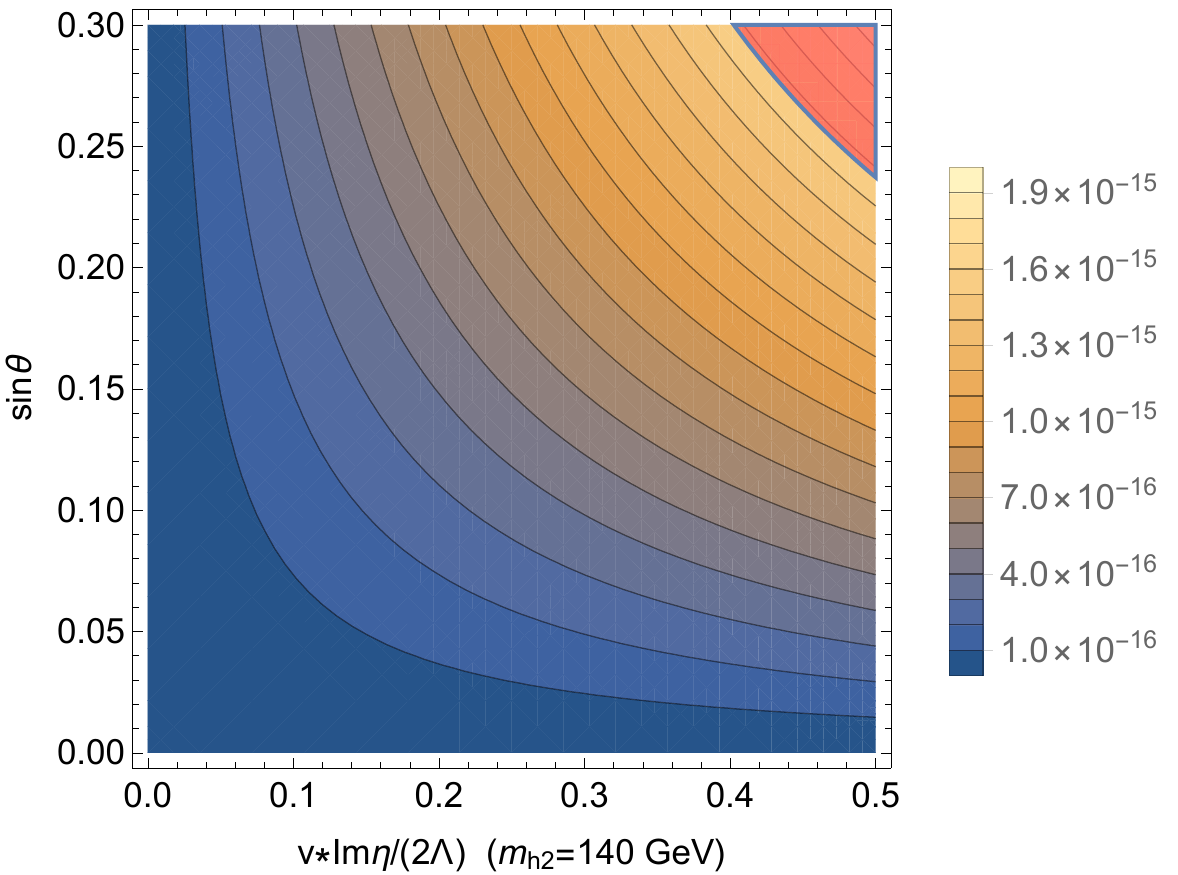}\\[4mm]
\includegraphics[scale=1]{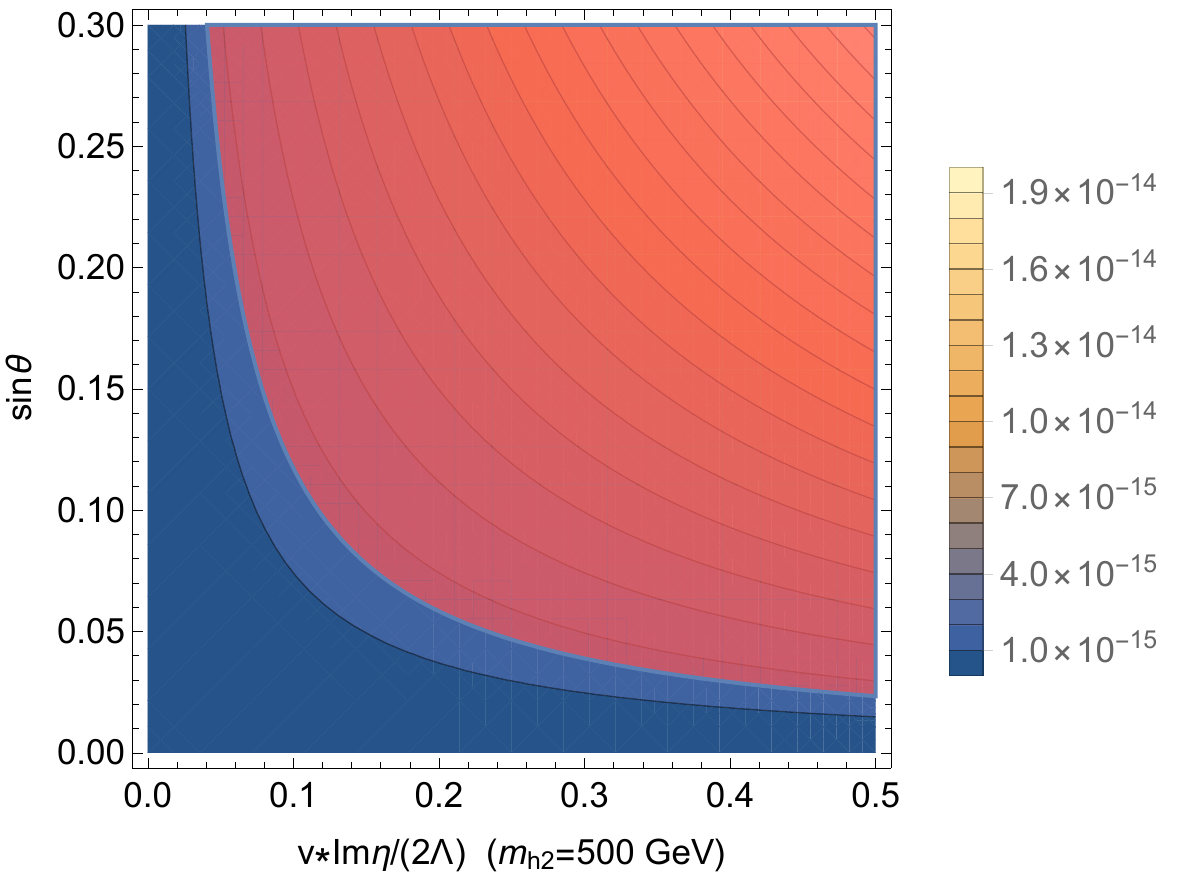}
\caption{The contours showing the $e$EDM contributions in the SM+RS model for $m_{h_2}=140$ (top) and  $m_{h_2}=500$ GeV (bottom). The red region is ruled out by experimental data.}
\label{SM+RS}
\end{center}
\end{figure}

In Figure~\ref{SM+RS}, we show contours of $d_e$ in the $\sin\theta$-$({v\Im\eta}/{2\Lambda})$-plane. The superimposed red regions are ruled out by the experimental bound in Eq.~\eqref{ACME}.
In the left panel of the figure $m_{h_2}=140$ GeV and in the right panel $m_{h_2}=500$ GeV. As predicted by Eq.~\eqref{de-SMRS}, when $m_{h_2} \approx m_{h_1} =125$ GeV,
and $g(z_{th_1})\approx g(z_{th_2})$, the 2-loop contributions to $d_e$ are reduced. Hence, as $m_{h_2}$ approaches $m_{h_1}$, a larger region of the parameter space will survive the $e$EDM bounds as shown by the smaller excluded red area in the left panel of Figure~\ref{SM+RS} in comparison to the right panel.

\subsection{Remark on complex singlet extension}
The results derived in the previous subsections are directly
applicable also to the case where SM is extended by a complex singlet
scalar (SM+CS).
It has been shown~\cite{Branco:1999fs} that an apparent CP violating phase
in a model with a scalar doublet and a complex singlet scalar can be rotated away, and no
explicit or spontaneous CP violation can be introduced in the SM+CS model. Similar to the SM+RS
model, the only CP violation would come from higher dimensional operators such as in
Eq.~\eqref{higherorder}.


The SM doublet $\Phi$ and SM singlet $S$ are defined as:
\be
\Phi = \left( \begin{array}{c} G^+ \\ \frac{1}{\sqrt{2}} \left( v + \phi_1 + i G^0 \right)\\ \end{array} \right),~~
S = \left( \frac{w  + \phi_2 + i \phi_3}{\sqrt{2}}  \right).
\label{dec-singlet}
\ee
with $v$ and $w$ as VEVs of the doublet and the singlet, respectively. Similar to the SM+RS case, $\Phi$ is the SM Higgs doublet with $G^\pm$ and $G^0$ as the Goldstone bosons.

The gauge eigenstates $\phi_{1,2,3}$ are rotated to the mass eigenstates $h_{1,2,3}$ through
 \be
 \label{rotation-SMCS}
 \phi_i=R_{ij} h_j, \quad
 \left(
 \begin{array}{c}
 \phi_1\\ \phi_2\\\phi_3\\
 \end{array} \right)
 =
 \left(
 \begin{array}{ccc}
 \cos\theta & \sin\theta & 0 \\
 -\sin\theta & \cos\theta & 0 \\
 0 & 0 & 1  \\
 \end{array} \right)
 \left(
 \begin{array}{c}
 h_1\\ h_2\\ h_3\\
 \end{array} \right).
 \ee

Due to its singlet nature, $\phi_3$ does not couple to the fermions and the $W$ boson\footnote{Through the higher order operator, $\mathcal{L}_{CPV}$, $\phi_3$ has a coupling to the top quark. However, it does not contribute to the Barr-Zee diagrams since it has no coupling to $e$ and $\mu$.}.
Therefore, $\phi_3$ does not influence the calculations of
the $d_e$ and $a_\mu$ in comparison to the SM+RS model and the results are identical to the ones presented in the preceding section.


\section{The Two Higgs Doublet Model (2HDM)}
\label{section-2hdm}

Extending the SM with one extra scalar doublet with the same
SM
quantum numbers as the SM-Higgs doublet\footnote{Note that extending the SM with a doublet with different charges, e.g. hypercharge, still leads to a 2HDM.}, one arrives at the well-studied
2HDM \cite{Lee:1973iz, Gunion:1989we, Branco:2011iw,Keus:2015hva, Zarikas:1995qb, Lahanas:1998wf, Aliferis:2014ofa}.
The most general 2HDM potential can be written in the following form:
\bea
\label{2hdm-pot}
V &=& -\mu^2_{1} (\Phi_1^\dagger \Phi_1) -\mu^2_2 (\Phi_2^\dagger \Phi_2)
-\biggl[ \mu^2_3(\Phi_1^\dagger \Phi_2) + {\rm{h.c.}} \biggr]
\\
&&
+ \lambda_{1} (\Phi_1^\dagger \Phi_1)^2
+ \lambda_{2} (\Phi_2^\dagger \Phi_2)^2
+ \lambda_{3} (\Phi_1^\dagger \Phi_1)(\Phi_2^\dagger \Phi_2)
 + \lambda_{4} (\Phi_1^\dagger \Phi_2)(\Phi_2^\dagger \Phi_1)
\nonumber\\
&&+ \biggl[\lambda_{5} (\Phi_1^\dagger \Phi_2)^2
+\lambda_{6}  (\Phi_1^\dagger \Phi_1) (\Phi_1^\dagger \Phi_2)
+\lambda_{7}  (\Phi_2^\dagger \Phi_2) (\Phi_1^\dagger \Phi_2)
+ {\rm{h.c.}} \biggr].\nonumber
\eea

In the general case, due to the existence of two scalar doublets to which fermions can couple, 2HDMs suffer from Flavour Changing Neutral Current interactions (FCNCs) at tree-level, which are strongly restricted experimentally.
It is known that imposing a softly broken $Z_2$ symmetry 
on the scalar potential, and extending it to the fermion sector can forbid these FCNCs \cite{Glashow:1976nt},\cite{Paschos:1976ay}.
Depending on the $Z_2$ charge assignment of the fermions, four independent types of Yukawa interactions are allowed, and these are known as Type-I, Type-II, Type-X (Lepton-specific) and Type-Y (Flipped) in the literature\cite{Gunion:1989we, Branco:2011iw, Barger:1989fj}, and references therein. These will be discussed in Section~\ref{section-yukawa}. In what follows, the transformation of the scalar doublets under this $Z_2$ symmetry is fixed to be $\Phi_1 \to + \Phi_1$ and $\Phi_2 \to - \Phi_2$.

Imposing the softly broken $Z_2$ symmetry forbids the $\lambda_{6,7}$ terms in the potential in Eq.~\eqref{2hdm-pot},
\be
\lambda_6 = \lambda_7 = 0.
\ee
The rest of the parameters are real with the exception of $\mu^2_3$ and $\lambda_{5}$ which are complex and defined as
\be
\mu_3^2=\Re\mu_3^2+ i\; \Im(\mu_3^2), \qquad \lambda_5=\Re\lambda_5 + i\; \Im\lambda_5 .
\ee
In this paper, we take the VEVs of the doublets to be real and positive and study explicit CP-violation which occurs when $\Im(\lambda_5^* [\mu_3^2]^2) \neq 0$~\cite{Haber:2006ue},
through the complex parameters of the potential.

In general, the decomposition of the scalar doublets is as follows
\be
\Phi_1= \doublet{$\begin{scriptsize}$ \phi^+_1 $\end{scriptsize}$}{\frac{v_1+h_1^0+ia^0_1}{\sqrt{2}}} ,\quad
\Phi_2= \doublet{$\begin{scriptsize}$ \phi^+_2 $\end{scriptsize}$}{\frac{v_2+h^0_2+ia^0_2}{\sqrt{2}}} ,
\label{fields-2hdm}
\ee
where $v_1$ and $v_2$ are taken to be real with $v^2=v_1^2+v_2^2=(246 \mbox{GeV})^2$ and, as usual, we define $\tan\beta=v_2/v_1$.

\subsection{Minimisation of the 2HDM potential}

The minimization of the potential implies
\bea
&&\mu^2_1 =
-\tan\beta \Re\mu^2_3
+v^2 s^2_\beta \Re\lambda_5
+\frac{v^2}{4}  \left(
2 \lambda_1+\lambda_3+\lambda_4
+c_{2 \beta} (2 \lambda_1-\lambda_3-\lambda_4)
\right)
\nonumber\\
&&\mu^2_2 =
-\cot\beta \Re\mu^2_3
+v^2 c^2_\beta \Re\lambda_5
+\frac{v^2}{4}  \left(
2 \lambda_2+\lambda_3+\lambda_4
+c_{2 \beta} (-2\lambda_2+\lambda_3+\lambda_4)
\right)
\nonumber\\
&&\Im\mu^2_3 =
v^2 s_\beta c_\beta \Im\lambda_5,
\eea
where $s_\beta$ and $ c_\beta$ stand for $\sin\beta$ and $\cos\beta$, respectively.

At this point, it is useful to rotate the doublets to the so called
Higgs basis~\cite{Davidson:2005cw},
\be
\label{higgs-basis}
\left(
\begin{array}{c}
\widehat{\Phi}_1\\ \widehat{\Phi}_2\\
\end{array} \right)
=
\left(
\begin{array}{ccc}
\cos\beta & \sin\beta  \\
-\sin\beta & \cos\beta \\
\end{array} \right)
\left(
\begin{array}{c}
\Phi_1\\ \Phi_2\\
\end{array} \right),
\ee
where only one of the doublets has a VEV
\be
\widehat{\Phi}_1= \doublet{$\begin{scriptsize}$ G^+ $\end{scriptsize}$}{\frac{v+\phi_1+iG^0}{\sqrt{2}}} ,\quad
\widehat{\Phi}_2= \doublet{$\begin{scriptsize}$ H^+ $\end{scriptsize}$}{\frac{\phi_2+i\phi_3}{\sqrt{2}}} ,
\label{fields}
\ee
and one can separate the Goldstone bosons, $G^\pm,G^0$, from the physical states. The mass of the charged Higgs is calculated to be
\be
m_{H^\pm}^2 =  \frac{\Re\mu_3^2}{s_\beta c_\beta}- \frac{v^2}{2} (\lambda_4
+2 \Re\lambda_5 ).
\ee
The neutral mass-squared matrix, $\mathcal{M}^2$, shown in detail in
Appendix~\ref{2hdm-details}, is a $3\times 3$ matrix which is diagonalised
by the rotation matrix $R$,
\be
\label{mass-2hdm}
R^T \mathcal{M}^2 R = \mathcal{M}^2_{\rm{diag}} = \mbox{diag}(m^2_{h_1}, m^2_{h_2}, m^2_{h_3}),
\ee
where we take $h_1$ to be the observed Higgs boson at the LHC with $m_{h_1}=125$ GeV.

The rotation matrix, $R$, depends on the three mixing angles, $\theta_{12},\theta_{13}$ and
$\theta_{23}$, where the latter two angles represent CP-violation and will vanish in the
CP-conserving limit. Therefore, we take these angles to be small since,
as it will be shown later, they prove to be very small in the interesting and allowed regions
of the parameter space. The angle $\theta_{12}$ represents the mixing of the SM-like Higgs with the
other CP-even state. As shown in Eq.~\eqref{sin-theta}, to agree with the observed Higgs
data, we take $\theta_{12}$ to be small.

With all mixing angles being small ($\cos\theta_i \simeq 1$ and $\sin\theta_i \simeq \theta_i$),
the rotation matrix, $R$, simplifies to the form
\be
\label{rotation-2hdm}
\phi_i=R_{ij} h_j,\qquad
\left(
\begin{array}{c}
\phi_1\\ \phi_2\\\phi_3\\
\end{array} \right)
=
\left(
\begin{array}{ccc}
1 & \theta_{12} & \theta_{13} \\
-\theta_{12} & 1 & \theta_{23} \\
-\theta_{13} & -\theta_{23} & 1 \\
\end{array} \right)
\left(
\begin{array}{c}
h_1\\ h_2\\ h_3\\
\end{array} \right).
\ee
With this simplified form, one can calculate the angles in terms of the parameters of the potential as shown in Appendix~\ref{2hdm-details}.

After minimisation, the 9 independent parameters of the model,
\be
\mu_1^2, ~\mu_2^2, ~\Re\mu_3^2, ~\lambda_1, ~\lambda_2, ~\lambda_3, ~\lambda_4, ~\Re\lambda_5, ~\Im\lambda_5,
\ee
can be expressed in terms of
\be
\tan\beta, ~v, ~m_{h_1}, ~m_{h_2}, ~m_{h_3}, ~m_{H^\pm}, ~\theta_{12}, ~\theta_{13}, ~\theta_{23},
\ee
which we take as input parameters for our numerical calculations.

\subsection{Yukawa and gauge couplings}
\label{section-yukawa}

In the general 2HDM, interactions of the scalar sector with SM fermions are defined as
\be
-\mathcal{L}_{Y} =
Y_u \bar{Q}'_L i \sigma_2 \Phi_u^* u'_R
+Y_d \bar{Q}'_L  \Phi_d d'_R
+Y_e \bar{L}'_L  \Phi_e e'_R +{\rm{h.c.}}
\label{yukawas}
\ee
where $\Phi_{u,d,e}$ are $\Phi_1$ and/or $\Phi_2$ depending on the type of Yukawa interactions. This correspondence is determined according to Table~\ref{Tab:type} after the $Z_2$ charge assignments for fermions have
been specified.

Starting from Eq.~\eqref{yukawas}, one rotates $\Phi_{1,2}$ to $\widehat{\Phi}_{1,2}$ in the Higgs basis using Eq.~\eqref{higgs-basis}.
The primed fermion gauge doublets and singlets, will have to be written in
terms of the unprimed mass eigenstates using the usual unitary matrices $U_L$
and $U_R$, which also diagonalise the fermion
mass and Yukawa matrices simultaneously. The Yukawa interactions can then be
written in the following
compact form
\bea
\label{yukawas-2hdm}
{\cal L}_{Y_{d}}
&&=
\bar d_L \frac{m_{d}}{v} d_R
\sum_i^3 \left( R_{1i}+\xi_{d} (R_{2i}+i\;R_{3i})\right)h_i, \\
{\cal L}_{Y_{l}}
&&=
\bar e_L
\frac{m_{l}}{v} e_R
\sum_i^3 \left( R_{1i}+\xi_{l} (R_{2i}+ i\;R_{3i})\right)h_i, \\
{\cal L}_{Y_{u}}
&&=
\bar u_L\frac{m_u}{v} u_R
\sum_i^3 \left( R_{1i}+\xi_{u} (R_{2i} -i\;R_{3i})\right)h_i,
\eea
where the $R_{ij}$ are the rotation matrix elements defined in Eq.~\eqref{rotation-2hdm}
and the coefficients $\xi_i$  are Type-specific as defined in Table~\ref{Tab:type}.

The scalar-gauge interactions are derived from the kinetic terms and are of the form
\bea
\mathcal{L}_{kin}
&=& \mid D_\mu \Phi_1 \mid^2 +
\mid D_\mu \Phi_2 \mid^2 \;
= \; \mid D_\mu \widehat{\Phi}_1 \mid^2 +
\mid D_\mu \widehat{\Phi}_2 \mid^2
\\
&\supset &
\frac{2m_W^2}{v}\;\phi_1 \; W_\mu W^\mu + \frac{m_Z^2}{v} \;\phi_1 \; Z_\mu Z^\mu
=
R_{1i} h_i \left( \frac{2m_W^2}{v} W_\mu W^\mu + \frac{m_Z^2}{v} Z_\mu Z^\mu \right) \nonumber
\eea
where, again, $R_{1i}$ are rotation matrix elements defined in Eq.~\eqref{rotation-2hdm}.

\begin{table}[h]
\begin{center}
\begin{tabular}{|c||c|c|c|c|c|c||c|c|c|}
\hline
& $\Phi_1$ & $\Phi_2$ & $u_R^{}$ & $d_R^{}$ & $e_R^{}$ & $Q_L$, $L_L$ & $\xi_d$ & $\xi_u$ & $\xi_l$
\\
\hline
Type-I  & $+$ & $-$ & $-$ & $-$ & $-$ & $+$  & $\cot\beta$ & $\cot\beta$ & $\cot\beta$
\\
Type-II & $+$ & $-$ & $-$ & $+$ & $+$ & $+$ & $-\tan\beta$ & $\cot\beta$ & $-\tan\beta$
\\
Type-X  & $+$ & $-$ & $-$ & $-$ & $+$ & $+$ & $\cot\beta$ & $\cot\beta$ & $-\tan\beta$
\\
Type-Y  & $+$ & $-$ & $-$ & $+$ & $-$ & $+$ & $-\tan\beta$ & $\cot\beta$ & $\cot\beta$
\\
\hline
\end{tabular}
\end{center}
\vspace{-5mm}
\caption{$Z_2$ charge assignment and $\xi$-coefficients in the Yukawa couplings of $d,u,l$ fermions in the four types of Yukawa interactions.}
\label{Tab:type}
\end{table}


In all the results that follow, we take into account theoretical and experimental bounds as shown in detail in Appendix \ref{constraints}. For our plots, we find it instructive to show a large region of $\cot\beta$, and point out, in each subsection, the regions that are ruled out experimentally.

\subsection{$a_\mu$ and $d_e$ in 2HDMs}

\subsubsection{General Type-independent formulas for $a_\mu$ and $d_e$}

The contribution from 1-loop diagrams to $a_\mu$ and $d_e$ are
\be
a^{1-loop}_\mu = -\frac{m_\mu^4}{8\pi^2 v^2}
\sum^3_{i=1}
\int_0^1 dx \int_0^x dy
\frac{ (y+1)(y-1) (R_{1i}+\xi_l R_{2i})^2 +(y-1)^2 (\xi_l R_{3i})^2}{m_\mu^2 [y(y-x)+(1-y) ] +m_{h_i}^2 y} ,
\ee
\be
\label{de-1loop-2hdm}
d_e^{1-loop}=
\frac{e\; m_e^3}{8\pi^2v^2}
\sum^3_{i=1}
\xi_l R_{3i} (R_{1i}+\xi_l R_{2i})
\int_0^1 dx
\int_0^x dy
\frac{(y-1) }{m^2_e [y(y-x)+(1-y)]+m^2_{h_i} y}.
\ee

The 2-loop contributions from up-Type and down-Type quarks, leptons  and the $W$ boson to $a_\mu$ are
\be
a_{\mu,~u}^{2-loop}=\frac{2}{3}
\left(
\frac{\alpha G_F m^2_\mu}{\sqrt{2}\pi^3 }\right)
\sum^3_{i=1}
\biggl[(R_{1i}+\xi_l R_{2i})(R_{1i}+\xi_u R_{2i})
f(z_{uh_i})
+\xi_l \xi_u R_{3i}^2
g(z_{uh_i})
\biggr],
\ee
\be
a_{\mu,~d}^{2-loop}=\frac{2}{3}
\left(
\frac{\alpha G_F m^2_\mu}{\sqrt{2}\pi^3 }\right)
\sum^3_{i=1}
\biggl[(R_{1i}+\xi_l R_{2i})(R_{1i}+\xi_d R_{2i})
f(z_{d h_i})
-\xi_l \xi_d R_{3i}^2
g(z_{d h_i})
\biggr],
\ee
\be
a_{\mu,~l}^{2-loop}=\frac{2}{3}
\left(
\frac{\alpha G_F m^2_\mu}{\sqrt{2}\pi^3 }\right)
\sum^3_{i=1}
\biggl[(R_{1i}+\xi_l R_{2i})^2
f(z_{l h_i})
- \xi_l^2 R_{3i}^2
g(z_{l h_i})
\biggr],
\ee
\be
a_{\mu,~W}^{2-loop}=-
\left(
\frac{\alpha G_F m^2_\mu}{4\sqrt{2}\pi^3}\right)
\sum^3_{i=1}
R_{1i}(R_{1i}+\xi_l R_{2i})
\biggl[
3 f(z_{Wh_i}) +\frac{23}{4} g(z_{Wh_i})+\frac{3}{4} h(z_{Wh_i}) + \frac{f(z_{Wh_i})-g(z_{Wh_i})}{2 z_{Wh_i}}
\biggr].
\label{amu-2loop-2hdm}
\ee

The 2-loop contributions from up-Type and down-Type quarks, leptons  and the $W$ boson to $d_e$ are
\be
d_{e,~u}^{2-loop}=
\frac{e \alpha G_F m_e}{3\sqrt{2}\pi^3 }
\sum^3_{i=1}
\biggl[
\xi_l R_{3i}(R_{1i}+\xi_u R_{2i})f(z_{u h_i})
- \xi_u R_{3i}(R_{1i}+\xi_l R_{2i})
g(z_{u h_i})
\biggr],
\ee
\be
d_{e,~d}^{2-loop}=
\frac{e \alpha G_F m_e}{3\sqrt{2}\pi^3 }
\sum^3_{i=1}
\biggl[
\xi_l R_{3i}(R_{1i}+\xi_d R_{2i})f(z_{d h_i})
+ \xi_d R_{3i}(R_{1i}+\xi_l R_{2i})
g(z_{d h_i})
\biggr],
\ee
\be
d_{e,~l}^{2-loop}=
\frac{e \alpha G_F m_e}{3\sqrt{2}\pi^3 }
\sum^3_{i=1}
\biggl[
\xi_l R_{3i}(R_{1i}+\xi_l R_{2i})f(z_{l h_i})
+ \xi_l R_{3i}(R_{1i}+\xi_l R_{2i})
g(z_{l h_i})
\biggr],
\ee
\be
d_{e,~W}^{2-loop}=-
\frac{e\alpha G_F m_e}{8\sqrt{2}\pi^3}
\sum^3_{i=1}
R_{1i}
\xi_l R_{3i}
\biggl[
3 f(z_{Wh_i}) +\frac{23}{4} g(z_{Wh_i})+\frac{3}{4} h(z_{Wh_i}) + \frac{f(z_{Wh_i})-g(z_{Wh_i})}{2 z_{Wh_i}}
\biggr].
\ee

Note that these results are \textit{type-independent}: each type of 2HDM
can be studied further numerically when the corresponding values of $\xi_l, \xi_d$
and $\xi_u$ presented in Table~\ref{Tab:type} are implemented.

\subsubsection{The numerical formulas for $a_\mu$ and $d_e$ for given masses}

To gain insight into how the constraints on $a_\mu$ and $d_e$ operate in
different models, it is instructive to look at the explicit numerical form of the total $a_\mu$ and $d_e$ contributions. Here, we present explicitly
the numerical formulas for exemplary values of $m_{h_{2,3}}=200,300$ GeV; of course,
the formulas corresponding to any other mass texture can be easily produced
from the general results presented in the preceding subsection.

The total contribution from the scalars to $a_\mu$ is
\bea
a_\mu &=& 10^{-11} \biggl[
-1.7
+\xi_l \xi_u(2.2+1.4\theta_{12}^2+2.0\theta_{13}^2+2.3\theta_{23}^2)
+ \xi_l(0.8\theta_{12}
-0.3\theta_{13}\theta_{23})
\nonumber\\
&&
+\xi_u(-0.3\theta_{12}
+ 0.8\theta_{13}\theta_{23})
-0.9\;\theta^2_{12}
-0.3\;\theta^2_{13}
\\
&&
+\xi_l \xi_d(0.5+3.2\theta_{12}^2-4.0\theta_{13}^2-1.1\theta_{23}^2)\times 10^{-2}
+\xi_d(-0.1\theta_{12}
+ 0.8\theta_{13}\theta_{23})\times 10^{-2}
\nonumber
\\
&&
+\xi^2_l (0.2+1.1\theta_{12}^2-1.3\theta_{13}^2-0.3\theta_{23}^2)
\times 10^{-2} \biggr],
\nonumber
\eea
where the last two lines are the contributions of the down-Type quarks (mostly $b$) and charged leptons (mostly $\tau$) to the Barr-Zee diagrams which clearly are sub-dominant.
Hence, Type-I and Type-Y (and similarly, Type-II and Type-X), whose only difference is in $\xi_d$, contribute almost identically to $d_e$ and $a_\mu$, especially when $\xi_{l,d}$ are not very large as it will be clarified further here.

\begin{figure}[h!]
\begin{center}
\hspace{-60mm}\includegraphics[height=70mm]{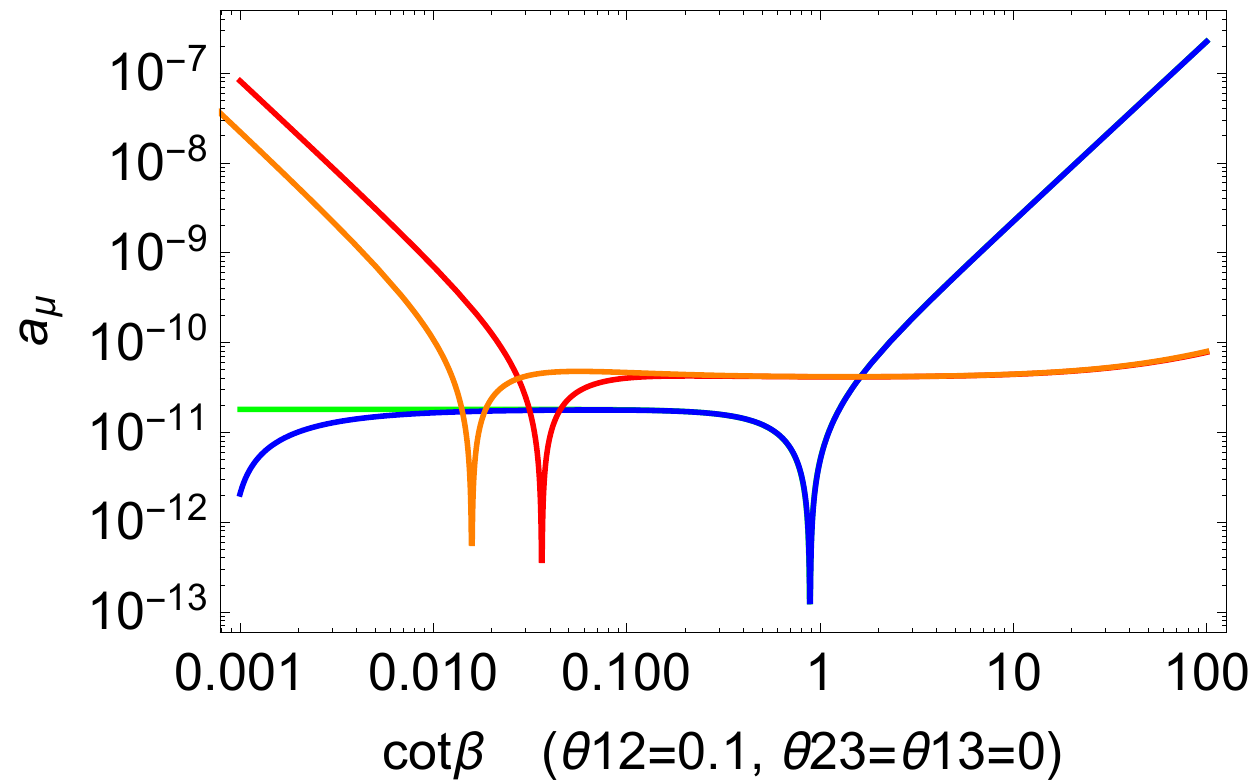}\\[5mm]
\hspace{-30mm}\includegraphics[height=70mm]{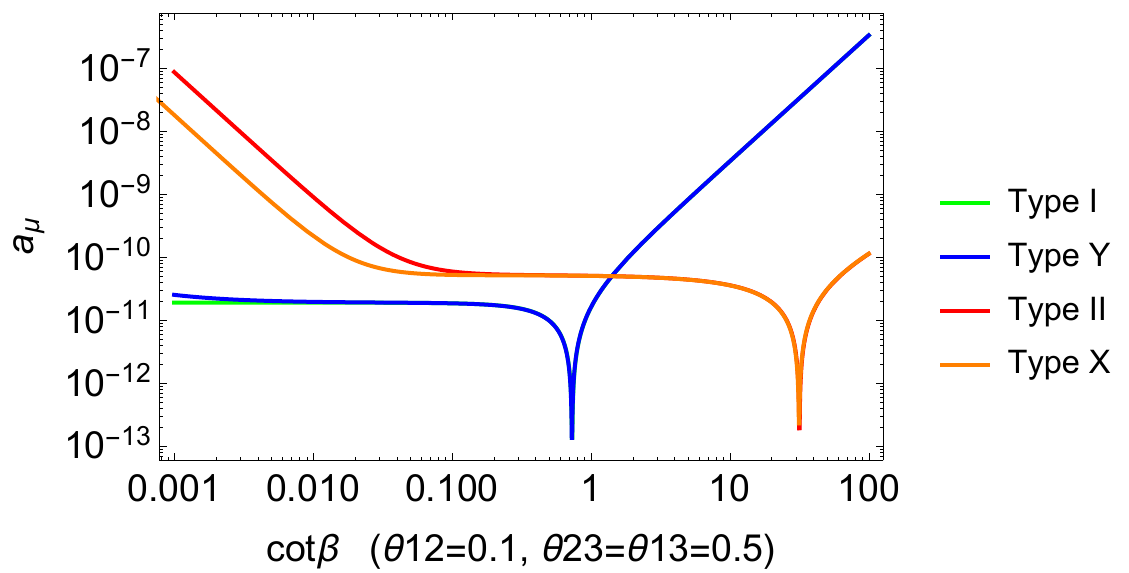}\\[5mm]
\includegraphics[height=70mm]{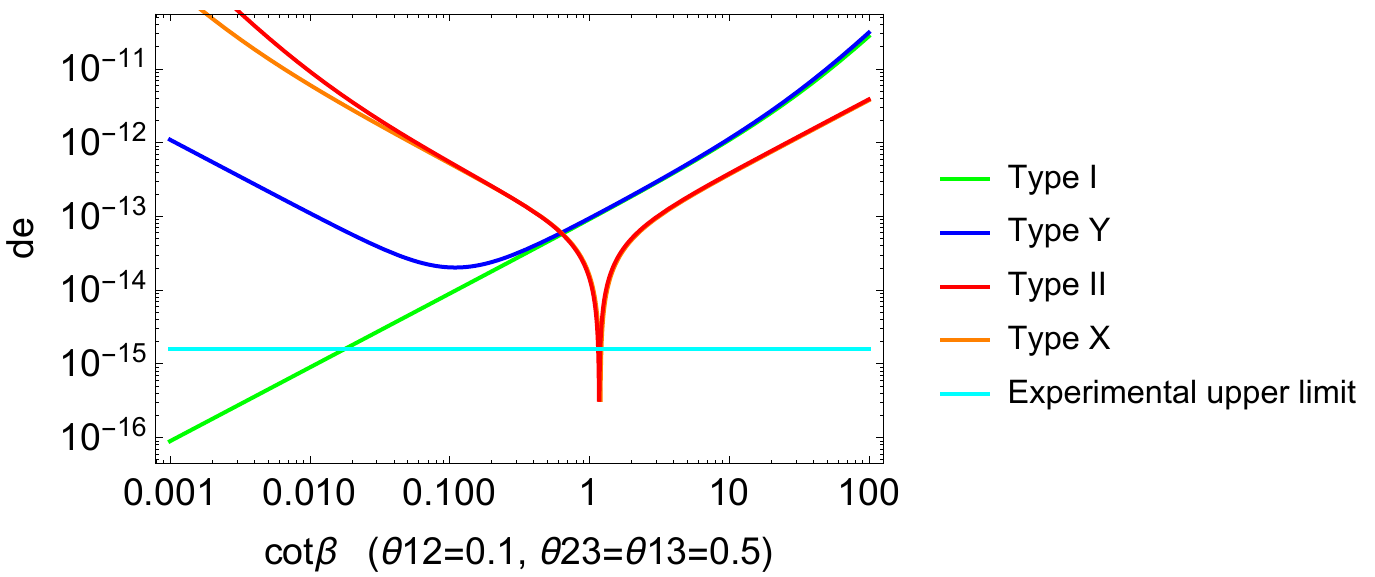}
\caption{$a_\mu$ (top) and $d_e$ (bottom) contribution in different 2HDM Types for fixed values of masses ($m_{h_{2,3}}=200,300$ GeV) in the CP-conserving limit (left) and in the presence of CP-violation (right).
}
\label{2HDM-amu-heavy}
\end{center}
\end{figure}

To see the exact difference between Type-I and Y (and similarly Type-II and X), we show the explicit numeric formulas in each case. For Type I,
\be
\label{numeric-amu-typeI}
a_\mu^{\mbox{I}} = 10^{-11} \biggl[
-1.7+
\cot^2\beta(2.2+1.4\theta_{12}^2+2\theta_{13}^2+2.3\theta_{23}^2)
+ \cot\beta(0.4\theta_{12}
+ 0.4\theta_{13}\theta_{23})
-0.9\theta^2_{12}
-0.3\theta^2_{13}
\biggr],
\ee
and for Type Y,
\be
\label{numeric-amu-typeY}
a_\mu^{\mbox{Y}} =
a_\mu^{\mbox{I}}
+10^{-11} \biggl[\biggl(-0.5+4\theta_{13}^2+\theta_{23}^2
+\cot^2\beta(-0.5+\theta_{23}^2)
+\tan\beta (\theta_{12}-0.8\theta_{13}\theta_{23})
\biggr)\times 10^{-2}
\biggr],
\ee
which has a subdominant correction with respect to Type-I. Note that when $\tan\beta < 10^{2}$, this correction is negligibly small, as it is shown in Figure~\ref{2HDM-amu-heavy}.
From these equations we see that Type-I (and Y) are capable of producing a large enough $a_\mu$ in the $\cot\beta \gtrsim 10$ region.

For Type II,
\bea
\label{numeric-amu-typeII}
a_\mu^{\mbox{II}} &=& 10^{-11} \biggl[
- 3.9
+\cot\beta(-0.3\theta_{12}
+ 0.8\theta_{13}\theta_{23})
+\tan\beta(-0.8\theta_{12}
+0.3\theta_{13}\theta_{23})
-2.3(\theta^2_{12}
+\theta^2_{13}+\theta_{23}^2)
\nonumber\\
&&
+\tan^2\beta\biggl(
0.7+4\theta_{12}^2-5\theta_{13}^2-1.3\theta_{23}^2
\biggr)\times 10^{-2}
\biggr],
\eea
and for Type X
\bea
\label{numeric-amu-typeX}
a_\mu^{\mbox{X}} &=&
a_\mu^{\mbox{II}}
+10^{-11} \biggl[\biggl(-0.5+\theta_{23}^2+
\tan^2\beta(-0.5-3\theta_{12}^2-6\theta_{13}^2+\theta_{23}^2)
\biggr)\times 10^{-2}
\biggr],
\eea
which again has a sub-dominant correction to Type-II. Note that when $\tan\beta < 10$, this correction is negligibly small.
From these equations we see that Type-II (and X) are capable of producing a large enough $a_\mu$ in the $\cot\beta \gtrsim 100$ and $\tan\beta \gtrsim 100$ regions.
These findings are summarised in Figure~\ref{2HDM-amu-heavy}, where we show the $a_\mu$ contributions in the CP-conserving limit ($\theta_{13}=\theta_{23}=0$) on the top panel and in the presence of CP-violation on the middle panel.

On the other hand, keeping only the leading terms, the total $e$EDM contributions are
\bea
d_e &=& 10^{-14}
\biggl|
\xi_l\biggl(
9.6 \theta_{13}
+6.6 \theta_{23} \theta_{12}
\biggr)
+ \xi_u \biggl(
6.6 \theta_{13}
+10.6 \theta_{23} \theta_{12}
\biggr)
+ \xi_l \xi_u \biggl(
0.8 \theta_{23}
-4.3 \theta_{13} \theta_{12} \biggr)
\nonumber\\
&&
+ \xi_d \biggl(
-0.2 \theta_{13}
-0.1 \theta_{23} \theta_{12}
\biggr)
+ \xi_l \xi_d \biggl(
-0.1 \theta_{23}
+0.5 \theta_{13} \theta_{12} \biggr)
+ \xi^2_l \biggl(
-0.03 \theta_{23}
+0.1 \theta_{13} \theta_{12} \biggr)
\biggr|
\nonumber\\
\eea
The bound in Eq.~\eqref{ACME} then gives the following constraints: For Type I,
\be
\label{numeric-de-typeI}
\biggl|
\cot\beta (16 \theta_{13}+17 \theta_{23}\theta_{12} )
+\cot^2\beta (0.7 \theta_{23} -3.6 \theta_{13}\theta_{12})
\biggr|
< 0.15
\ee
and for Type Y,
\bea
\label{numeric-de-typeY}
&&\biggl|
\cot\beta (16.2 \theta_{13}+17.2 \theta_{23}\theta_{12} )
+\cot^2\beta (0.8 \theta_{23} -4.1 \theta_{13}\theta_{12})
\nonumber\\
&&~~
+\tan\beta (0.2 \theta_{13}+0.1 \theta_{23}\theta_{12} )
+0.1 \theta_{23}-0.5 \theta_{12}\theta_{13}
\biggr|
< 0.15.
\eea
Note that the difference between the $d_e$ contribution in Type-I and Y is proportional to $\tan\beta$ whose effect is visible in the low $\cot\beta$ region in Figure~\ref{2HDM-amu-heavy}.

To satisfy these constraints, in both Type-I and Y, one requires small $\cot\beta$. Note also that when $\cot\beta$ is small, $\tan\beta$ is large which makes the $d_e$-surviving region in Type-Y more constrained when
compared to Type-I.

For Type II,
\bea
\label{numeric-de-typeII}
&&\biggl|
\cot\beta
(6.6 \theta_{13}+10.6 \theta_{23}\theta_{12})
+(-0.8 \theta_{23} +4.3\ \theta_{13}\theta_{12})
\nonumber\\
&&~~
+\tan\beta
(- 9.4 \theta_{13}-6.4 \theta_{23}\theta_{12} )
+\tan^2\beta
(- 0.1 \theta_{23}+0.6 \theta_{13}\theta_{12} )
\biggr|
< 0.15
\eea
For Type X,
\bea
\label{numeric-de-typeX}
&&\biggl|
\cot\beta
(6.4 \theta_{13}+10.4 \theta_{23}\theta_{12})
+(-0.7 \theta_{23} +3.8\ \theta_{13}\theta_{12})
\nonumber\\
&&~~
+\tan\beta
(- 9.6 \theta_{13}-6.6 \theta_{23}\theta_{12} )
+\tan^2\beta
(- 0.03 \theta_{23}+0.1 \theta_{13}\theta_{12} )
\biggr|
< 0.15
\eea
whose contributions are very similar to each other, with both types surviving the $d_e$ constraints in the $\tan\beta \approx \cot\beta \approx 1$ region.
The similarities of Type II and X are also visible in Figure~\ref{2HDM-amu-heavy} where the two types only differ slightly in the low $\cot\beta$ region. 

Superimposing the $a_\mu$ and $d_e$ plots, one can see that with heavy scalars, it is not possible to have a large enough $a_\mu$ contribution with the amount of CP-violation that is allowed by the $e$EDM data.

We emphasize that the above numerical formulas are presented for exemplary values of masses $m_{h_{2,3}}=200,300$ GeV. In the next subsection, we will
analyse different mass hierarchies in more detail.

\subsection{2HDM Results}
\label{2HDM-results}
We divide this section into three subsections dealing with heavy ($m_{h_{2,3}} \gtrsim m_{h_1}$), medium ($m_{h_{2,3}} \approx m_{h_1}$) and light ($m_{h_{2,3}} \lesssim m_{h_1}$) mass regions.

\subsubsection{Heavy mass region}

\begin{figure}[h!]
\begin{center}
\hspace{-52mm}\includegraphics[height=50mm]{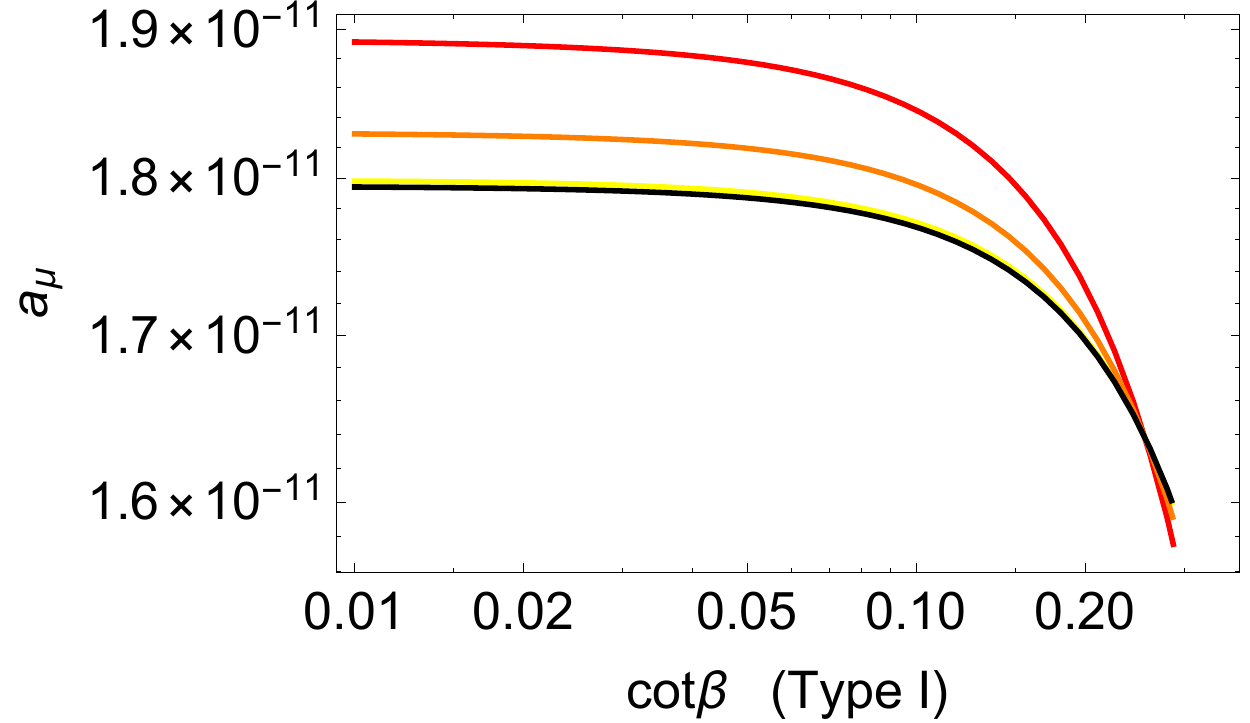}\\[4mm]
\hspace{-2mm}\includegraphics[height=50mm]{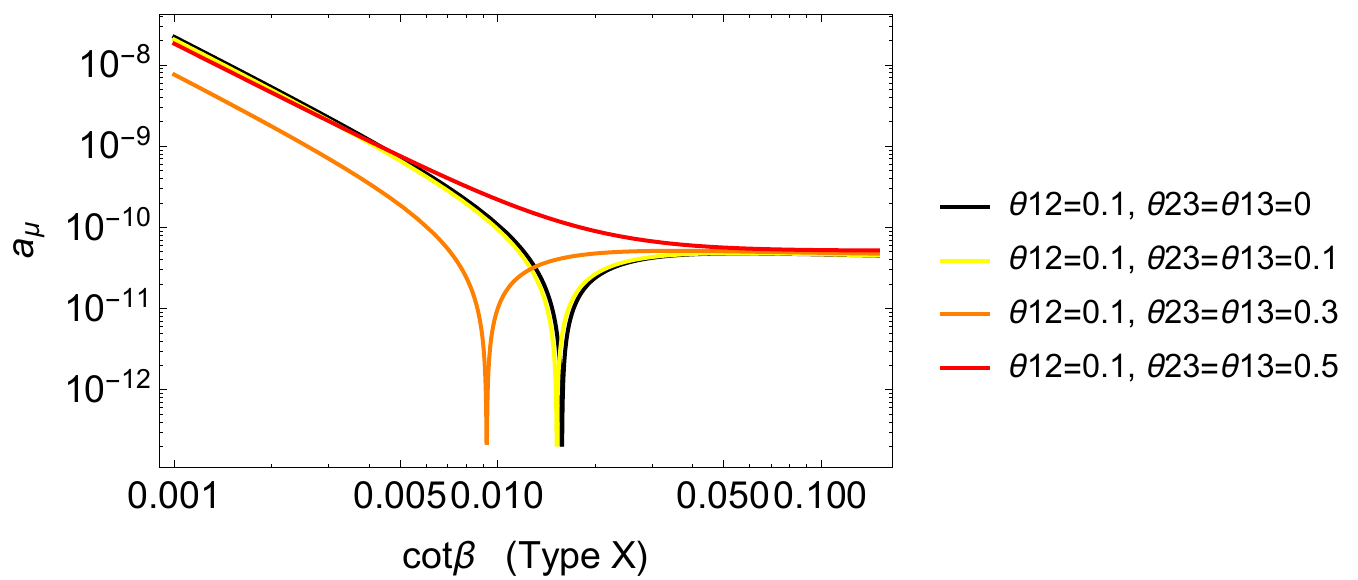}\\[4mm]
\hspace{-47mm}\includegraphics[height=50mm]{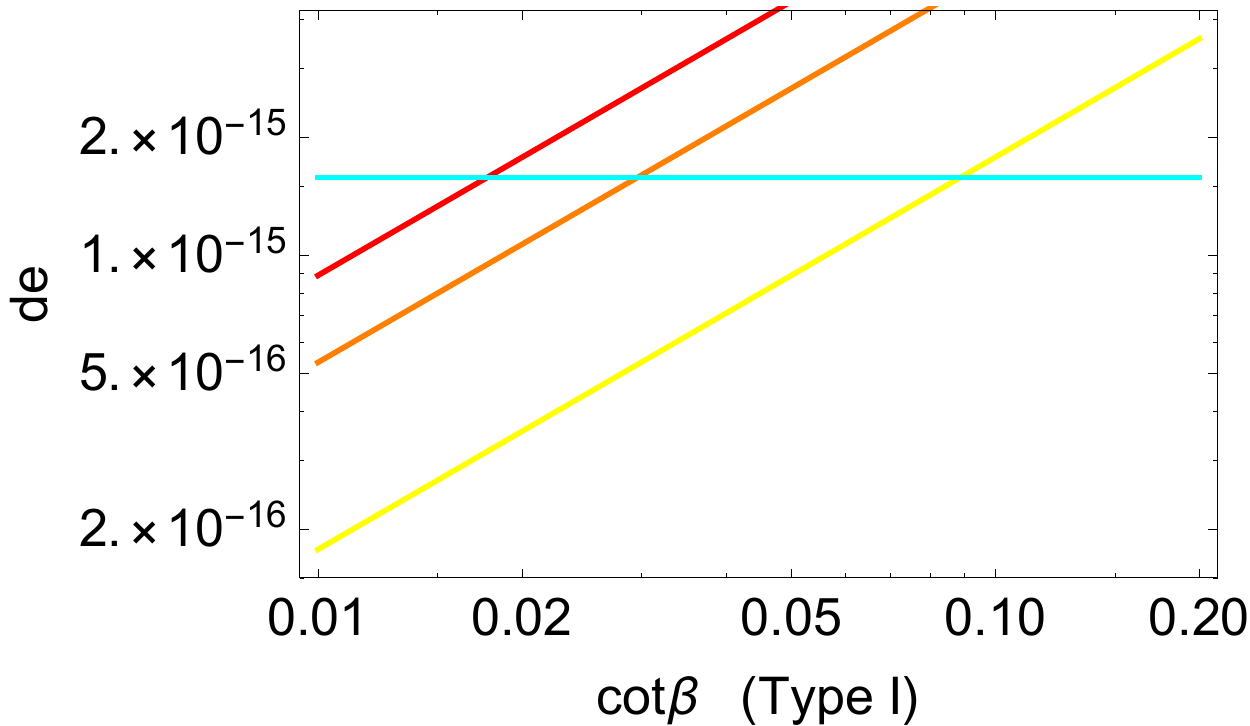}\\[4mm]
\includegraphics[height=50mm]{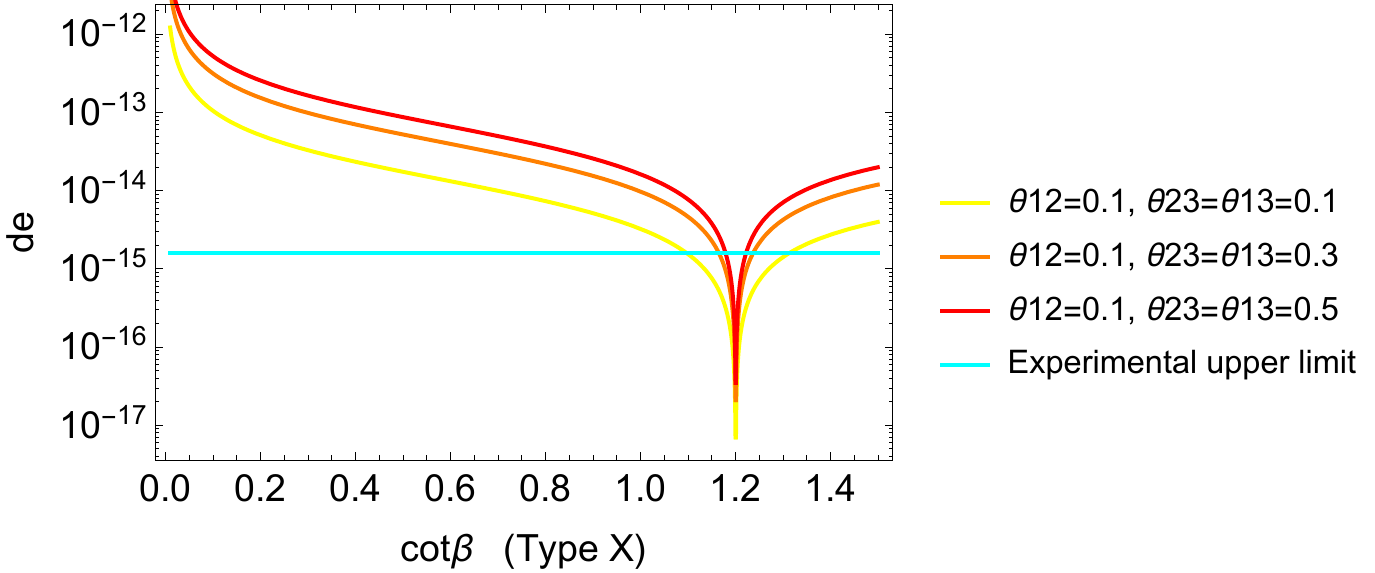}
\caption{$a_\mu$ (top) and $d_e$ (bottom) values in Type I (left) and Type X (right) for different values of angles and fixed values of masses ($m_{h_{2,3}}=200,300$ GeV). The behaviour of Type Y and II are similar to Type I and X, respectively.}
\label{amu-250200}
\end{center}
\end{figure}

To investigate the effect of CP-violation more closely, in Figure~\ref{amu-250200}, we show the $a_\mu$ and $d_e$ contributions for different values of the CP-violating angles, $\theta_{13}$ and $\theta_{23}$ for Type I and Type X for fixed scalar masses, $m_{h_{2,3}}=200, 300$ GeV. The black lines show the $a_\mu$ contribution of each model in the CP-conserving limit and the cyan line shows the experimental upper limit on the $d_e$ contribution.
Note that in Type I, larger CP-violating angles lead to larger $a_\mu$ values while the effect is more complicated and $\cot\beta$-dependent in Type X.
Clearly with increasing CP-violation, the $d_e$ contribution increases and the surviving region of the parameter space shrinks.
As mentioned before, the behaviour of Type Y and II are, respectively, similar to Type I and X.

\begin{figure}[t!]
\begin{center}
\includegraphics[scale=0.55]{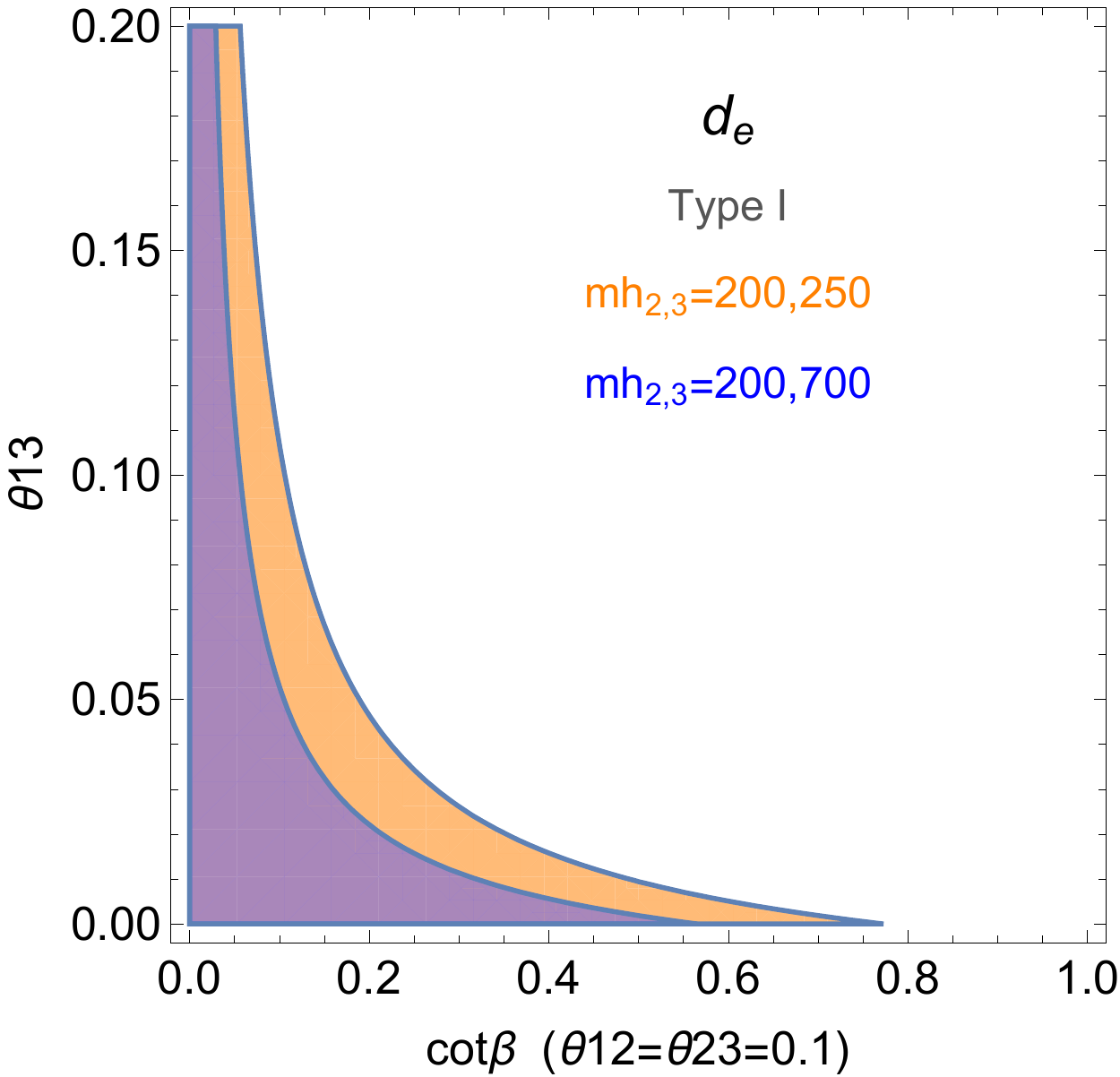} \hspace{10mm}
\includegraphics[scale=0.55]{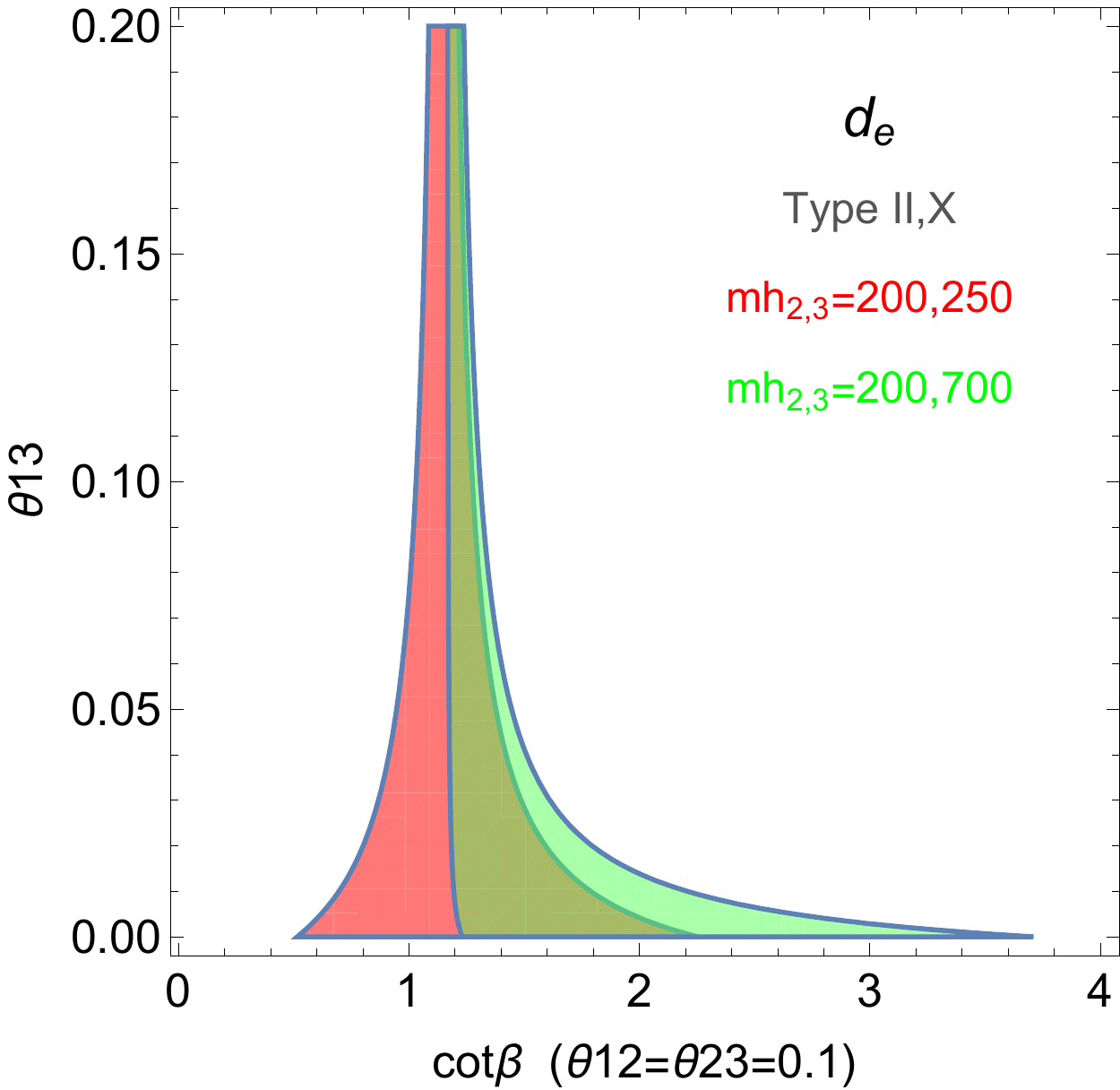}\\[3mm]
\includegraphics[scale=0.55]{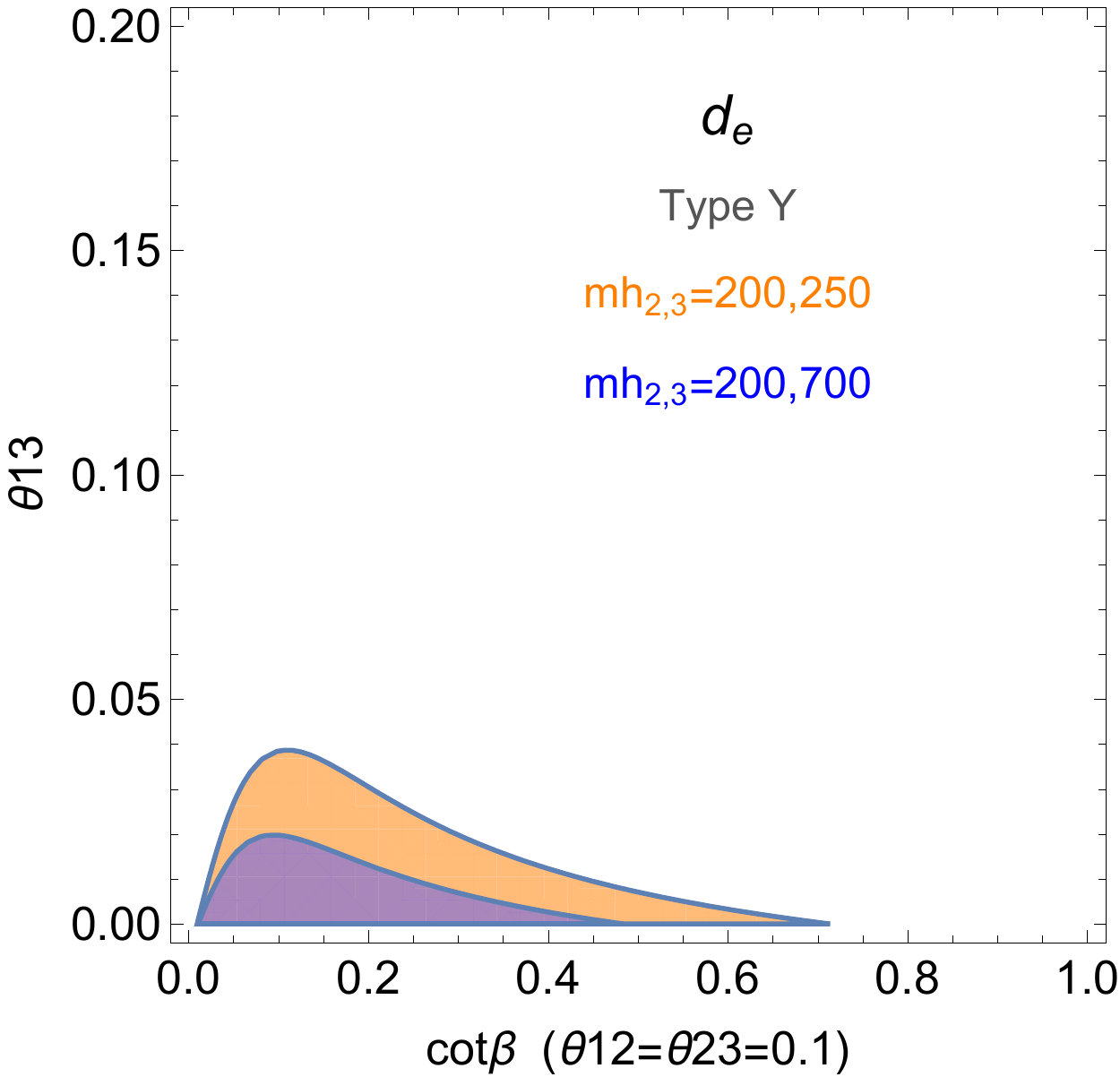} \hspace{10mm}
\includegraphics[scale=0.55]{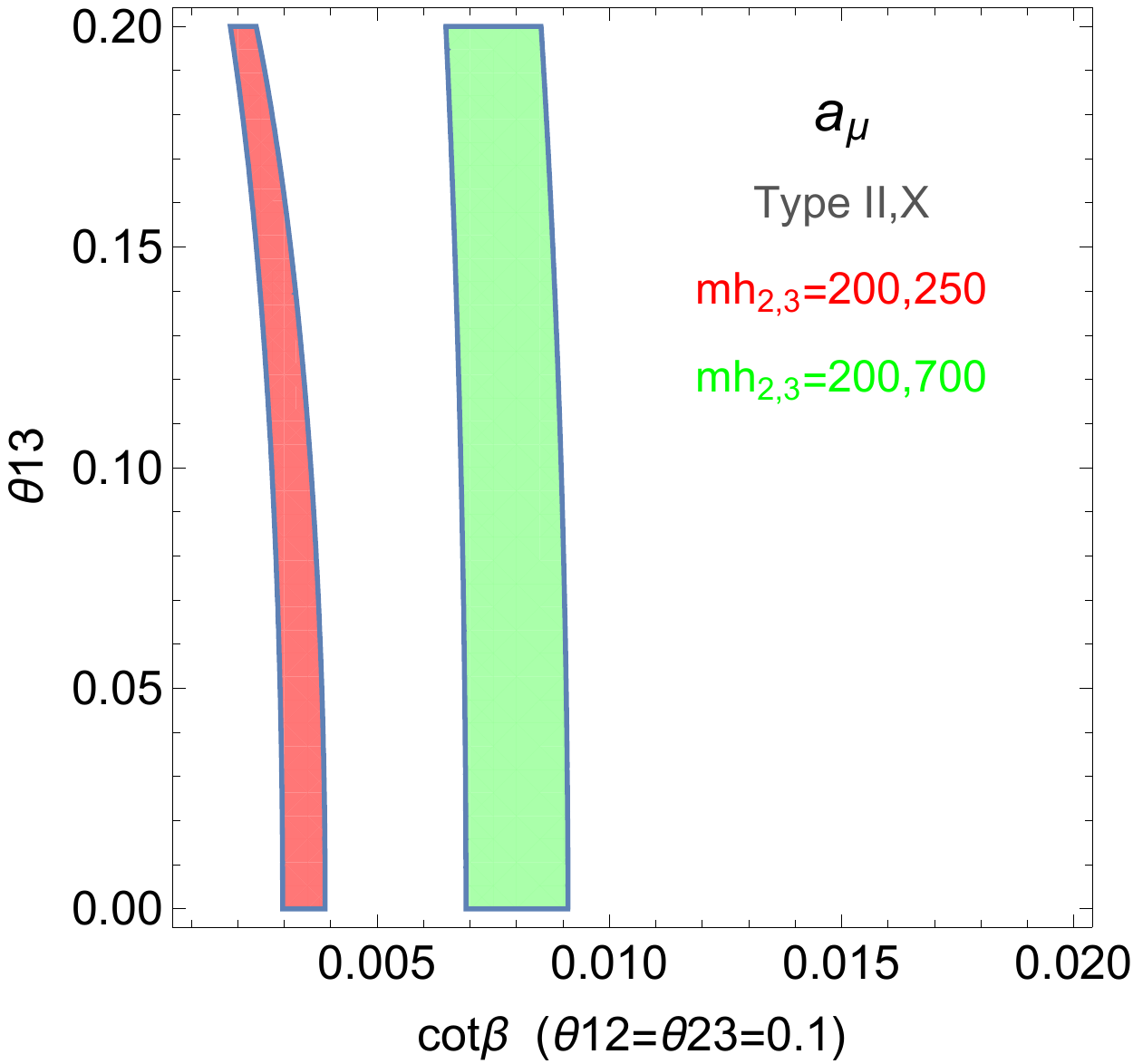}\\[3mm]
\includegraphics[scale=0.55]{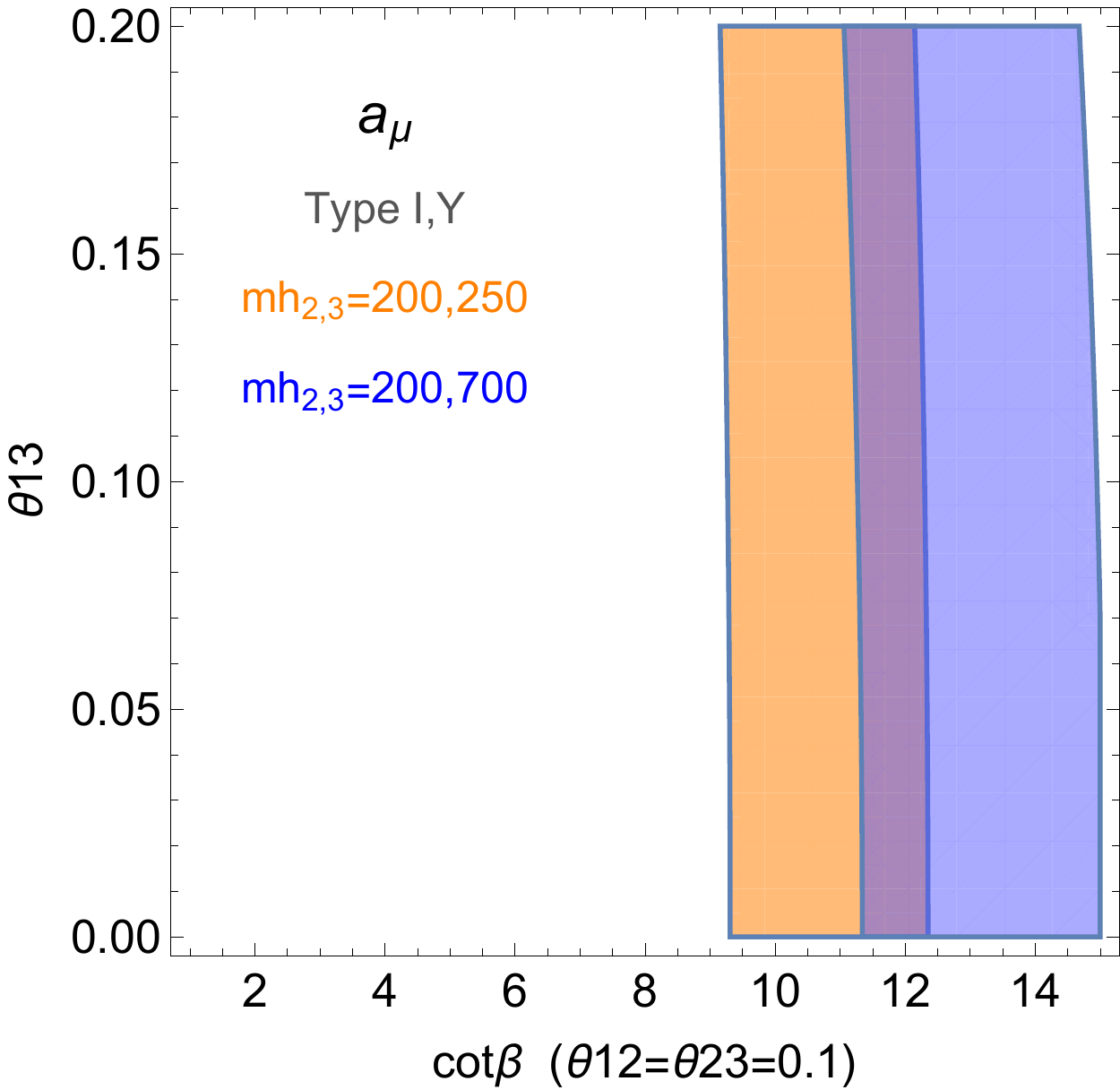} \hspace{10mm}
\includegraphics[scale=0.55]{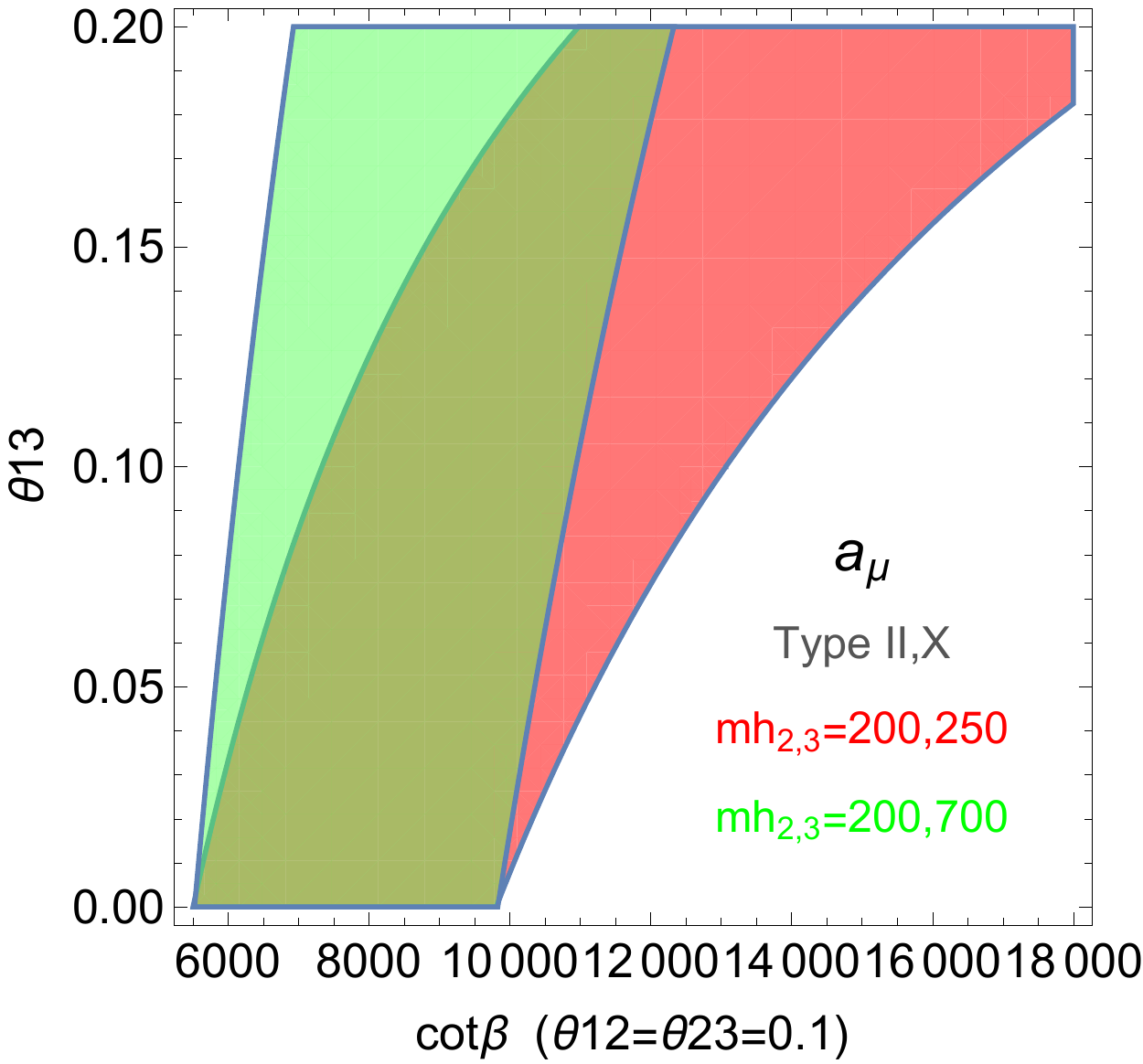}
\caption{Regions surviving $d_e$ bounds vs. regions producing $a_\mu$ with the deviation observed for Type I,Y (left) and Type II,X (right),
for different $m_{h_{2,3}}$ masses (in GeV) and fixed values of $\theta_{12}$ and $\theta_{23}$.}
\label{IY-de-amu-heavy}
\end{center}
\end{figure}

To study the effect of the scalar masses, in Figure~\ref{IY-de-amu-heavy}, we show the regions surviving the $d_e$ constraint and regions producing $a_\mu$ within the observed band in Type I, Y, II and X 2HDMs for two sets of scalar masses and fixed
values of $\theta_{12}=\theta_{23}=0.1$.
Types I and Y show the expected
behaviour in agreement with Figure~\ref{2HDM-amu-heavy}: Type Y is more constrained by $d_e$ in comparison to Type I due to the contribution proportional to $\tan\beta$
(see Eq.~\eqref{numeric-de-typeY}), which is large in
small $\cot\beta$ region. However, the contributions to $a_\mu$ are
almost identical in both types of models. Clearly the $a_\mu$ bands do not overlap with the $d_e$ surviving regions in this case.
Type II and X contribute almost identically to both $a_\mu$ and $d_e$: there are two regions, very small $\cot\beta$ and very large $\cot\beta$ which lead to the correct value for $a_\mu$ in agreement with Figure~\ref{2HDM-amu-heavy}. However, none of these regions pass the $d_e$ bounds which are satisfied in the $\cot\beta \approx 1$ as also confirmed by Fig.~\ref{2HDM-amu-heavy}.

Aside from the $e$EDM constraints, note that large values of $\cot\beta$ lead to large scalar-fermion couplings which are ruled out due to flavour and/or collider constraints. It has been shown in the CP-conserving limit in \cite{Mahmoudi:2009zx, Cherchiglia:2016eui, Broggio:2014mna, Abe:2015oca} that due to these constraints only Type I and X models survive in the low $\cot\beta$ region.

\subsubsection{Medium mass region}

\begin{figure}[h!]
\begin{center}
\hspace{-60mm}\includegraphics[height=70mm]{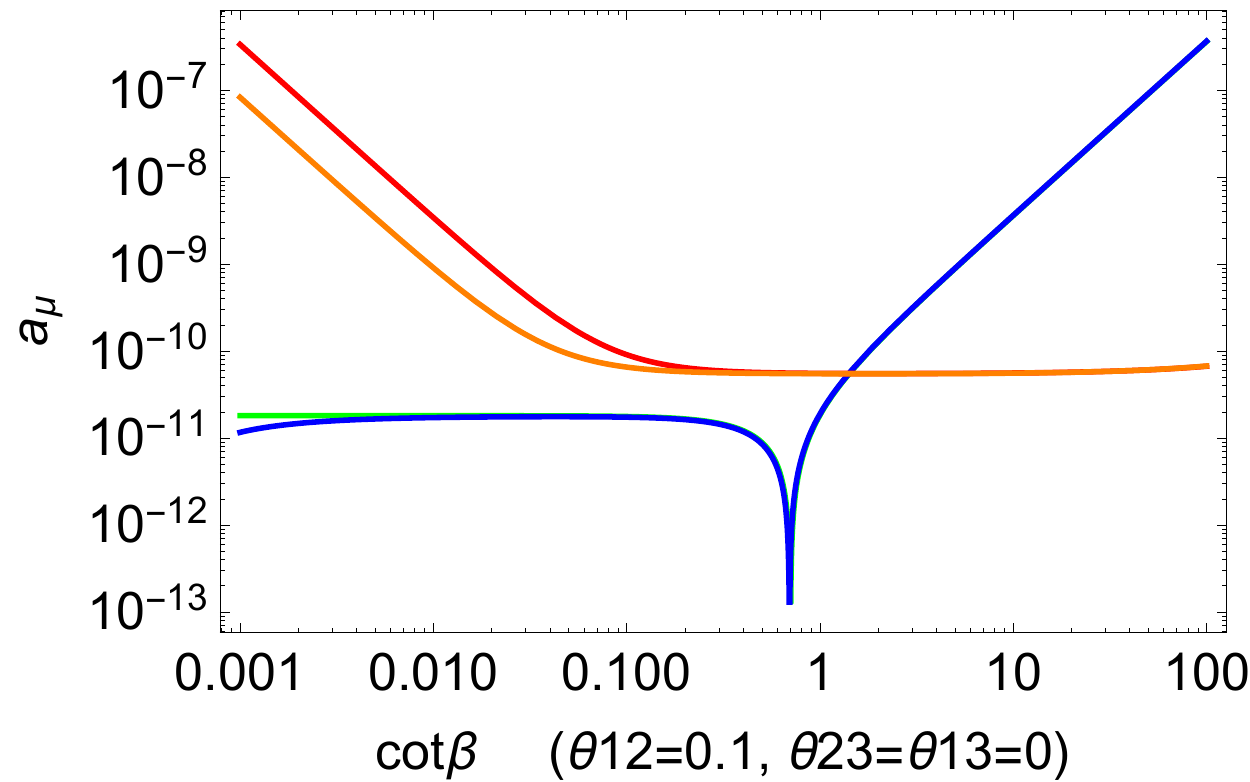}\\[5mm]
\hspace{-30mm}\includegraphics[height=70mm]{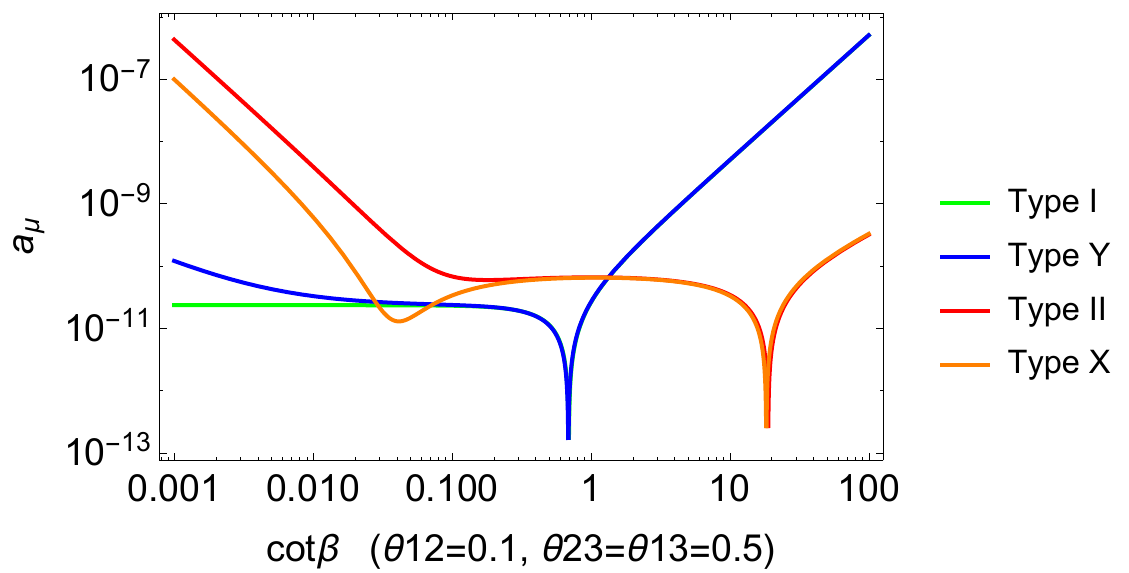}\\[5mm]
\includegraphics[height=70mm]{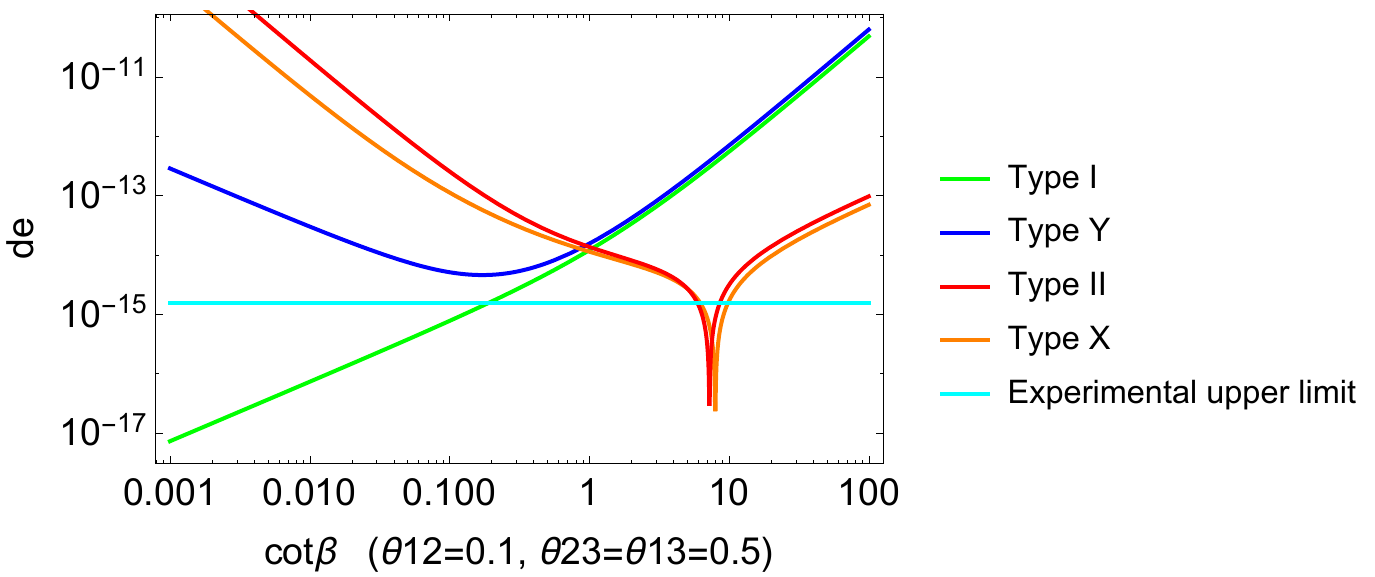}
\caption{$a_\mu$ (top) and $d_e$ (bottom) values in different 2HDM Types for fixed values of angles and masses ($m_{h_{2,3}}=145,105$ GeV). }
\label{amu-de-145105}
\end{center}
\end{figure}

Next we turn to the medium mass region where all scalars have masses comparable with $m_{h_1}$. In Figure~\ref{amu-de-145105}, we show the behaviour of all four types of 2HDM over a large range of $\cot\beta$ values for fixed values of the angles. The behaviour is similar to the heavy mass region with a significant contribution to $a_\mu$ in the large $\cot\beta$ region in Type I, Y and in the small $\cot\beta$ region in Type II, X, both with and without CP-violation.
The $d_e$ contributions are also similar to the heavy scalar case with Type I, Y favouring the low $\cot\beta$ region while Type Y is more constrained, and with Type II, X leaning towards the $\cot\beta \approx 10$ region.

\begin{figure}[h!]
\begin{center}
\hspace{-52mm}\includegraphics[height=50mm]{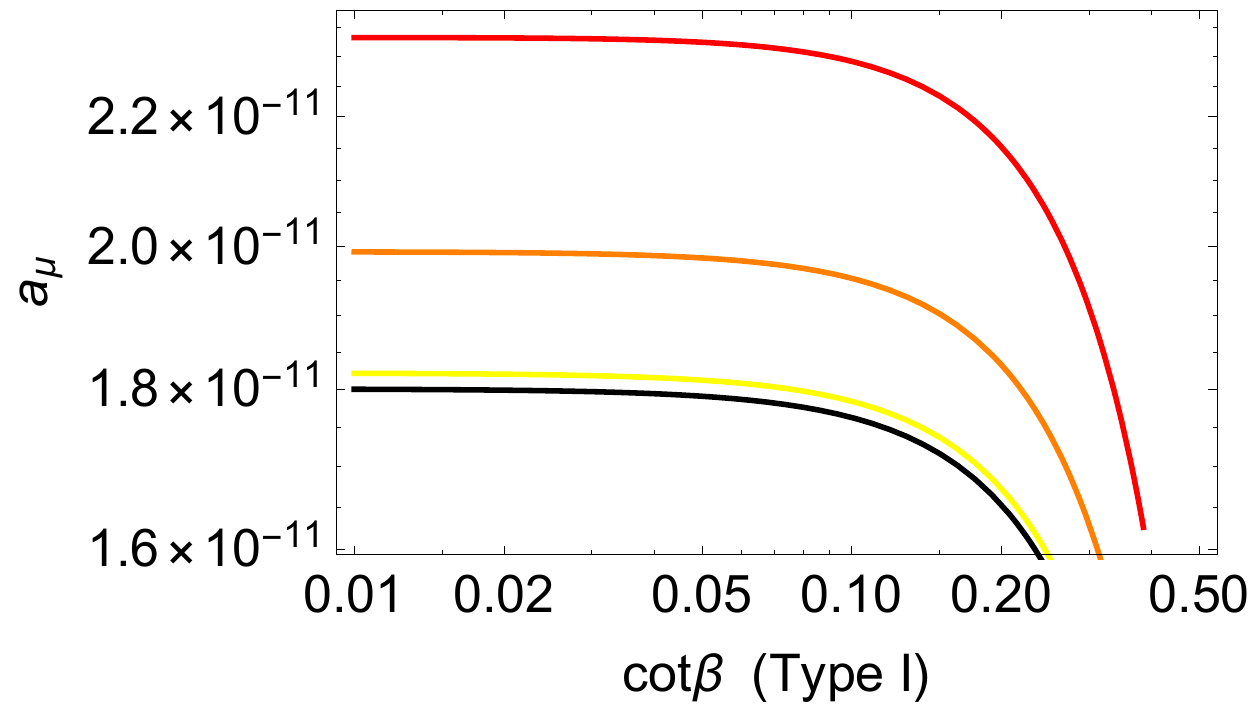}\\[4mm]
\hspace{-2mm}\includegraphics[height=50mm]{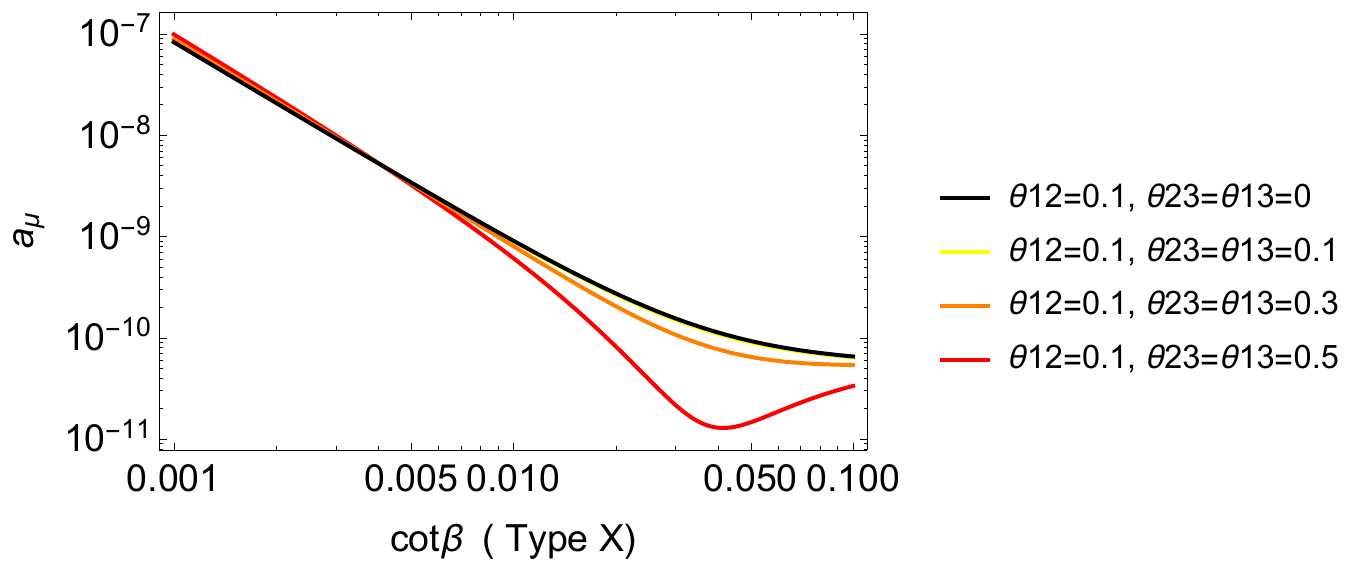}\\[4mm]
\hspace{-47mm}\includegraphics[height=50mm]{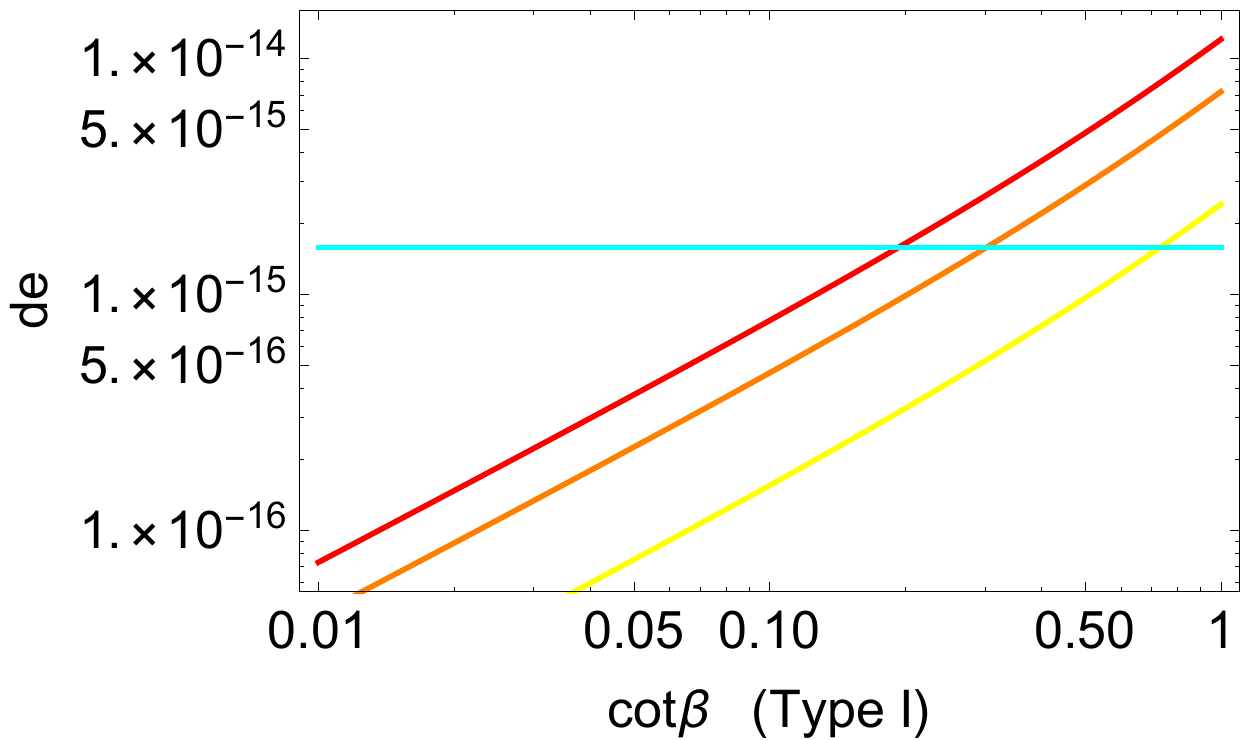}\\[4mm]
\includegraphics[height=50mm]{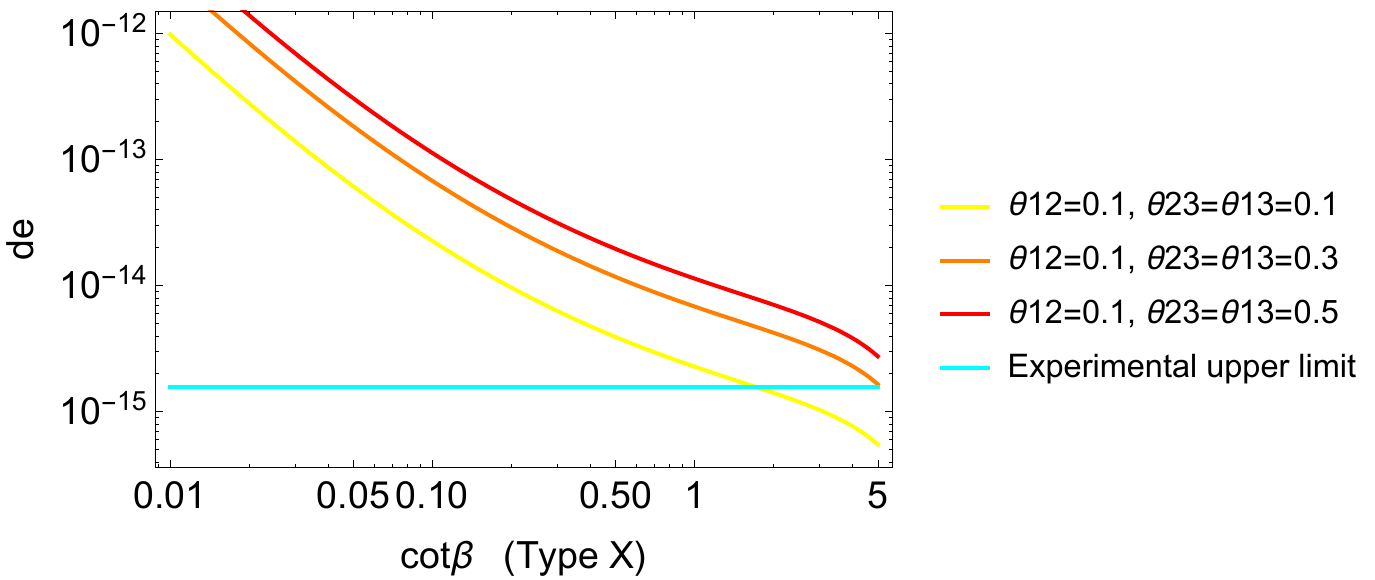}
\caption{$a_\mu$ (top) and $d_e$ (bottom) values in Type I and Type X for different values of angles and fixed values of masses ($m_{h_{2,3}}=145,105$ GeV). The behaviour of Type Y and II are similar to Type I and X, respectively.}
\label{amu-145105}
\end{center}
\end{figure}

To get a closer look at the effect of CP-violation, we present in Figure~\ref{amu-145105}, $a_\mu$ and $d_e$ contributions of Type I and X for fixed scalar masses and varying angles.
Note that in Type I, larger CP-violating angles lead to larger $a_\mu$ values while the effect is more complicated and $\cot\beta$-dependent in Type X.
Clearly with increasing CP-violation, the $d_e$ contribution increases and the surviving region of the parameter space shrinks.
As mentioned before, the behaviour of Type Y and II are similar to Type I and X, respectively.

\begin{figure}[t!]
\begin{center}
\includegraphics[scale=0.55]{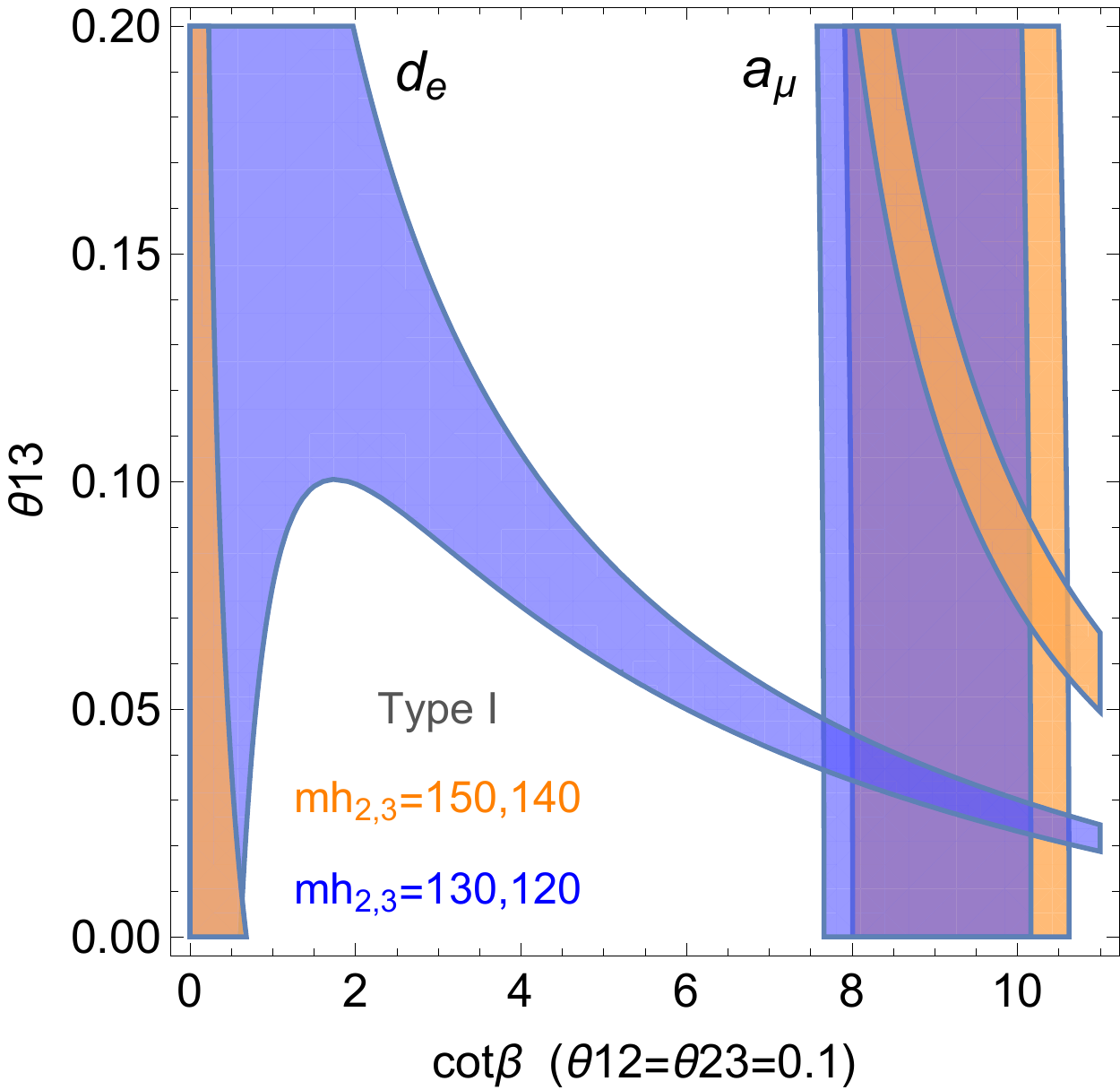} \hspace{10mm}
\includegraphics[scale=0.55]{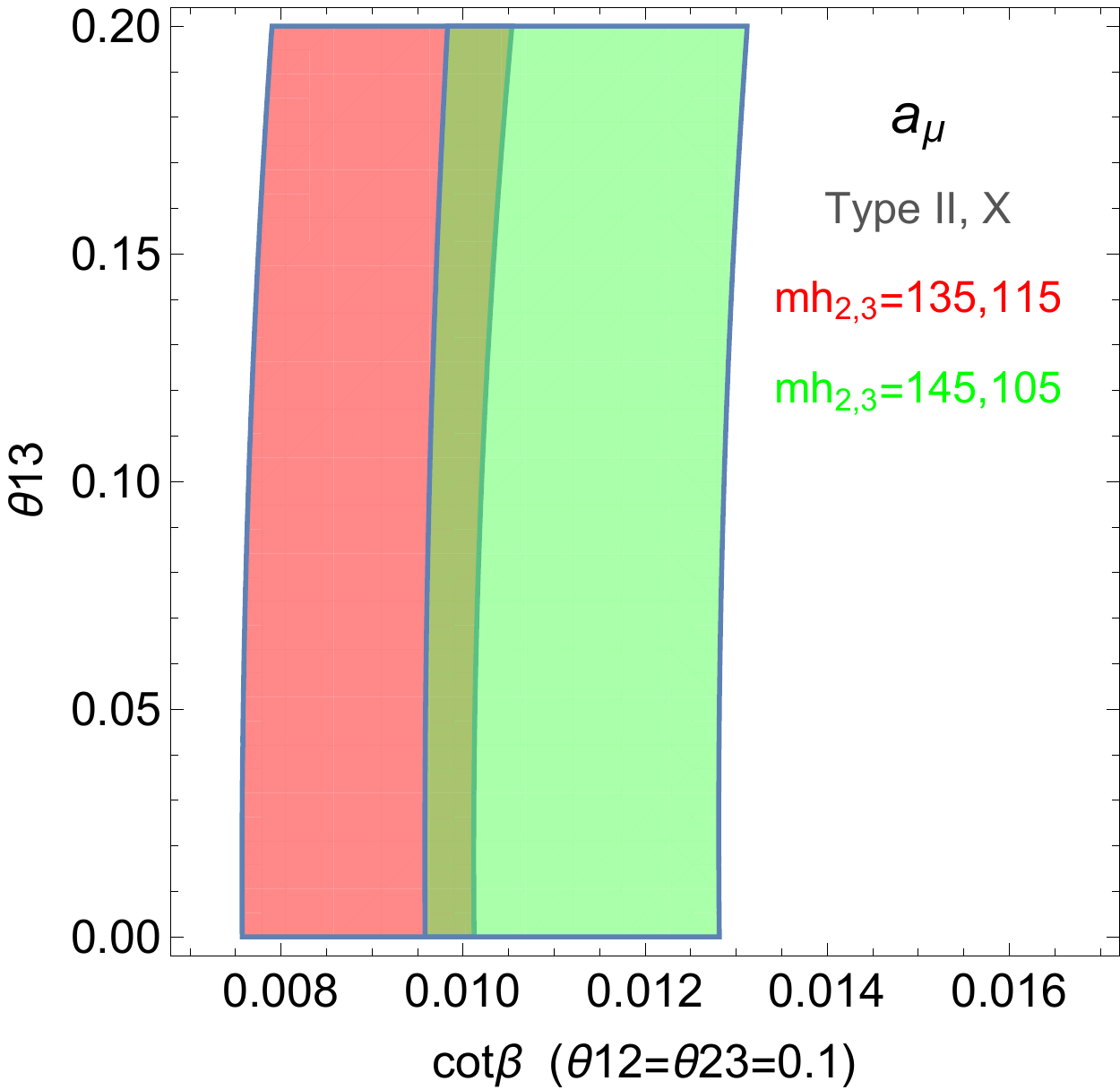}\\[3mm]
\includegraphics[scale=0.55]{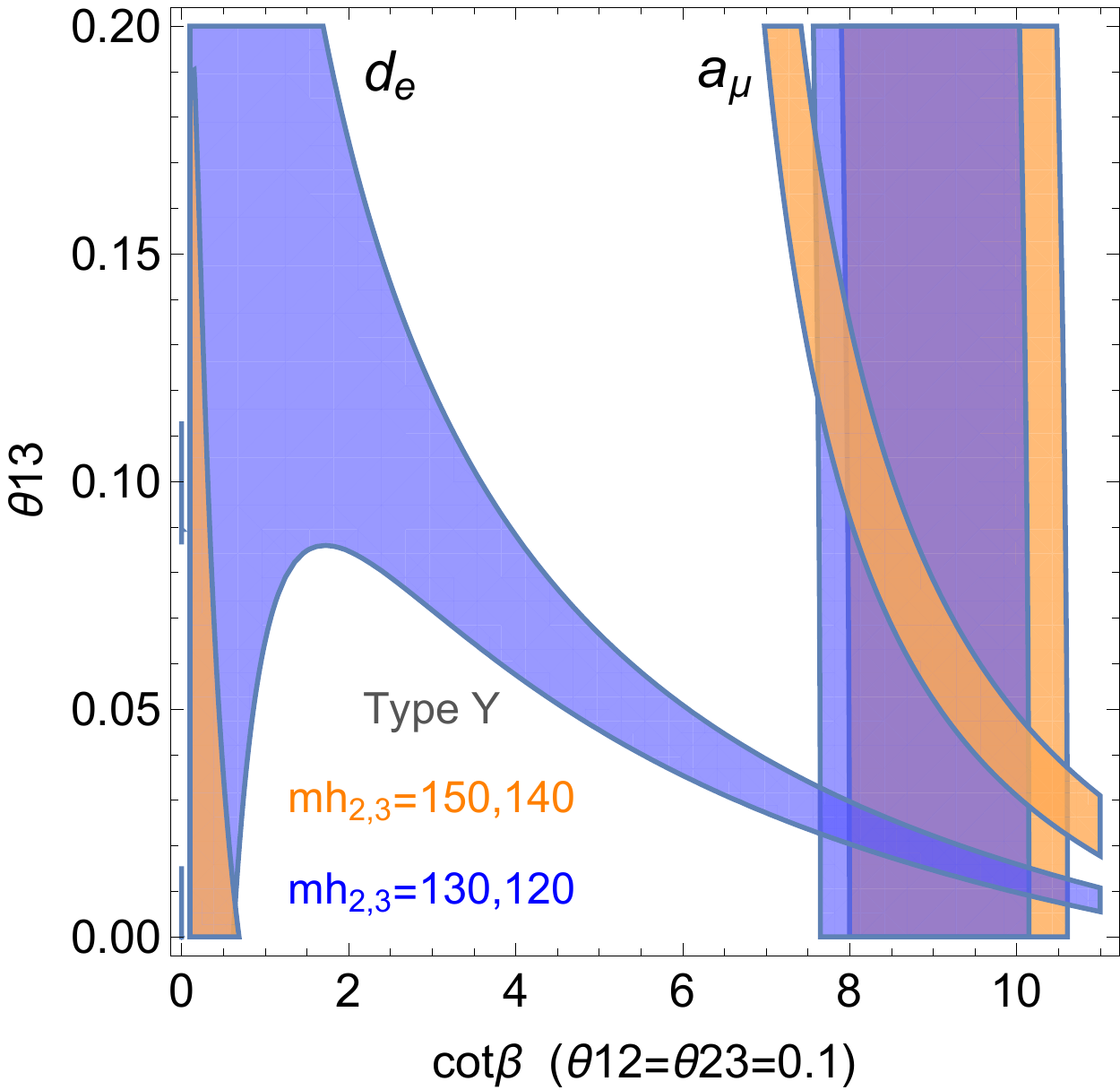} \hspace{10mm}
\includegraphics[scale=0.55]{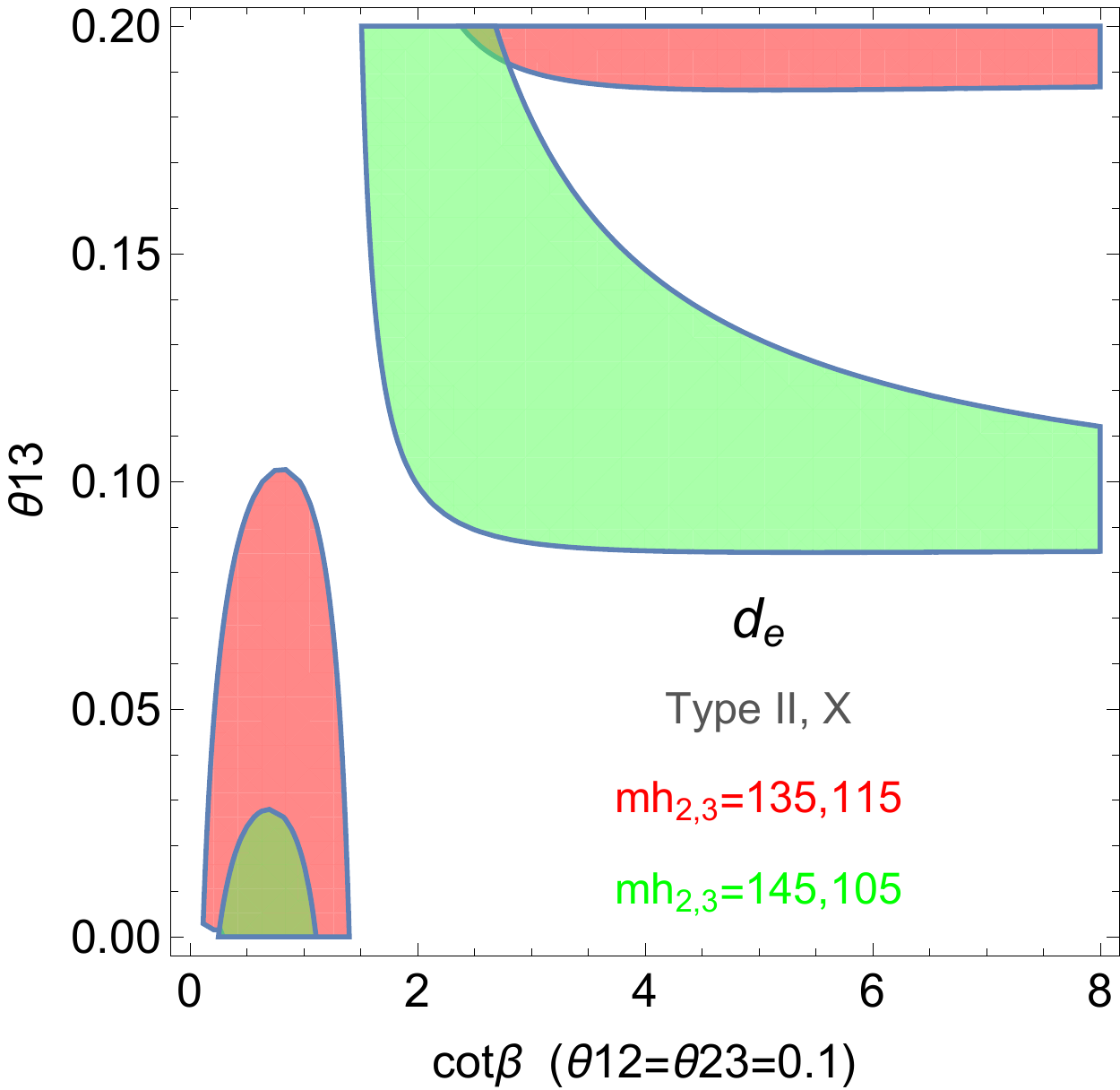} \\[3mm]
\hspace{85mm}
\includegraphics[scale=0.55]{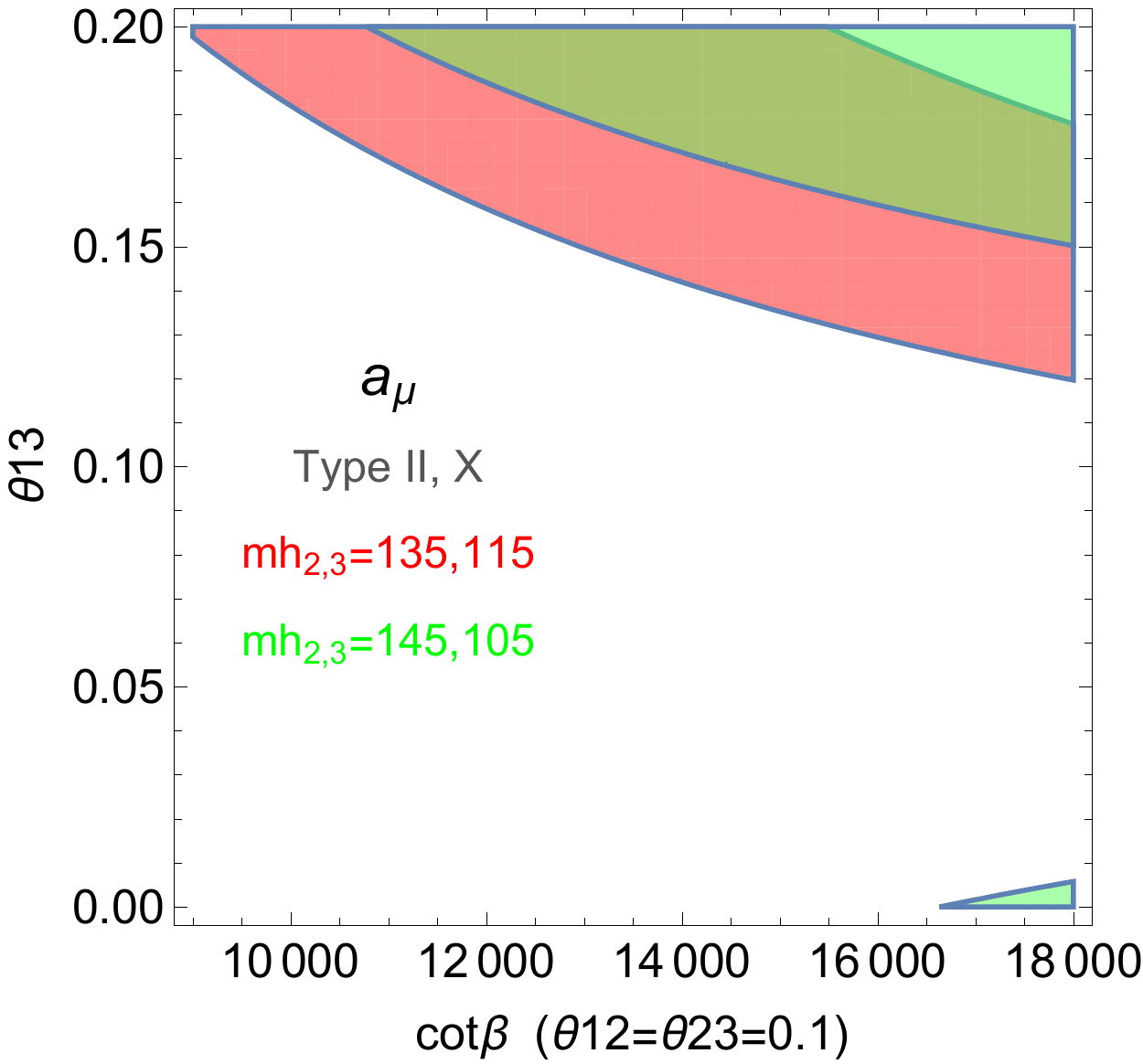}
\caption{Regions surviving $d_e$ bounds while producing $a_\mu$ in Type-I,Y (left) and Type II,X (right) for mid-range $m_{h_{2,3}}$ masses (in GeV) and fixed values of $\theta_{12}$ and $\theta_{23}$. Note that in in Type-I,Y plots by changing $m_{h_{2,3}}$ masses, it is possible to cover the whole plane.}
\label{IY2-de-amu-medium}
\end{center}
\end{figure}

To see the effect of the scalar masses, in Figure~\ref{IY2-de-amu-medium}, we show regions producing $a_\mu$ within the $3.6 \sigma$ band and regions surviving the $d_e$ limits for mid-range scalar masses and fixed values of $\theta_{12}= \theta_{23}=0.1$.
In Type I and Y, the regions corresponding to different constraints overlap when $\cot\beta \approx 9$, and therefore in this mass range one can explain $a_\mu$ and remain compatible with the $e$EDM experiments.
Let us stress that in Type I and Y one can
satisfy the constraint on $d_e$ at any point
of the plane shown in Fig.~\ref{IY2-de-amu-medium} by changing the masses and the mass splittings. However, the region where the observed $a_\mu$ can be produced lies robustly at large $\cot\beta$, which is ruled out by
too large scalar-fermion couplings.
The corresponding plots for Type II and X show that in neither types there is a region where the $d_e$ and $a_\mu$ plots overlap, with $a_\mu$ preferring the very small and very large $\cot\beta$ values while $d_e$ bounds are satisfied in the $\cot\beta \sim \mathcal{O}(1)$.

\begin{figure}[h!]
\begin{center}
\hspace{-30mm}\includegraphics[height=70mm]{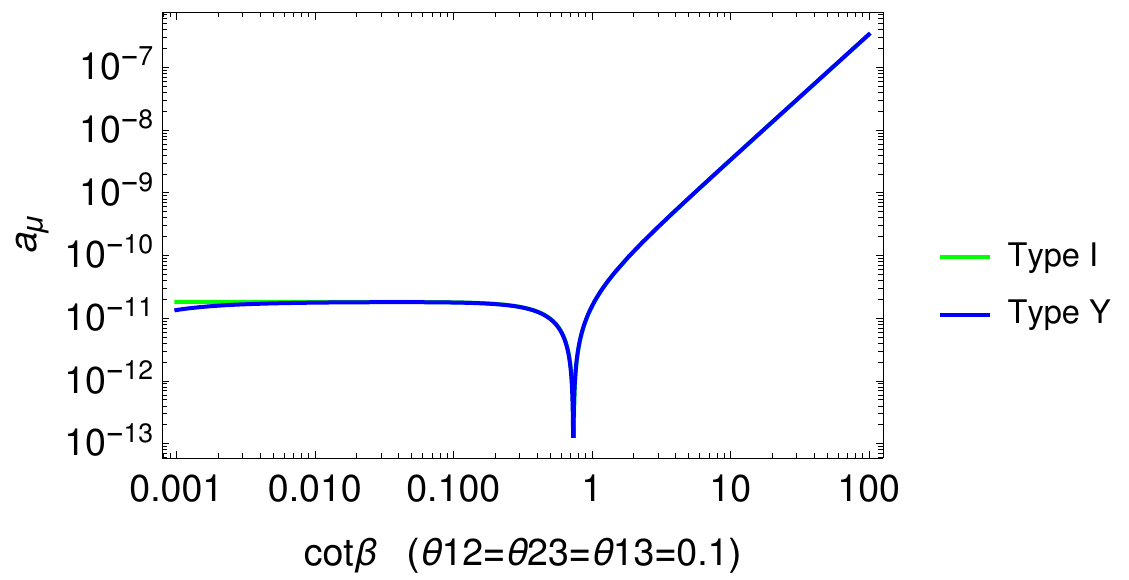}\\[5mm]
\includegraphics[height=70mm]{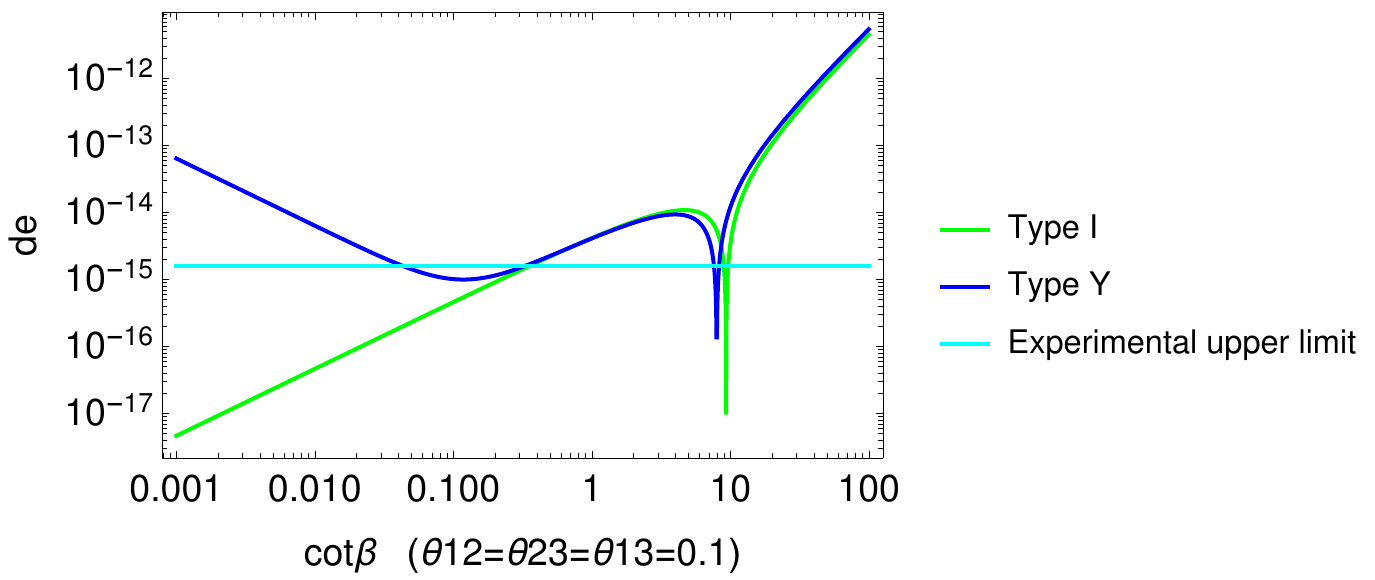}
\caption{$a_\mu$ (top) and $d_e$ (bottom) values in Type I,Y for fixed values of angles and masses ($m_{h_{2,3}}=150,140$ GeV). Notice the $d_e$-surviving region $\cot\beta \approx 10$ which contributes to $a_\mu$ sufficiently.}
\label{amu-de-150140-new}
\end{center}
\end{figure}

Note that in Fig.~\ref{amu-145105} the value for the CP-violating angles $\theta_{23},\theta_{13}$ is chosen to be $0.5$ for the $a_\mu$ plot to show the enhanced effect of CP-violation. Such a high value of CP-violation is strongly constrained by eEDMs as shown in the same figure in the $d_e$ plot with the same $\theta_{23},\theta_{13}$ values. In Fig.~\ref{IY2-de-amu-medium} where we claim that the $a_\mu$ and $d_e$ favourable regions overlap, the CP-violating angles are very small $\theta_{23},\theta_{13} \simeq 0.1$ and well within the $d_e$ bounds as shown in Figure~\ref{amu-de-150140-new}.

\subsubsection{Light mass region}

It has been shown \cite{Cheung:2001hz, Cherchiglia:2017uwv} that in the CP-conserving limit, Type-X 2HDM can produce a large enough $a_\mu$ due to the positive contribution from a very light CP-odd scalar, $m_A \approx 30$ GeV, and a large $\tan\beta \approx 60$. Our calculations, when taken to the CP-conserving limit ($\theta_{13}=\theta_{23}=0$), confirm these results.

\begin{figure}[h!]
\begin{center}
\hspace{-60mm}\includegraphics[height=70mm]{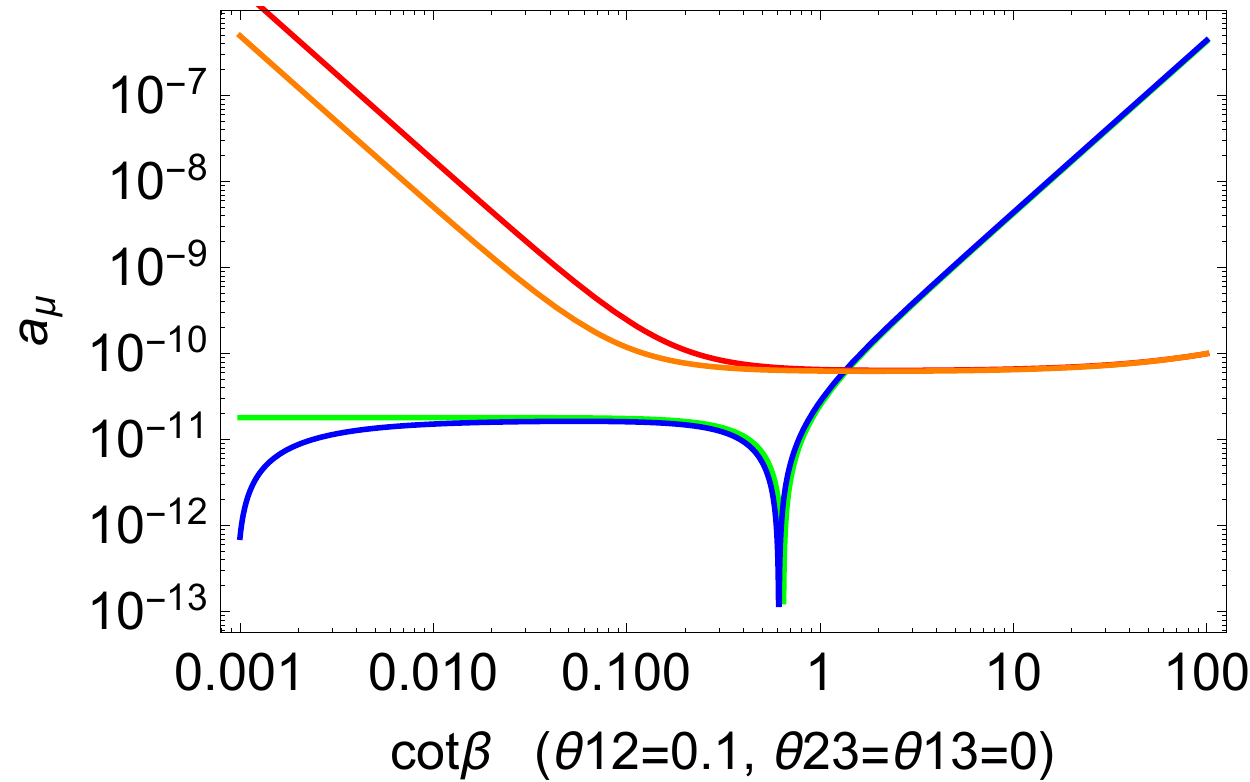}\\[5mm]
\hspace{-30mm}\includegraphics[height=70mm]{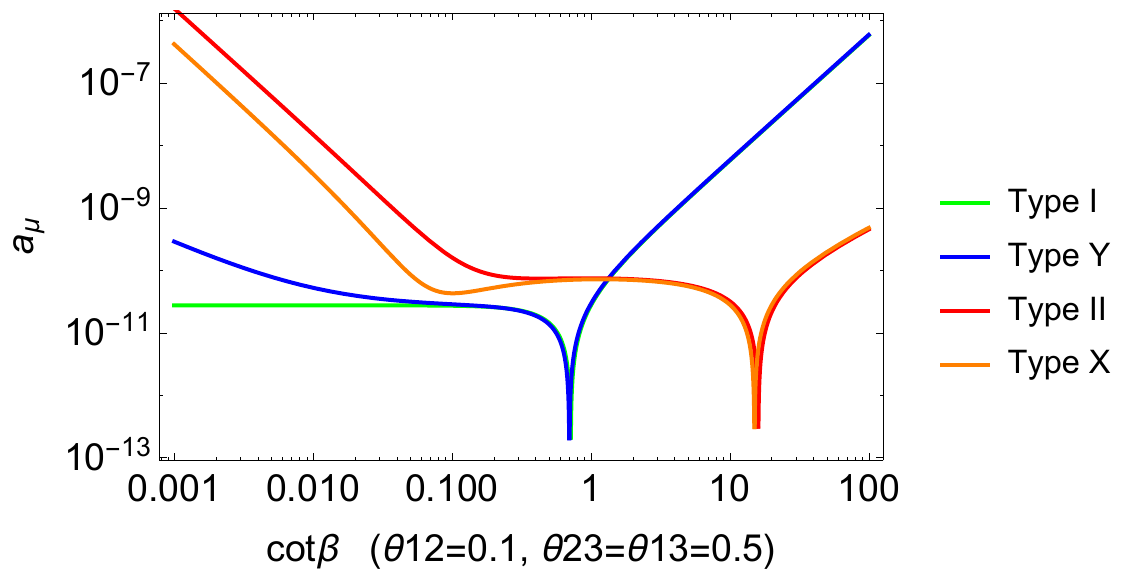}\\[5mm]
\includegraphics[height=70mm]{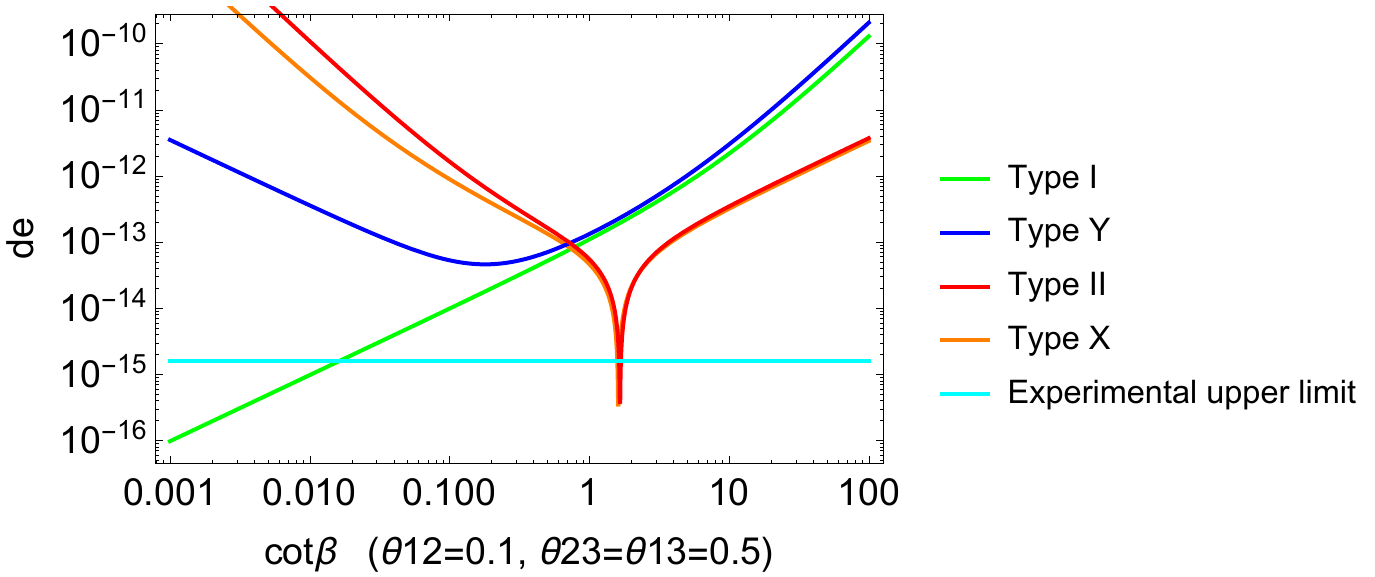}
\caption{$a_\mu$ (top) and $d_e$ (bottom) values in different 2HDM Types for fixed values of angles and masses ($m_{h_{2,3}}=200,50$ GeV).}
\label{amu-light}
\end{center}
\end{figure}

In Figure~\ref{amu-light}, we show the effect of CP-violation on the $a_\mu$ contribution in different 2HDM Types for light scalars masses. We also show the $d_e$ contribution in different 2HDM Types in this mass region and the upper limit imposed by the $e$EDM experiments.

\begin{figure}[h!]
\begin{center}
\hspace{-52mm}\includegraphics[height=50mm]{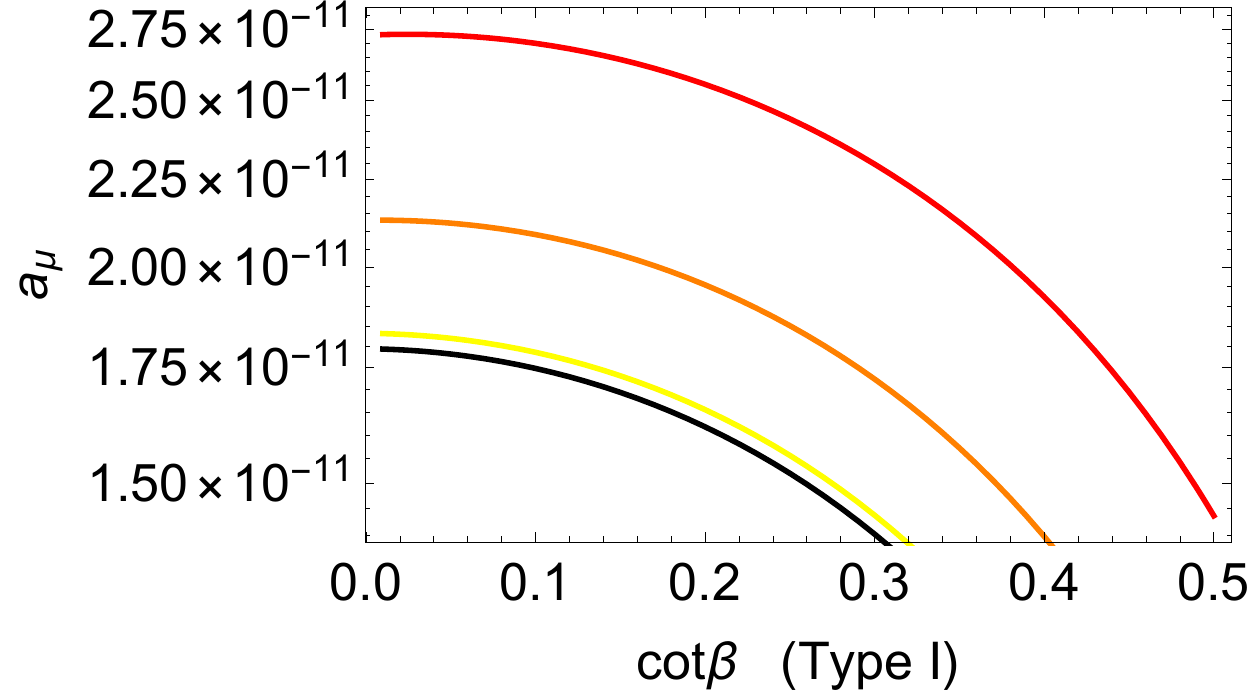}\\[4mm]
\hspace{-2mm}\includegraphics[height=50mm]{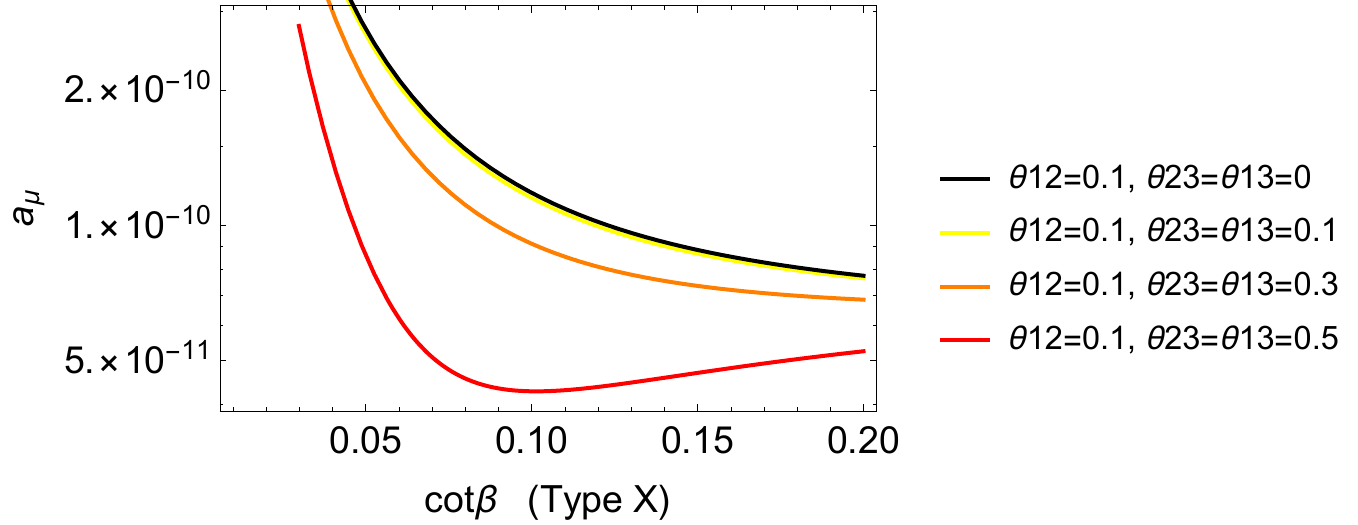}\\[4mm]
\hspace{-47mm}\includegraphics[height=50mm]{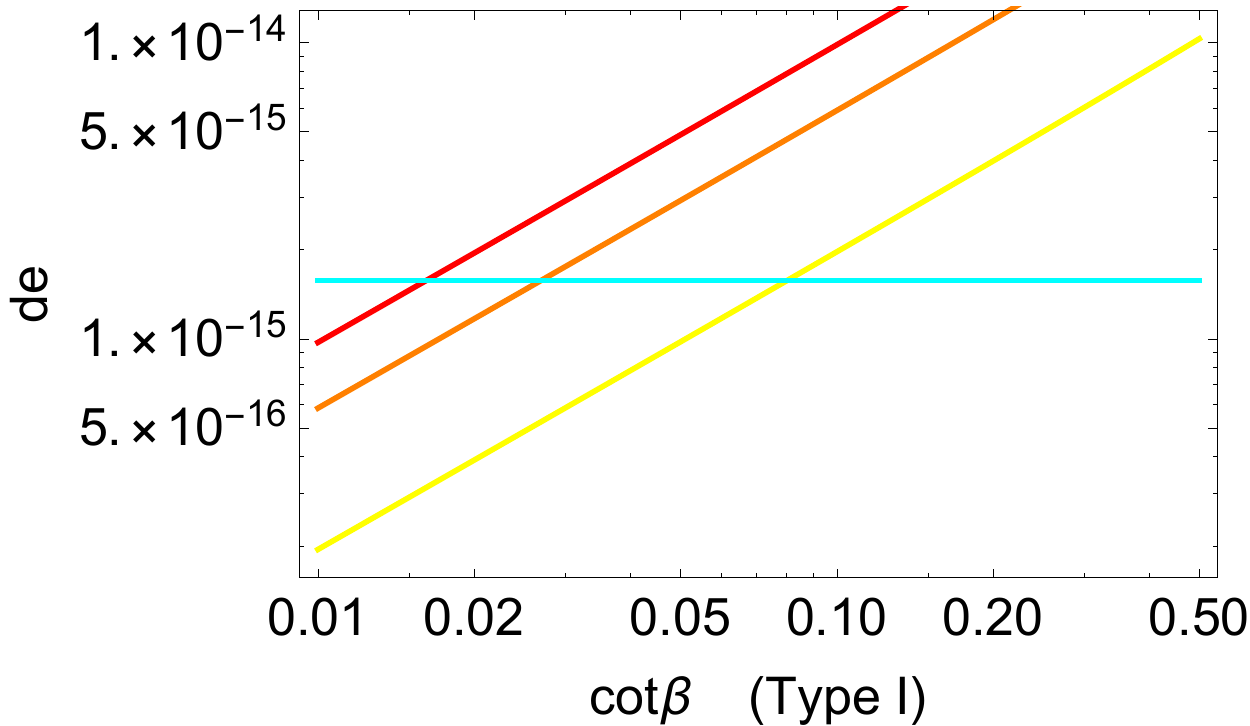}\\[4mm]
\includegraphics[height=50mm]{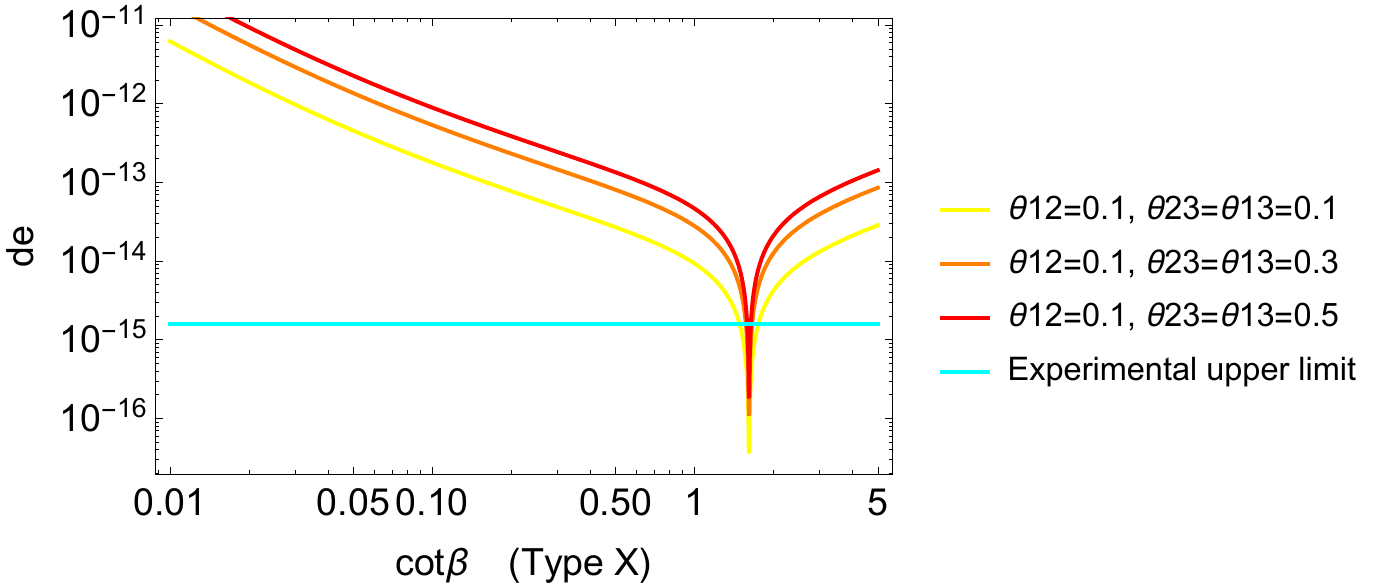}
\caption{$a_\mu$ (top) and $d_e$ (bottom) values in Type I (top) and Type X (bottom) for different values of angles and fixed values of masses ($m_{h_{2,3}}=200,50$ GeV). The behaviour of Type Y and II are similar to Type I and X, respectively.}
\label{details-light}
\end{center}
\end{figure}

To clarify the effect of CP-violation, in Figure~\ref{details-light}, we plot $a_\mu$ and $d_e$ contributions in Type I and X for fixed scalar mass values while varying the CP-violating angles. The corresponding plots for Type Y and II are similar to Type I and X, respectively, as discussed in detail before.

\begin{figure}[t!]
\begin{center}
\includegraphics[scale=0.55]{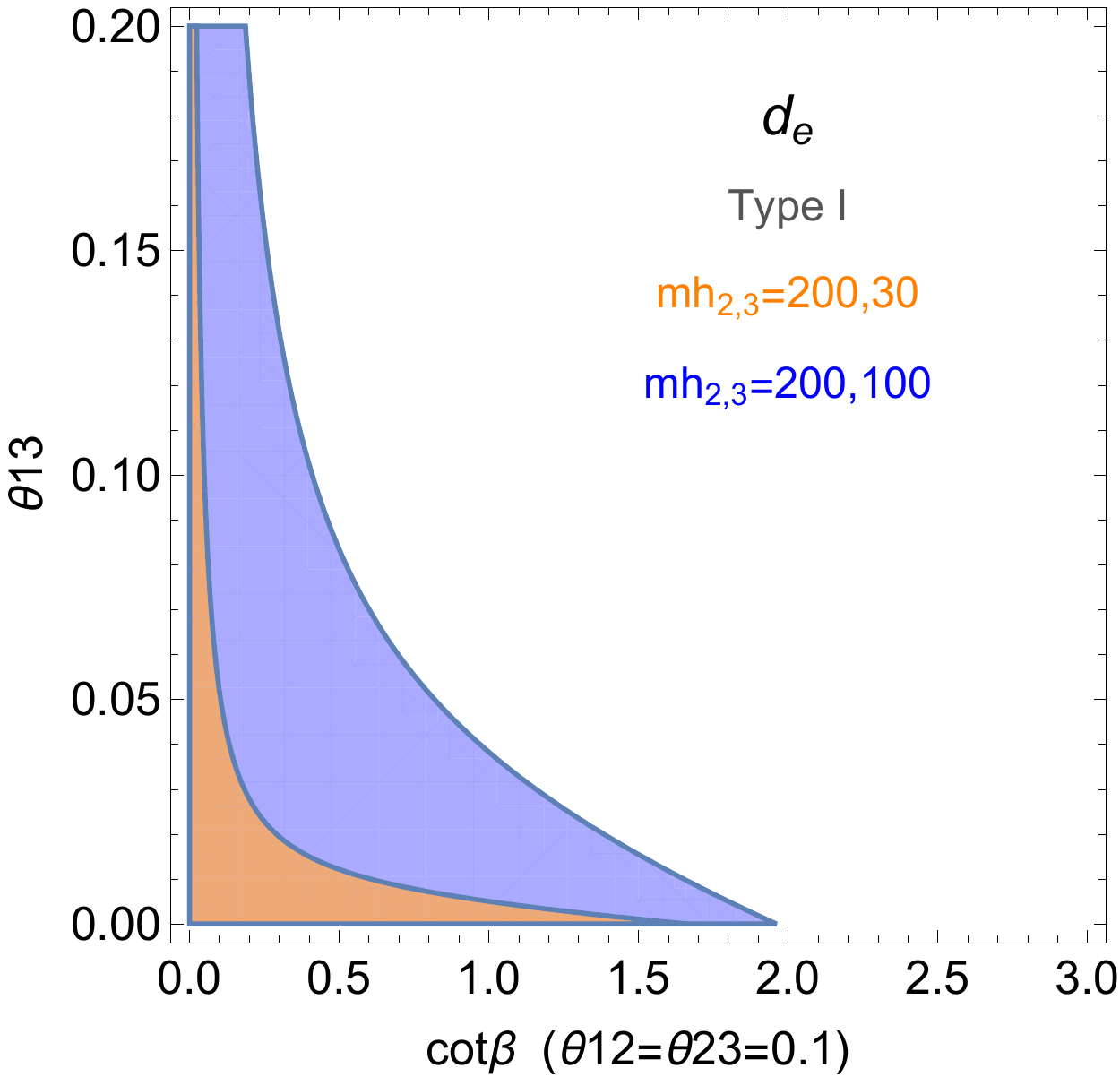} \hspace{10mm}
\includegraphics[scale=0.55]{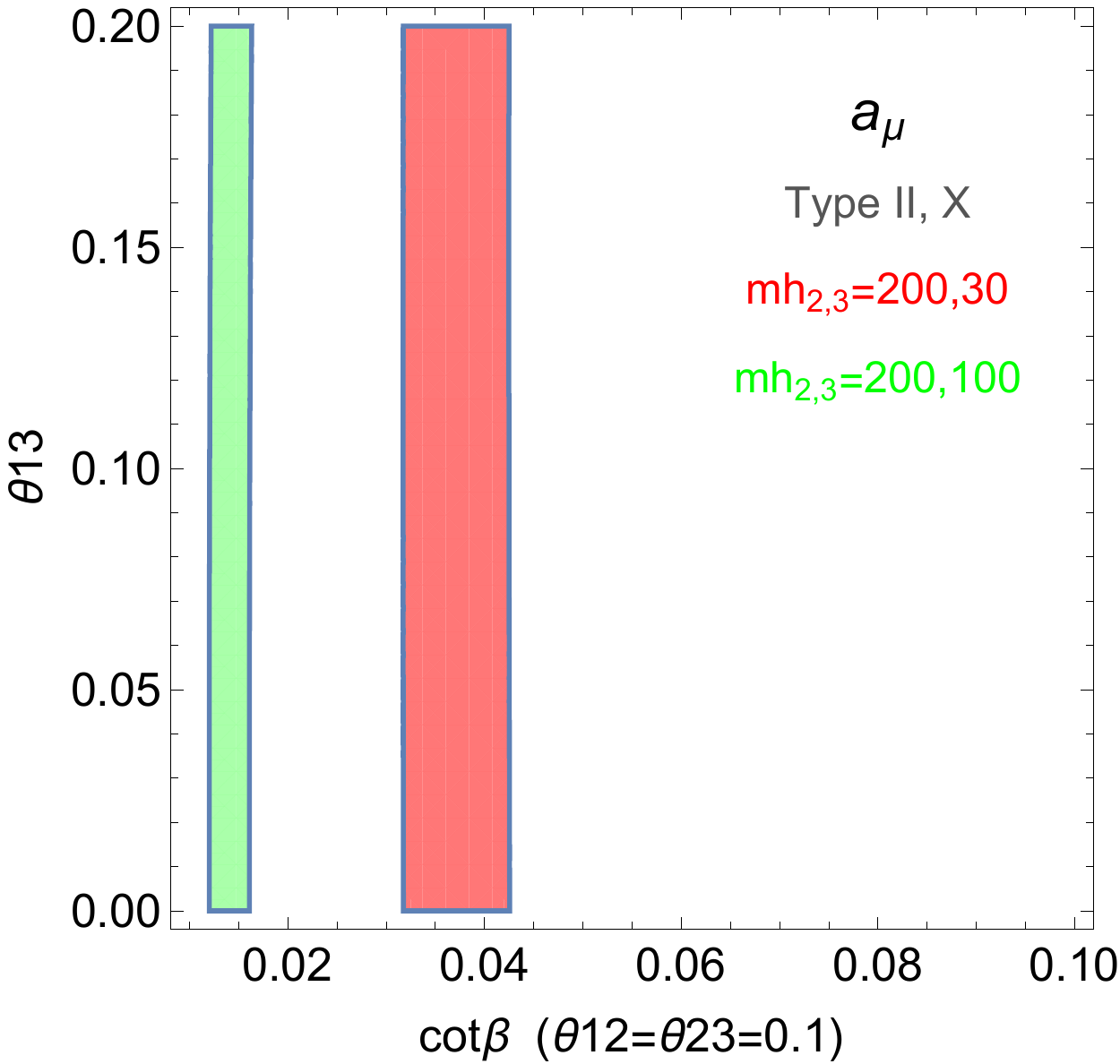}\\[3mm]
\includegraphics[scale=0.55]{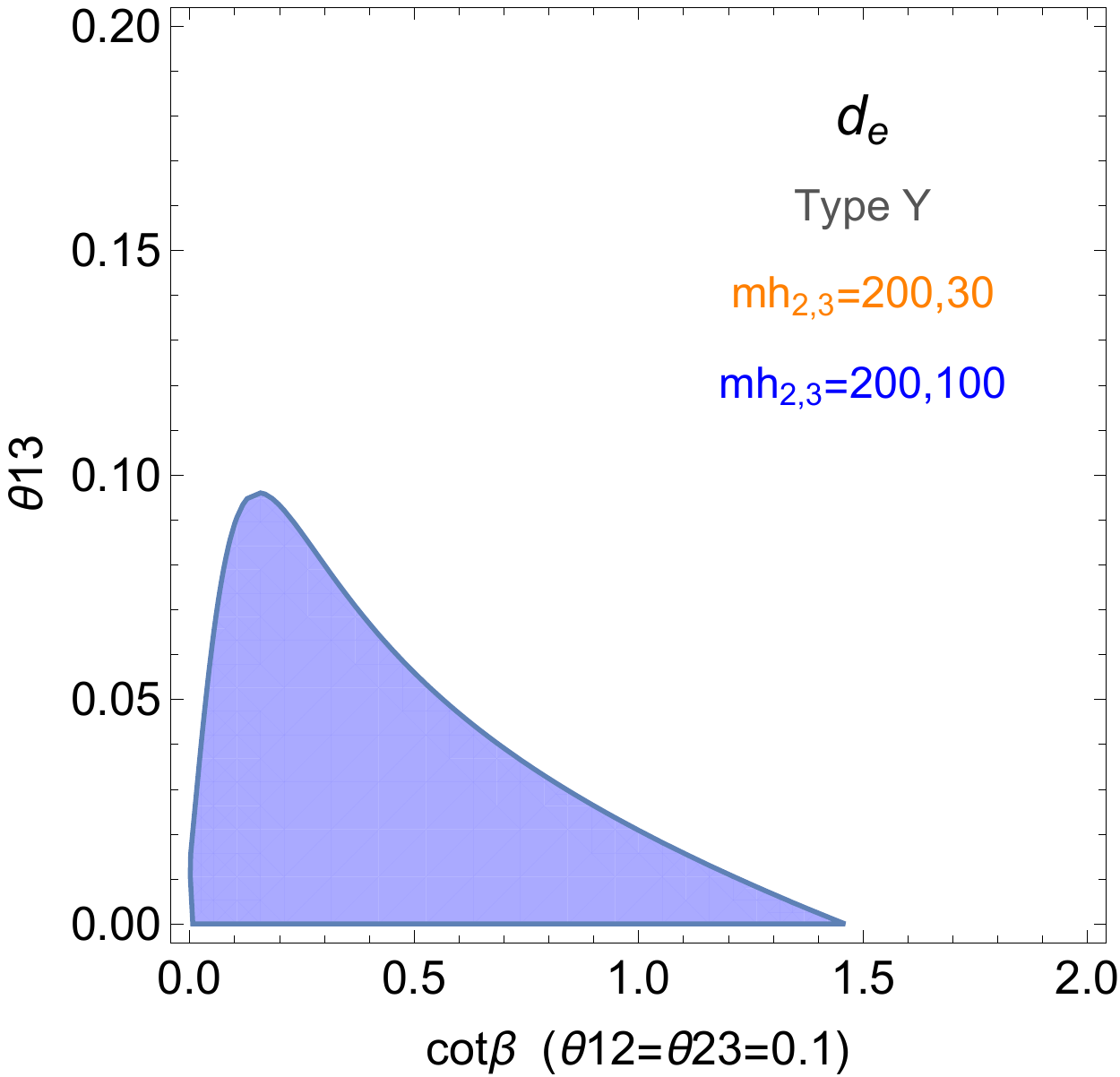} \hspace{10mm}
\includegraphics[scale=0.55]{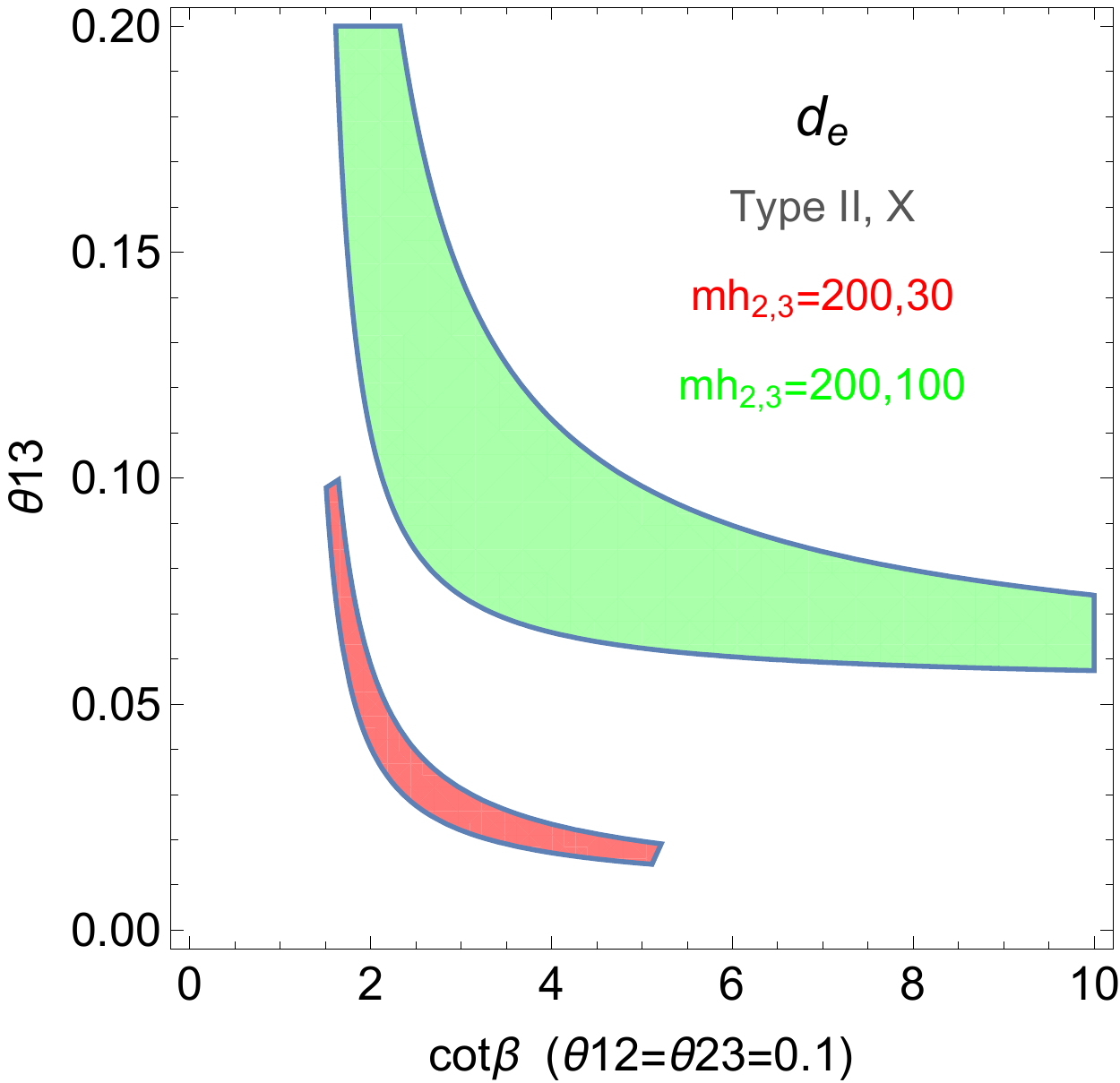} \\[3mm]
\includegraphics[scale=0.55]{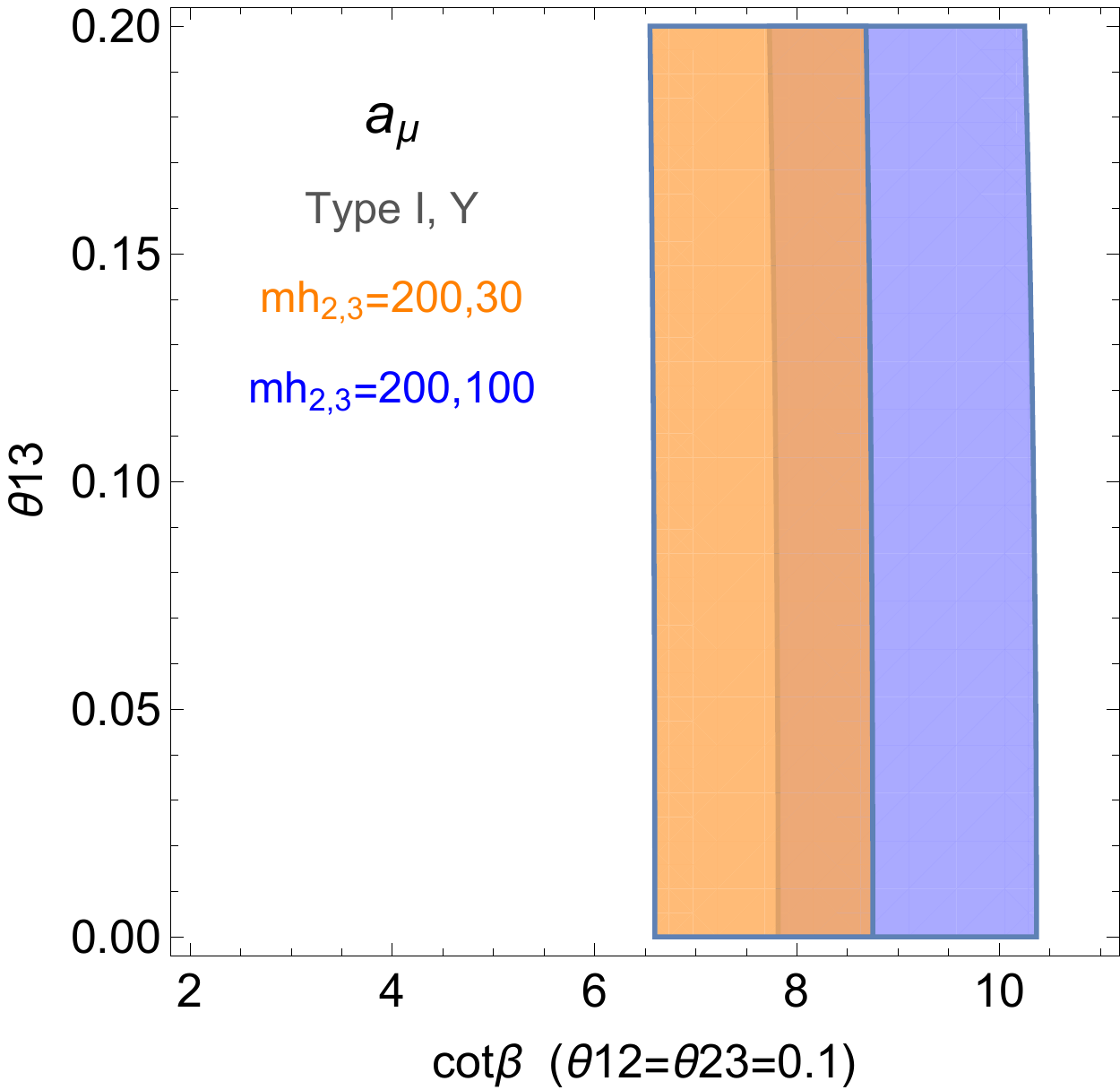} \hspace{10mm}
\includegraphics[scale=0.55]{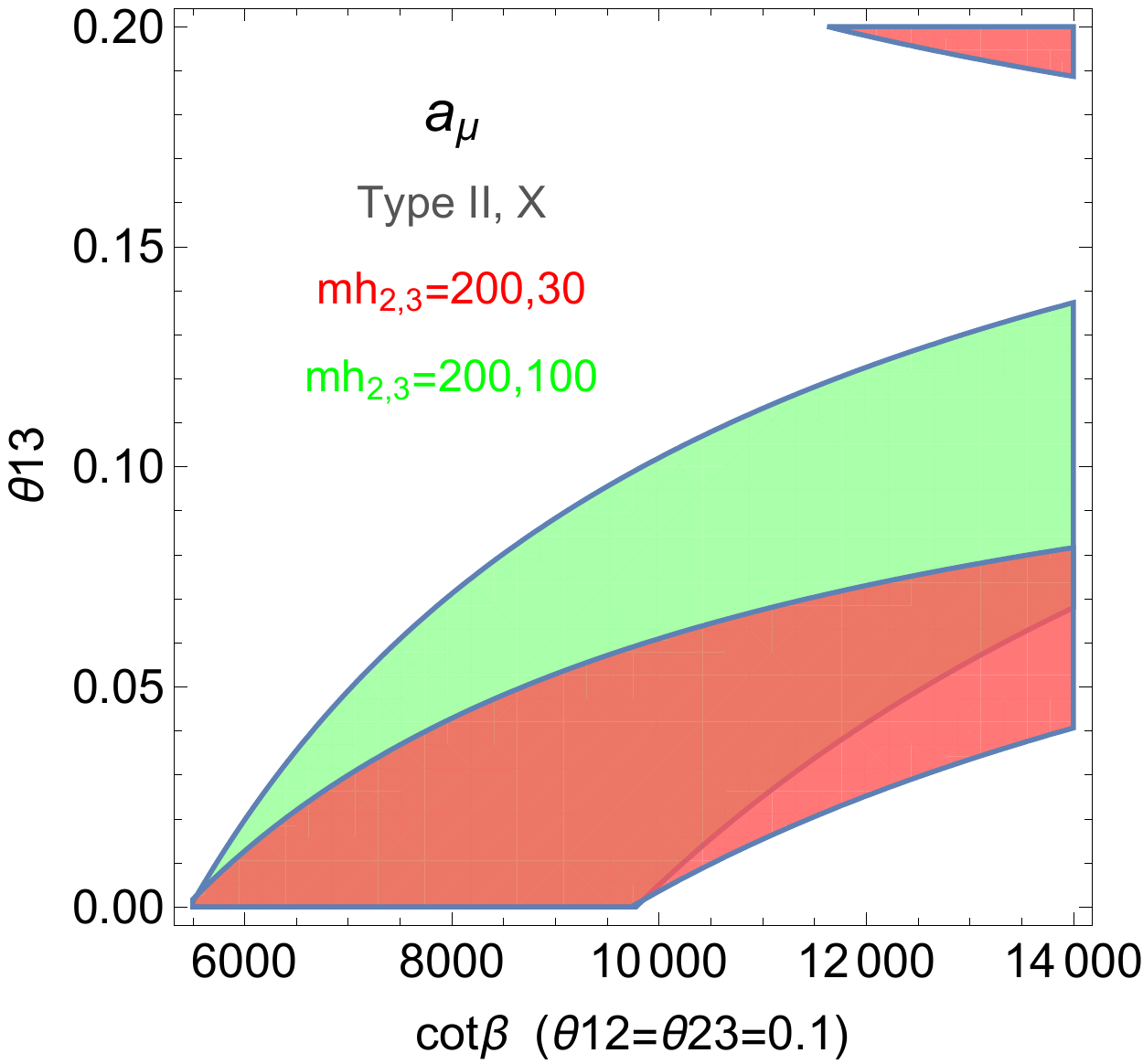}
\caption{Regions surviving $d_e$ bounds vs. regions producing $a_\mu$ within the deviation observed in 2HDM Type I,Y (left) and Type II,X (right) for light scalar masses (in GeV) and fixed $\theta_{{12,23}}$ values.}
\label{IY-de-amu-light}
\end{center}
\end{figure}

To see how different scalar masses affect the $a_\mu$ and $d_e$ contributions, in Figure~\ref{IY-de-amu-light}, we show regions surviving $d_e$ bounds and regions producing $a_\mu$ within the $3.6 \sigma$ observed value in different 2HDM Types for different $m_{h_{2,3}}$ masses with fixed $\theta_{12}, \theta_{23}$ values.
This figure confirms our statement in Figs.~\ref{amu-light}: in Type I and Y, the $a_\mu$ behaviour is very similar while Type Y is more constrained by $d_e$ data. Clearly, $a_\mu$ requires $\cot\beta \approx 10$, while $d_e$ constrains $\cot\beta$ to be less than $1$ in Type I and Y with no overlap between the two regions.
Type II and X, clearly showing no overlap between the $d_e$ and $a_\mu$ regions with $a_\mu$ preferring the very small and very large $\cot\beta$ values while $d_e$ bounds are satisfied for $\cot\beta \sim \mathcal{O}(1)$.

Similar to the other mass regions, large $\cot\beta$ values are ruled out due to flavour and/or collider constraints in Type II and Y, and only Type I and X survive in the low $\cot\beta$ region.

To summarise the findings in these subsections:
In all mass ranges, Type I and Y contribute efficiently to $a_\mu$ in the large $\cot\beta$ region. In Type II and X, very small $\cot\beta$ values lead to the correct $a_\mu$ values.
On the other hand, bounds from $d_e$ experiments are satisfied in the small $\cot\beta$ region for Type I and Y with Type Y more constrained, and in the $\cot\beta \gtrsim 1$ for Type II and X.

Super-imposing the $d_e$ and $a_\mu$ plots reveal that only in Type I and Y and only in the medium mass region, one can simultaneously produce the observed
contribution to $a_\mu$ and remain compatible with the results from $e$EDM experiments. However, this happens only at relatively large values of $\cot\beta$ ruled out by current experimental data.
In the low $\cot\beta$ region we find that only Type I and Type X models remain viable: Type X is a suitable choice for obtaining $a_\mu$ contributions in the absence of CP-violation, while Type I has the smallest contribution to $d_e$ and therefore a suitable choice for a CP-violating model, but yielding only the SM contribution to $a_\mu$.


\section{2HDM + singlet extension}
\label{section-2hdm+CS}
The singlet extension of the 2HDM is a relatively popular model~\cite{Bonilla:2014xba, Kakizaki:2016dza, Alanne:2016wtx} for several reasons:
First, the scalar sector of 2HDM+S resembles that of the Next to Minimal Supersymmetric SM (NMSSM). Second, it is understood that the singlet extension of the SM and the 2HDM are incapable of providing a viable DM candidate and allow for CP-violation simultaneously~\cite{Keus:2016orl}. Therefore, going beyond the simplest extensions of the SM seems inevitable.

In general, the decomposition of the scalar multiplets is as follows.
\be
\Phi_1= \doublet{$\begin{scriptsize}$ \phi^+_1 $\end{scriptsize}$}{\frac{v_1+h_1^0+ia^0_1}{\sqrt{2}}} ,\quad
\Phi_2= \doublet{$\begin{scriptsize}$ \phi^+_2 $\end{scriptsize}$}{\frac{v_2+h^0_2+ia^0_2}{\sqrt{2}}} ,\quad
S = \frac{1}{\sqrt{2}} (w  + \phi_4 + i \phi_5).
\label{fields-2hdms}
\ee

Here, we discuss directly the complex singlet extension of 2HDM. The results could easily be translated to the real singlet case by setting the imaginary component of the singlet, $\phi_5$, and the related parameters ($\theta_{15},\theta_{25},\theta_{35},\theta_{45}$ in the rotation matrix in Eq.~(\ref{rotation-2hdms})) to zero. Also, in the calculation of the $d_e$ and $a_\mu$, the sums will be over $i=1,\dots, 4$, corresponding to the four scalar mass eigenstates, $h_{1,2,3,4}$.

The most general 2HDM+CS potential has the the form $V=V^{d}+V^{s}+V^{ds}$, where
\bea
\label{pot2hdms}
V^{d} &=& -\mu^2_{1} (\Phi_1^\dagger \Phi_1) -\mu^2_2 (\Phi_2^\dagger \Phi_2) - \mu^2_3(\Phi_1^\dagger \Phi_2)   \notag\\
&&+  \lambda_{1} (\Phi_1^\dagger \Phi_1)^2+ \lambda_{2} (\Phi_2^\dagger \Phi_2)^2  + \lambda_{3} (\Phi_1^\dagger \Phi_1)(\Phi_2^\dagger \Phi_2)
 + \lambda_{4} (\Phi_1^\dagger \Phi_2)(\Phi_2^\dagger \Phi_1) \notag\\
&&+   \lambda_{5}  (\Phi_1^\dagger \Phi_2)^2 + \lambda_{6} (\Phi_1^\dagger \Phi_1)(\Phi_1^\dagger \Phi_2)
+ \lambda_{7} (\Phi_2^\dagger \Phi_2)(\Phi_1^\dagger \Phi_2),
\\
V_{s} =&& - \mu_4^2 (S^* S) - \mu_5^2 ( S^2)
\nonumber\\&&
+ \lambda_{8} (S^*S)^2 + \lambda_{9} (S^*S)(S^2) + \lambda_{10} (S^4 )
\nonumber\\&&
+ \kappa_1 (S ) + \kappa_2 (S^3 ) + \kappa_3(S)(S^*S),
\\[1mm]
V_{ds} = &&\lambda_{11}(\Phi_1^\dagger\Phi_1)(S^* S) + \lambda_{12} (\Phi_1^\dagger\Phi_1)(S^2)
+ \kappa_4 (\Phi_1^\dagger\Phi_1) ( S)
\nonumber\\
 &&
+\lambda_{13}(\Phi_2^\dagger\Phi_2)(S^* S) + \lambda_{14} (\Phi_2^\dagger\Phi_2)( S^2)
+ \kappa_5 (\Phi_2^\dagger\Phi_2) ( S )
\nonumber\\
 &&
+\lambda_{15}(\Phi_1^\dagger\Phi_2)(S^* S) + \lambda_{16} (\Phi_1^\dagger\Phi_2)(S^2)
 + \lambda_{17} (\Phi_1^\dagger\Phi_2)(S^{*2})
 \nonumber\\
 &&
+ \kappa_6 (\Phi_1^\dagger\Phi_2) ( S ) +
\kappa_7 (\Phi_1^\dagger\Phi_2) ( S^* ).
\eea
Similarly as in the case of SM+RS, the linear term $\kappa_1$ can be removed
by a translation of $S$.

Similar to the 2HDM, the 2HDM+CS suffers from tree-level FCNCs due to the existence of more than one scalar doublet which in general could couple to fermions. As in 2HDM, this can be alleviated by imposing a $Z_2$ symmetry on the scalar sector and extending it to the fermion sector in a similar manner. The transformation of the scalar multiplets under this $Z_2$ symmetry is fixed to be
\be
\Phi_1 \to + \Phi_1, ~~\Phi_2 \to - \Phi_2, ~~S \to +S.
\label{Z2-2hdms}
\ee
The fermionic $Z_2$ charges are as shown in Table~\ref{Tab:type} which define the 2HDM type of the model.
Imposing this symmetry on the potential while allowing for a soft breaking term $\mu_3^2$, forbids the following parameters,
\be
\lambda_6 =\lambda_7= \lambda_{15}= \lambda_{16}= \lambda_{17} =\kappa_6 =\kappa_7 =0.
\label{zero-params}
\ee
CP-violation is introduced explicitly through the following complex parameters,
\be
\mu_3^2,~\mu_5^2,~\kappa_{1},~\kappa_{2}, ~\kappa_{3},~\kappa_{4}, ~\kappa_{5}, ~ \lambda_5,\lambda_9,~\lambda_{10},~\lambda_{12},
~\lambda_{14}.
\ee
We take the VEVs, $v_1$, $v_2$ and $w$ to be real.

\subsection{Minimisation of the 2HDM+CS potential}

The minimisation conditions are presented in Appendix~\ref{2hdms-details} under which the minimum of the potential is realised at $\langle \Phi_1 \rangle = v_1$, $\langle \Phi_2 \rangle = v_2$, $\langle S \rangle = w$.

Similar to the 2HDM, it is useful to rotate the doublets to the Higgs basis while the singlet remains unchanged,
\be
\label{higgs-basis-2hdms}
\left(
\begin{array}{c}
\widehat{\Phi}_1\\ \widehat{\Phi}_2\\
\widehat{S}\\
\end{array} \right)
=
\left(
\begin{array}{ccc}
\cos\beta & \sin\beta & 0  \\
-\sin\beta & \cos\beta & 0 \\
0 & 0 & 1 \\
\end{array} \right)
\left(
\begin{array}{c}
\Phi_1\\ \Phi_2\\ S \\
\end{array} \right),
\ee
with $\tan\beta=v_2/v_1$. 
Then, only one of the doublets has a VEV,
\be
\widehat{\Phi}_1= \doublet{$\begin{scriptsize}$ G^+ $\end{scriptsize}$}{\frac{v+\phi_1+iG^0}{\sqrt{2}}} ,\quad
\widehat{\Phi}_2= \doublet{$\begin{scriptsize}$ H^+ $\end{scriptsize}$}{\frac{\phi_2+i\phi_3}{\sqrt{2}}} ,\quad
\widehat{S} = \frac{1}{\sqrt{2}} (w  + \phi_4 + i \phi_5),
\label{2hdms-fields}
\ee
and one can separate the Goldstone bosons, $G^\pm,G^0$, from the physical states. The charged Higgs mass is calculated to be
\be
m_{H^\pm}^2 =  \frac{\Re\mu_3^2}{\sin\beta \cos\beta}- \frac{v^2}{2} (\lambda_4
+2 \Re\lambda_5 ).
\ee
The neutral mass-squared matrix, $\mathcal{M}^2$, shown in detail in Appendix~\ref{2hdms-details}, is a $5\times 5$ matrix which is diagonalised by the rotation matrix $R$,
\be
\label{mass-2hdms}
R^T \mathcal{M}^2 R = \mathcal{M}^2_{diag} = \mbox{diag}(m^2_{h_1}, m^2_{h_2}, m^2_{h_3}, m^2_{h_4}, m^2_{h_5})
\ee
where, as before, we take $h_1$ to be the observed Higgs boson at the LHC.

The rotation matrix, $R$, contains ten mixing angles, $\theta_{12-15},\theta_{23-25}$, $\theta_{34,35}$ and $\theta_{45}$ among which five represent CP-violation, namely $\theta_{13,15,23,25,34}$, and will vanish in the CP-conserving limit.
We, therefore, take these angles to be small since, as it will be shown later, they prove to be very small in the interesting and allowed regions of the parameter space. Of the remaining
angles $\theta_{12}$ and $\theta_{14}$ represent the mixing of the SM-like Higgs with the other CP-even states. To keep this state mostly doublet-like and to agree with the observed Higgs data, we take these angles to be small.

The remaining angles, $\theta_{24}$, $\theta_{35}$ and $\theta_{45}$ do not contribute
to the observables in which we are interested here, and therefore, to simplify the analysis,
we assume all mixing angles to be small ($\cos\theta_i \simeq 1$ and $\sin\theta_i \simeq \theta_i$). As a result, the rotation matrix, $R$, simplifies to the form
\be
\label{rotation-2hdms}
\phi_i=R_{ij} h_j  \mbox{ ; } \quad
\left(
\begin{array}{c}
\phi_1\\ \phi_2\\\phi_3\\ \phi_4\\ \phi_5
\end{array} \right)
=
\left(
\begin{array}{ccccc}
1 & \theta_{12} & \theta_{13} & \theta_{14} &\theta_{15}\\
-\theta_{12} & 1 & \theta_{23} &\theta_{24}&\theta_{25} \\
-\theta_{13} & -\theta_{23} & 1 &\theta_{34} &\theta_{35} \\
-\theta_{14} & -\theta_{24} & -\theta_{34} &1 &\theta_{45} \\
-\theta_{15} & -\theta_{25} & -\theta_{35} & -\theta_{45} & 1 \\
\end{array} \right)
\left(
\begin{array}{c}
h_1\\ h_2\\ h_3\\ h_4\\ h_5\\
\end{array} \right).
\ee
With this simplified form, one can calculate the angles in terms of the parameters of the potential as shown in Appendix~\ref{2hdms-details}.

After minimisation, the 32 independent parameters of the model,
\be
\mu_{1,2,4}^2,  ~\Re\mu_{3,5}^2,
~\lambda_{1-4}, ~\lambda_{8,11,13},
~\Re\kappa_{1-5}, ~\Im\kappa_{1-5}, ~\Re\lambda_{5,9,10,12,14}, ~\Im\lambda_{5,9,10,12,14},
\ee
can be expressed in terms of
\be
\tan\beta, ~v,~w, ~m_{h_{1-5}}, ~m_{H^\pm}, ~\theta_{12-15}, ~\theta_{23-25}, ~\theta_{34,35,45},
~\Re\mu_{3,5}^2, ~\Im\kappa_{1-5}, ~\Re\lambda_{5}, ~\Im\lambda_{5,9,10,12,14},
\ee
which we take as input parameters for our numerical calculations.
In all the results that follow, we take into account the same theoretical and experimental bounds as in section \ref{section-2hdm} translated to fit the 2HDM+CS model accordingly.

\subsection{$a_\mu$ and $d_e$ in 2HDM+CS}
Due to the singlet nature of $\phi_4$ and $\phi_5$, they do not directly couple to the SM fermions and gauge bosons.
As a result,
the Yukawa and kinetic terms are similar to the 2HDM discussed in Section \ref{section-yukawa} with $i=1,\cdots,5$, corresponding to the five scalar mass eigenstates, $h_{1,2,3,4,5}$.
The last two rows of the rotation matrix in Eq.~\eqref{rotation-2hdms} do not appear in the calculations of $d_e$ and $a_\mu$ whose contributions are very much 2HDM-like with the sums running over all five scalar mass eigenstates, $h_{1,2,3,4,5}$, in Eqs.~\eqref{de-1loop-2hdm}-\eqref{amu-2loop-2hdm}.

We have studied numerically all four 2HDM-like types with an exemplary value of $0.1$ for all sub-dominant angles. Similar to the 2HDM scenario, we plot the constraints on $d_e$ and $a_\mu$ in the ($\theta_{13}$,$\cot\beta$)-plane for various values of scalar masses.

\begin{figure}[t!]
\begin{center}
\includegraphics[scale=0.60]{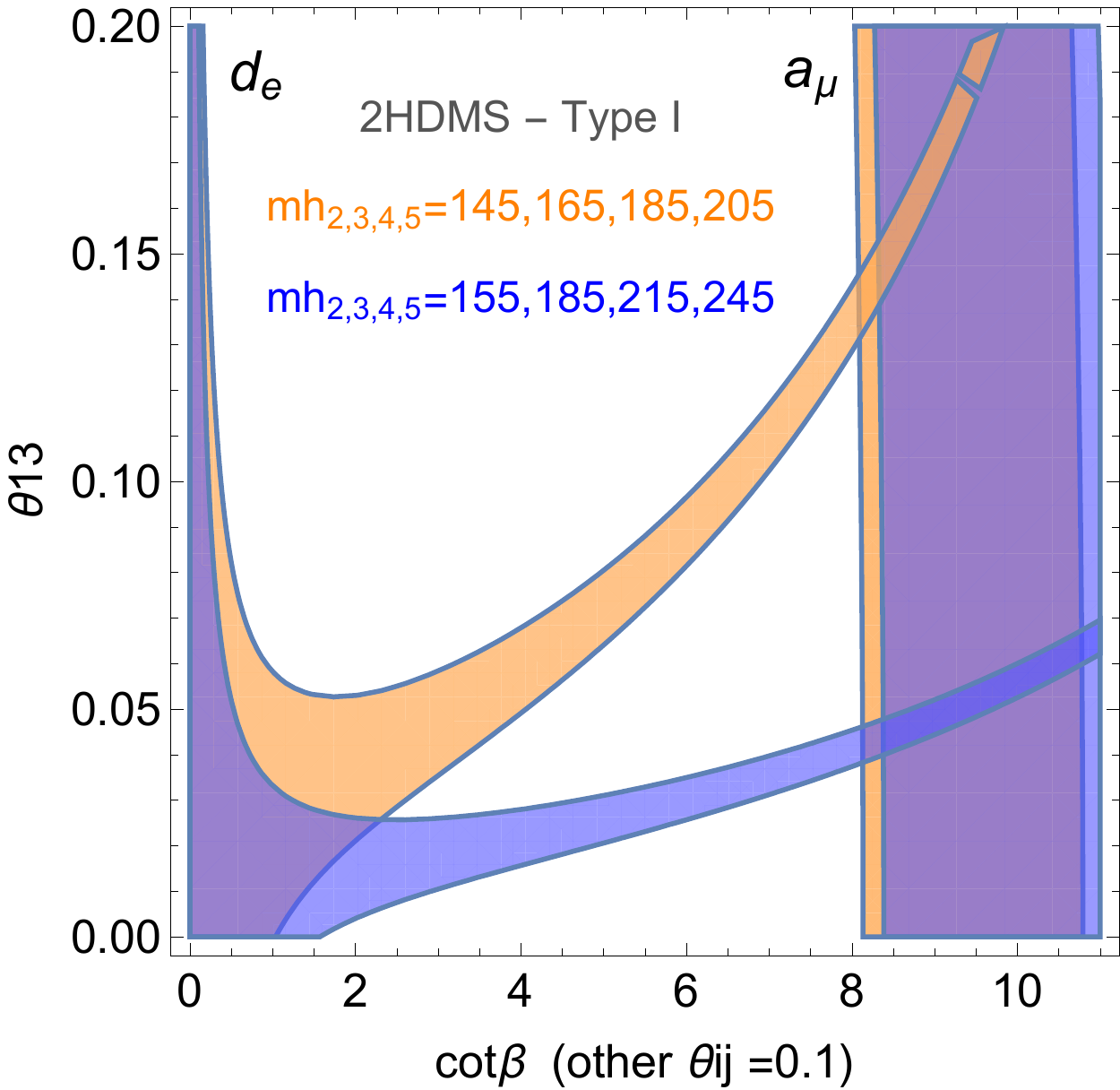}~~
\includegraphics[scale=0.60]{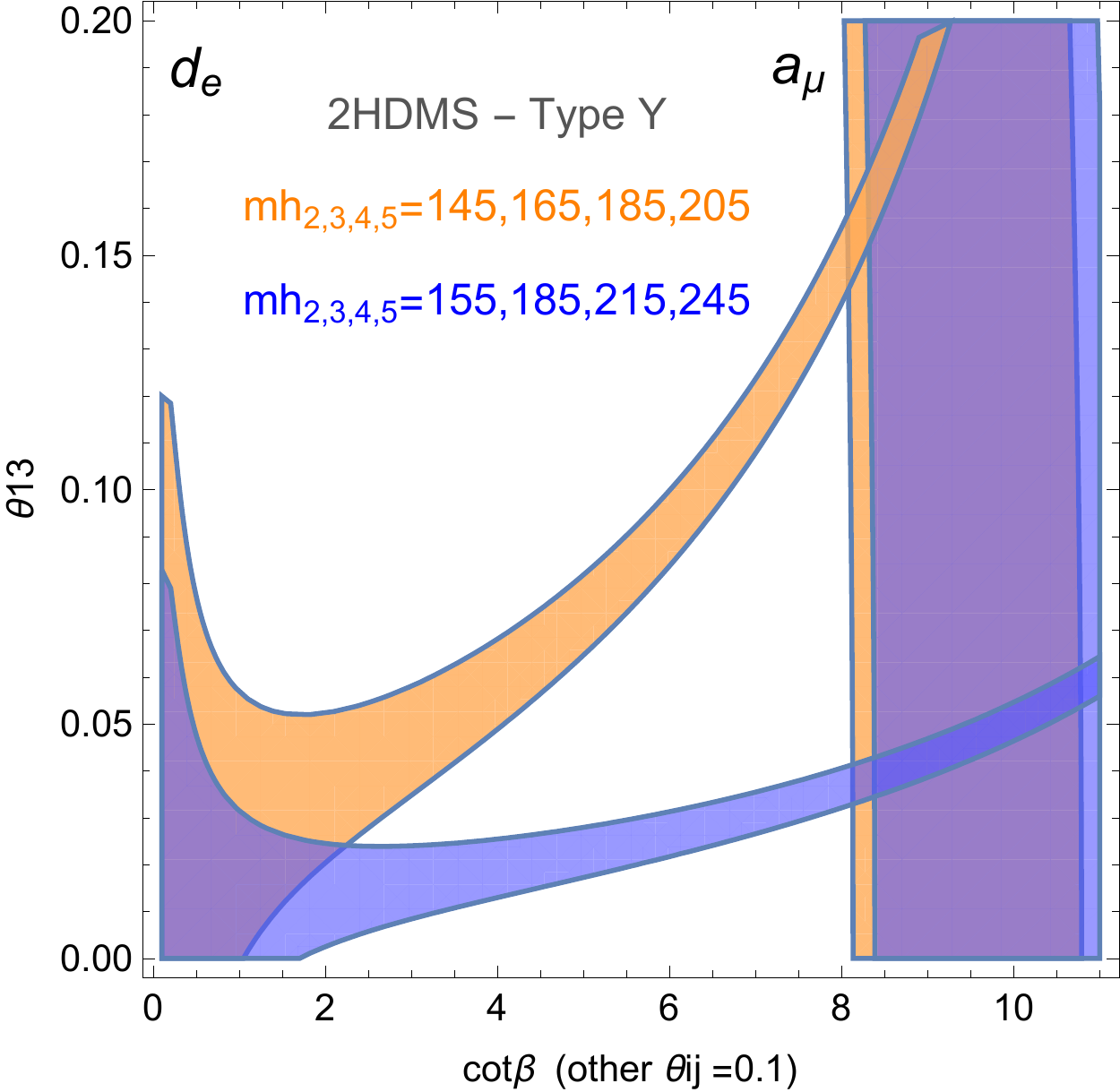}
\caption{Regions surviving $d_e$ bounds while producing $a_\mu$ in 2HDM+CS Type-I and Y for mid-range $m_{h_{2,3,4,5}}$ masses (in GeV) and fixed values of angles. Note that by changing $m_{h_{2,3,4,5}}$ masses, it is possible to cover the whole plane.}
\label{2HDMS-de-amu-medium}
\end{center}
\end{figure}

The main conclusion here is very similar to what was found in the 2HDM
subsection~\ref{2HDM-results}: only in Type I and Y and only in the
case of medium range masses $m_{h_{2,3,4,5}} \approx m_{h_1}$ is one able to
produce the observed value of $a_\mu$ while remaining consistent with
the $d_e$ constraints as shown in Figure \ref{2HDMS-de-amu-medium}.
However, as mentioned before, such large values of $\cot\beta$ are ruled out due to flavour and/or collider constraints.

\section{Conclusions and outlook}
\label{section-conclusion}

We have studied popular scalar extensions of the Standard Model and their contributions to the muon $(g-2)$, and electric dipole moment of the electron, $e$EDM.
Concretely, we studied first the real and complex singlet extension of the SM,
and second, as the main part of our analysis, we considered the 2-Higgs doublet model (2HDM) of Types I, II, X and Y and the 2HDM extended with the inclusion of a singlet scalar.

In the singlet extension of the SM, CP-violation is introduced by a
non-trivial higher dimensional operator connecting the singlet scalar with
the Higgs field and the top quark. We found that while the imaginary part
of this coupling can be compatible with the $e$EDM bounds,
a very large real part must be introduced in order to explain $a_\mu$.


In the CP-conserving limit, extensive $(g-2)_\mu$ studies have been done in different types of 2HDM with varying scalar masses, mixing angles and $\tan\beta$s in the literature. 
It has been shown that at 1-loop level contributions to $a_\mu$ are positive for the CP-even scalars and negative for the CP-odd scalars (and the charged scalar whose negative effect is negligible). 
The dominant contribution is from the 2-loop processes to which CP-even scalars contribute negatively and CP-odd scalars contribute possitively. 
As a result, very light CP-odd scalars have been shown to produce a large enough $a_\mu$ in large $\tan\beta$ regions for Type II,X and large $\cot\beta$ regons in Type I,Y. 
It is, therefore, intuitive to expect that by introducing CP-violation, less dramatic values of $\tan\beta$/$\cot\beta$ or scalar masses are required to produce an adequate $a_\mu$ contribution. 

All our calculations, confirm the known results in the CP-conserving limit and show the well-understood $(g-2)_\mu$ behaviour of different 2HDM types for varying masses in a wide range of $\tan\beta$s. Incremental changes in the scalar masses do not change the $a_\mu$ drastically.
By introducing CP-violation, which only adds the two CP-violating angles $\theta_{13}$ and $\theta_{23}$ to the known input parameters, we show that indeed the value of $a_\mu$ is affected. 
This effect, however small, is enhanced by increasing the amount of CP-violation (manifested in the values of $\theta_{13},\theta_{23}$ angles).
On the other hand, $\theta_{13},\theta_{23}$ angles are strongly constrained by eEDM experiments which only leave a small window in the parameter space to be explored.
We provide the detailed formulas for $a_\mu$ and $d_e$ contributions to show the subdominant effect of the parameters for which an exemplary value has been chosen in the plots.

In particular, we found that only when all scalars are relatively close in mass, CP-violating Type I and Y 2HDMs explain the muon anomalous magnetic moment in $\cot\beta \sim 10$ which is also allowed by the $e$EDM constraints. However, such large values of $\cot\beta$ are already ruled out by flavour/collider experiments.
Therefore, given the current status of the global set of constraints applied on all values of $\cot\beta$, in the CP-violating 2HDM, there exist no
viable parameter space in agreement with both $a_\mu$ and $e$EDM bounds.
In the low $\cot\beta$ region, Type X remains the only 2HDM type which has a large enough contribution to $a_\mu$, while Type I is the preferred type when introducing CP-violation as it contributes minimally to $d_e$.

In the singlet extension of the 2HDM, we show that the 2HDM behaviour is repeated and the model is capable of explaining the $(g-2)_\mu$ within the $d_e$ bounds when all scalars are relatively close in mass and the Yukawa interactions are of Type I and Y. However, this only occurs in the $\cot\beta \approx 10$ region which, again, is ruled out by flavour and collider experiments.

We have presented a robust way to implement the constraints on the muon anomalous magnetic moment and electric dipole moment of the electron on these models. Our central finding is that this allows one to categorically
exclude different types of 2HDMs and 2HDM+CS and, consequently, identify most viable
Yukawa interaction patterns which would be useful for more general model building based e.g. on the paradigm of minimal flavour violation.


The scalar extensions we have studied are applied in attempts to
explain the BAU via electroweak baryogenesis. Since then also
CP violation in the new sector needs to be introduced, our results help in establishing phenomenological viability of these models.

\section*{Acknowledgements}
The authors would like to thank Dominik Stoeckinger for useful discussions.
We acknowledge support from the Academy of Finland projects 274503, 267842 and 310130. VK also acknowledges support from
the H2020-MSCA-RISE-2014 grant no. 645722 (NonMinimalHiggs).
NK is supported by Vilho, Yrj{\"o} and Kalle V{\"a}is{\"a}l{\"a}
Foundation.

\appendix
\section{Loop functions}
\label{loop-functions}
The loop functions are:
\bea
f(z)&=&\frac{1}{2}z\int_{0}^{1}dx\frac{1-2x(1-x)}{x(1-x)-z}\log\left(\frac{x(1-x)}{z}\right),\\
g(z)&=&\frac{1}{2}z\int_{0}^{1}dx\frac{1}{x(1-x)-z}\log\left(\frac{x(1-x)}{z}\right),\\
h(z)&=&z^{2}\frac{\partial}{\partial z}\left(\frac{g(z)}{z}\right)=\frac{z}{2}\int_{0}^{1}\frac{dx}{z-x(1-x)}\left[1+\frac{z}{z-x(1-x)}\log\left(\frac{x(1-x)}{z}\right)\right].
\eea

\section{Details of minimisation of the 2HDM potential}
\label{2hdm-details}
The elements of the symmetric neutral mass-squared matrix, $\mathcal{M}^2$ in Eq. (\ref{mass-2hdm}), are of the form
\bea
\mathcal{M}^2_{11}&=&
\frac{1}{8} v^2 \biggl(4 \cos (2 \beta ) (\lambda_1-\lambda_2)+\lambda_1 \cos (4 \beta )+\lambda_2 \cos (4
   \beta )-\lambda_3 \cos (4 \beta )-\lambda_4 \cos (4 \beta )
\nonumber\\
&&
~~~~~~~
+3 \lambda_1+3 \lambda_2+\lambda_3+\lambda_4+4 \sin ^2(2 \beta ) \Re\lambda_5
\biggr)
\nonumber
\\
\mathcal{M}^2_{12}&=&
-\frac{1}{4} v^2 \sin (2 \beta ) \left[\cos (2 \beta ) \lambda_{1234}
+\lambda_1-\lambda_2-2 \cos (2 \beta ) \Re\lambda_5\right]
\nonumber
\\
\mathcal{M}^2_{13}&=&
-v^2 \sin (\beta ) \cos (\beta ) \Im(\lambda_5)
\nonumber
\\
\mathcal{M}^2_{22}&=&
\frac{1}{4} \left[2 \csc (\beta ) \sec (\beta ) \Re\mu_3^2+v^2 \sin ^2(2 \beta ) (\lambda_{1234}-2 \Re\lambda_5)\right]
\nonumber
\\
\mathcal{M}^2_{23}&=&
-\frac{1}{2} v^2 \cos (2 \beta ) \Im(\lambda_5)
\nonumber
\\
\mathcal{M}^2_{33}&=&
\csc (2 \beta ) \Re\mu_3^2-v^2 \Re\lambda_5,
\label{mass-matrix-elements-2hdm}
\eea
where $\lambda_{1234}=\lambda_1+\lambda_2-\lambda_3-\lambda_4$.

The angles defining the rotation matrix, $R$ in Eq. (\ref{rotation-2hdm}), are calculated to be
\begin{eqnarray}
\theta_{12}&=&
\frac{v^2 (2 \sin (2 \beta ) (\lambda_1-\lambda_2)+\sin (4 \beta ) (\lambda_{1234}-2 \Re\lambda_5))}
{\rho_2
-4 \csc (\beta ) \sec (\beta ) \Re\mu_3^2}
\\
\theta_{13}&=&
\frac{8 v^2 \sin ^2(2 \beta ) \Im(\lambda_5) (\cos (2 \beta ) (\lambda_1-\lambda_2)+\lambda_{1234}+2 \Re\lambda_5)}{
\rho_3
-2 v^2 \sin (2\beta )
\left[
\rho_4
+\rho_5
-4 (\cos (4 \beta )+7) \Re\lambda_5^2
\right]}
\\
\theta_{23}&=&
\frac{2 \Im(\lambda_5) \left[\rho_6-16 \cos (2 \beta ) \Re\mu_3^2\right]}{
\rho_3
-2 v^2 \sin (2 \beta )
\left[
\rho_4
+\rho_5
-4 (\cos (4 \beta)+7) \Re\lambda_5^2
\right]},
\end{eqnarray}
where
\bea
\rho_1&=& \cos (2 \beta ) (\lambda_1-\lambda_2)\\
\rho_2&=& 2 v^2 \left[2 \rho_1+\lambda_1+\lambda_2+\lambda_3+\lambda_4+\cos (4 \beta ) (\lambda_{1234}-2 \Re\lambda_5)+2 \Re\lambda_5\right]\\
\rho_3&=&8\Re\mu_3^2\left[\left(\cos (4 \beta )-1\right)\lambda_{1234}-2 \left(\cos (4 \beta )+3\right) \Re\lambda_5\right]\\
\rho_4&=&(\cos (4 \beta )-1) \left(4 \lambda_1 \lambda_2-(\lambda_3+\lambda_4)^2\right)\\
\rho_5&=&-4 \Re\lambda_5 \left[4\rho_1+\cos (4 \beta )(\lambda_3+\lambda_4)+4 \lambda_1+4 \lambda_2-\lambda_3-\lambda_4\right]\\
\rho_6&=&2 v^2 \sin (2 \beta ) \left[4 \cos (2 \beta ) (\lambda_1+\lambda_2)+(\cos (4 \beta )+3)
   (\lambda_1-\lambda_2)+8 \cos (2 \beta ) \Re\lambda_5\right].
\eea

\section{Constraints on the parameters}
\label{constraints}

\subsection{Theoretical bounds}

\begin{enumerate}

\item
\textbf{Stability of the potential}
\\[2mm] 
The scalar potential stability requires the potential to be bounded from below in any direction of the scalar space whose necessary and sufficient conditions are \cite{Grzadkowski:2009bt}
\be
\lambda_1 > 0,\quad \lambda_2 > 0, \quad \sqrt{\lambda_1\lambda_2} + \lambda_3 + \text{MIN}(0,~\lambda_4-|\lambda_5|) >0.   \label{vs3}  
\ee

\item
\textbf{Positive-definiteness of the Hessian - positivity of mass eigenvalues}
\\[2mm]
For the point $\langle \Phi_1 \rangle =\frac{v_1}{\sqrt{2}}, ~\langle \Phi_2 \rangle = \frac{v_2}{\sqrt{2}}$ to be a minimum of the potential, the second order derivative matrix must have a positive definite determinant.

Similar constrains are achieved by requiring the mass eigenvalues to be positive.

\item
\textbf{Perturbative unitarity}
\\[2mm] 
S-matrix unitarity for 2 to 2 elastic scattering, constrains the value of combinations of $\lambda$s in the potential \cite{Kanemura:1993hm}, \cite{Akeroyd:2000wc}.

\item
\textbf{Electroweak precision data}
\\[2mm]
Extra scalars affect the gauge boson propagators, parametrized by the oblique parameters $S$, $T$, $U$ \cite{Peskin:1991sw}-\cite{Haber:2010bw} by contributing to the neutral and charged current processes at low energies ($T$), or to neutral current processes at different energy scales ($S$). $U$ is generally small in new physics models. These parameters are constrained to be 
\begin{equation}
S = 0.05\pm0.11, \quad T = 0.09\pm0.13, \quad U = 0.01\pm0.11.
\end{equation}
determined from a fit with reference mass values of top and Higgs boson $m_{t}=173~\GeV$ and $m_{h}=125~\GeV$ are \cite{Baak:2014ora},\cite{Dolle:2009fn}.

\end{enumerate}

\subsection{Experimental bounds}

\begin{enumerate}

\item
\textbf{Flavour constraints}
\\[2mm]
The $B$ physics data provides constraints on $m_{H^\pm}$ and $\tan\beta$ in 2HDMs \cite{Misiak:2015xwa,Geng:1988bq,Deschamps:2009rh,Amhis:2016xyh}. Ref.~\cite{Mahmoudi:2009zx} provides a comprehensive study on various 
$B$ physics observables such as $b\to s\gamma$, $B^0$-$\bar{B}^0$ mixing, $B \to \tau\nu$ in 2HDMs.

Recently, the BaBar Collaboration has reported a measured ratios $\text{BR}(B \to D^{*} \tau \nu) / \text{BR}(B \to D^{*} \ell \nu)$ and $\text{BR}(B \to D \tau \nu) / \text{BR}(B \to D \ell \nu)$ ($\ell = e, \mu$) 
to deviate from the SM predictions by $2.7\;\sigma$ and $2.0\;\sigma$, respectively, and their combined deviation is $3.4\;\sigma$ \cite{Lees:2012xj}. Note that these deviations cannot be simultaneously explained by a $Z_2$ symmetric 2HDM which is flavour conserving, with or withour CP-violation.

\item
\textbf{Direct searches for extra Higgs bosons at the LHC}
\\[2mm] 
The search for extra neutral Higgs bosons decaying into $bb$, $\tau\tau$, $\gamma \gamma$, $Z\gamma$, $ZZ$, $WW$, $hh$ and $hZ$ \cite{Aad:2014fha,Aad:2014ioa,Aad:2014vgg,Aad:2015agg,Aad:2015kna,Aad:2015wra,Aad:2015xja,ATLAS-CONF-2016-056,ATLAS-CONF-2016-062,ATLAS-CONF-2016-059,ATLAS-CONF-2016-085,ATLAS-CONF-2016-074,ATLAS-CONF-2016-044,ATLAS-CONF-2016-082,ATLAS-CONF-2016-079,ATLAS-CONF-2016-071,ATLAS-CONF-2016-004,ATLAS-CONF-2016-015,ATLAS-CONF-2016-017,Khachatryan:2015cwa,Khachatryan:2015lba,Khachatryan:2015tha,Khachatryan:2015tra,Khachatryan:2015yea,Khachatryan:2016sey,CMS-PAS-HIG-14-029,CMS-PAS-HIG-16-014,CMS-PAS-EXO-16-035,CMS-PAS-EXO-16-027,CMS-PAS-HIG-16-033,CMS-PAS-HIG-16-023,CMS-PAS-HIG-16-029,CMS-PAS-HIG-16-025,CMS-PAS-EXO-16-034,CMS-PAS-HIG-16-011,CMS-PAS-HIG-16-037,CMS-PAS-HIG-16-002,CMS-PAS-HIG-16-032,CMS-PAS-HIG-15-013,CMS-PAS-HIG-16-034} 
using the LHC Run-I and II data, excludes $\tan\beta \gtrsim 10~(30)$ for $m_A=300$ (700) GeV in MSSM.
A similar bound is expected in the non-supersymmetric Type-II 2HDM, since the structure of the Yukawa interactions are the same. 
In Type-I 2HDM, there is no $\tan\beta$ enhancement in the Yukawa couplings since the Yukawa couplings are suppressed by the factor of $\cot\beta$. The production cross section is, therefore, suppressed by $\cot^2\beta$.

\item
\textbf{$A \to Zh$ searches}
\\[2mm]
Using LHC Run-I data, an upper limit on the $\sigma(gg \to A)\times \text{BR}(A \to Zh) \times \text{BR}(h \to f\bar{f})$
has been given \cite{Aad:2015wra} for $m_A=220$-1000 GeV. 
The upper limit for $f=\tau\,(b)$ is measured to be $0.098-0.013$ pb ($0.57-0.014$ pb). 
Our typical $gg \to H, A$ cross section is $\simeq 1$ pb for $m_{H, A}=200$ GeV and $\tan\beta \gtrsim 2$, and 
the $A\to Zh$ branching ratio is $\lesssim 10^{-2}$. 
Considering that the decay rate of the SM-like Higgs boson does not change much from the SM prediction,
the $h \to \tau\tau (b\bar{b})$ branching ratio is $\sim 7\% (60\%)$, meaning that our cross section is well below the upper limit.

\item 
\textbf{Gauge bosons width} 
\\[2mm]
The contribution of the extra scalars to the total gauge bosons widths~\cite{Agashe:2014kda} constrain the scalar masses:
\begin{equation}\label{eq:gwgz}
m_{H,A}+m_{H^\pm}\,\geq\,m_W,~\,m_{H}+m_{A}\,\geq\,m_Z,~\,2\,m_{H^\pm}\,\geq\,m_Z, 
\end{equation}

\item 
\textbf{Direct searches for charged scalars and their lifetime} 
\\[2mm]
A conservative lower limit for the mass of charged scalars is taken to be:
 $m_{H^\pm}\,\geq\,70\,\GeV$~\cite{Pierce:2007ut},\cite{Aad:2014kga,Khachatryan:2015qxa,Aad:2015typ,ATLAS-CONF-2016-088,CMS-PAS-HIG-16-031,ATLAS-CONF-2016-089,ATLAS-CONF-2016-104}. 

Moreover, to satisfy the bounds from long-lived charged particle searches, an upper limit is set on their lifetime to be $\tau_{H^\pm}\,\leq\,10^{-7}$ s, to guarantee their decay within the detector, which translates to an upper bound on their total decay width $\Gamma^\text{tot}_{H^\pm}\,\geq\,6.58\,\times\,10^{-18}\,\GeV$.

\item 
\textbf{Higgs signal strength} 
\\[2mm]
The signal strength, $\mu_{XY}^{}$, of the SM-like Higgs boson $h_1$ \cite{Khachatryan:2016vau,ATLAS-CONF-2016-063,ATLAS-CONF-2016-080,ATLAS-CONF-2016-081,ATLAS-CONF-2016-091,ATLAS-CONF-2016-112,CMS-PAS-HIG-16-003,CMS-PAS-HIG-16-020,CMS-PAS-HIG-16-033,CMS-PAS-HIG-16-043,CMS-PAS-HIG-16-038,CMS-PAS-HIG-17-003}, 
defined as
\bea
\mu_{XY}^{}& = &
\frac{\sigma(gg\to h_1)}{\sigma(gg\to h_{\text{SM}} )}\times \frac{\text{BR}(h_1\to XY)}{\text{BR}(h_{\text{SM}}\to XY)},\quad
XY = W^+W^-,~ZZ,~gg,~\gamma\gamma,~Z\gamma,~\tau^+\tau^-, 
\nonumber\\[2mm]
\mu_{b\bar{b}} &=& \frac{\sigma(q\bar{q}\to h_1 V)}{\sigma(q\bar{q}\to h_{\text{SM}} V)}\times \frac{\text{BR}(H_1\to  b\bar{b})}{\text{BR}(h_{\text{SM}} \to b\bar{b})}.
\eea
limits the contribution from new scalars to the Higgs observables.

\end{enumerate}

\section{Details of minimisation of the 2HDM+CS potential}
\label{2hdms-details}
The minimum of the potential is realised at
\bea
\label{minimisation-2hdms}
\mu_1^2 &=&
\frac{1}{8} \biggl(-8 t_\beta \Re\mu_3^2-2 v^2 c_{3 \beta} c^{-1}_\beta \Re\lambda_5+2 v^2 \Re\lambda_5+8 w^2
\Re\lambda_{12}+8 \sqrt{2} w \Re\kappa_4
\\
&&~~~~
+2 \lambda_1 v^2 c_{3 \beta} c^{-1}_\beta-\lambda_3 v^2 c_{3 \beta} c^{-1}_\beta-\lambda_4 v^2 c_{3 \beta} c^{-1}_\beta +6 \lambda_1 v^2+\lambda_3 v^2+\lambda_4 v^2+4 \lambda_{11}
w^2\biggr)
\nonumber
\\
\mu_2^2 &=&
\frac{1}{8} \biggl(-8 t^{-1}_\beta \Re\mu_3^2+2 v^2 s_{3 \beta} s^{-1}_\beta \Re\lambda_5+2 v^2 \Re\lambda_5+8 w^2
\Re\lambda_{14}+8 \sqrt{2} w \Re\kappa_5
\nonumber\\
&&~~~~
-2 \lambda_2 v^2 s_{3 \beta} s^{-1}_\beta +\lambda_3 v^2 s_{3 \beta} s^{-1}_\beta +\lambda_4 v^2 s_{3 \beta} s^{-1}_\beta +6 \lambda_2 v^2+\lambda_3 v^2+\lambda_4 v^2+4 \lambda_{13}
w^2\biggr)
\nonumber
\\
\Im\mu_3^2 &=&
v^2 s_{ \beta} c_{ \beta} \Im\lambda_5
\nonumber
\\
\mu_4^2 &=&
\frac{1}{2 w}
\biggl( 2\sqrt{2} \Re\kappa_1+\sqrt{2} v^2 c^2_\beta  \Re\kappa_4+\sqrt{2} v^2 cs^2_\beta \Re\kappa_5
+2 v^2 w c^2_\beta
\Re\lambda_{12}
\nonumber\\
&&~~~~~~
+2 v^2 w s^2_\beta \Re\lambda_{14}+4 w^3 \Re\lambda_{10}+4 w^3 \Re\lambda_9+3 \sqrt{2} w^2
\Re\kappa_2
\nonumber\\
&&~~~~~~
+3 \sqrt{2} w^2 \Re\kappa_3-4 w \Re\mu_5^2+\lambda_{11} v^2 w c^2_\beta +\lambda_{13} v^2 w s^2_\beta+2 \lambda_8 w^3
\biggr)
\nonumber
\\
\Im\mu_5^2 &=&
\frac{1}{4 w} \biggl(
2 \sqrt{2} \Im\kappa_1+\sqrt{2} v^2 c^2_\beta \Im\kappa_4+\sqrt{2} v^2 s^2_\beta \Im\kappa_5+2 v^2 w c^2_\beta
\Im\lambda_{12}
\nonumber\\
&&~~~~~
+2 v^2 w s^2_\beta \Im\lambda_{14}+4 w^3 \Im\lambda_{10}+2 w^3 \Im\lambda_9+3 \sqrt{2} w^2
\Im\kappa_2+\sqrt{2} w^2 \Im\kappa_3
\biggr).
\nonumber
\eea

The elements of the symmetric neutral mass-squared matrix, $\mathcal{M}^2$ in Eq. (\ref{mass-2hdms}), are of the form
\bea
\mathcal{M}^2_{11}&=&
 \frac{1}{8} v^2 (4 \cos (2 \beta ) (\lambda_1-\lambda_2)+3 \lambda_1+3 \lambda_2+\lambda_3+\lambda_4+\cos (4
   \beta ) (\lambda_{1234}-2 \Re\lambda_5)+2 \Re\lambda_5)\nonumber
\\
\mathcal{M}^2_{12}&=&
-\frac{1}{4} v^2 \sin (2 \beta ) (\cos (2 \beta ) \lambda_{1234}
+\lambda_1-\lambda_2-2 \cos (2 \beta ) \Re\lambda_5)\nonumber
\\
\mathcal{M}^2_{13}&=&
-v^2 \sin (\beta ) \cos (\beta ) \Im(\lambda_5)\nonumber
\\
\mathcal{M}^2_{14}&=&
\frac{1}{2} v \left(\cos ^2(\beta ) \left(\sqrt{2} \Re\kappa_4+2 w \Re\lambda_{12}+\lambda_{11} w\right)+\sin ^2(\beta )
   \left(\sqrt{2} \Re\kappa_5+2 w \Re\lambda_{14}+\lambda_{13} w\right)\right)\nonumber
\\
\mathcal{M}^2_{15}&=&
-\frac{1}{2} v \left(\cos ^2(\beta ) \left(\sqrt{2} \Im(\kappa_4)+2 w \Im(\lambda_{12})\right)+\sin ^2(\beta ) \left(\sqrt{2}
   \Im(\kappa_5)+2 w \Im(\lambda_{14})\right)\right)\nonumber
\\
\mathcal{M}^2_{22}&=&
\frac{1}{4} \left(2 \csc (\beta ) \sec (\beta ) \Re\mu_3^2+v^2 \sin ^2(2 \beta ) (\lambda_{1234}-2 \Re\lambda_5)\right)\nonumber
\\
\mathcal{M}^2_{23}&=&
-\frac{1}{2} v^2 \cos (2 \beta ) \Im(\lambda_5)\nonumber
\\
\mathcal{M}^2_{24}&=&
-\frac{1}{4} v \sin (2 \beta ) \left(\sqrt{2} \Re\kappa_4-\sqrt{2} \Re\kappa_5+w (\lambda_{11}-\lambda_{13}+2
   \Re\lambda_{12}-2 \Re\lambda_{14})\right)\nonumber
\\
\mathcal{M}^2_{25}&=&
\frac{1}{4} v \sin (2 \beta ) \left(\sqrt{2} \Im(\kappa_4)-\sqrt{2} \Im(\kappa_5)+2 w (\Im(\lambda_{12})-\Im(\lambda_{14}))\right)
\nonumber
\eea
\bea
\mathcal{M}^2_{33}&=&
\csc (2 \beta ) \Re\mu_3^2-v^2 \Re\lambda_5\nonumber
\\
\mathcal{M}^2_{34}&=&
0\nonumber
\\
\mathcal{M}^2_{35}&=&
0\nonumber
\\
\mathcal{M}^2_{44}&=&
\frac{1}{4 w}\bigg(-2 \sqrt{2} \Re(\kappa_1)-\sqrt{2} v^2 \cos ^2(\beta ) \Re\kappa_4-\sqrt{2} v^2 \sin ^2(\beta ) \Re\kappa_5\nonumber\\
&&~~~~~~+8 w^3 \Re\lambda_{10}+8 w^3 \Re\lambda_9+3 \sqrt{2} w^2 \Re\kappa_2+3 \sqrt{2} w^2 \Re\kappa_3+4 \lambda_8 w^3\bigg)\nonumber
\\
\mathcal{M}^2_{45}&=&
\frac{1}{4 w}\bigg(2 \sqrt{2} \Im(\kappa_1)+\sqrt{2} v^2 \cos ^2(\beta ) \Im(\kappa_4)+\sqrt{2} v^2 \sin ^2(\beta ) \Im(\kappa_5)-8 w^3
   \Im(\lambda_{10})\nonumber\\
&&~~~~~~-4 w^3 \Im(\lambda_9)-3 \sqrt{2} w^2 \Im(\kappa_2)-\sqrt{2} w^2 \Im(\kappa_3)\bigg)\nonumber
\\
\mathcal{M}^2_{55}&=&
-\frac{1}{4 w}\bigg(2 \sqrt{2} \Re(\kappa_1)+\sqrt{2} v^2 \cos ^2(\beta ) \Re\kappa_4+\sqrt{2} v^2 \sin ^2(\beta ) \Re\kappa_5\nonumber\\
&&~~~~~~+4 v^2 w \cos ^2(\beta ) \Re\lambda_{12}+4 v^2 w \sin ^2(\beta ) \Re\lambda_{14}+16 w^3 \Re\lambda_{10}\nonumber\\
&&~~~~~~+4 w^3
   \Re\lambda_9+9 \sqrt{2} w^2 \Re\kappa_2+\sqrt{2} w^2 \Re\kappa_3-8 w \Re\mu_5^2\bigg).
\eea

The zero entries in the mass-squared matrix are the result of the imposed $Z_2$ symmetry and vanishing parameters in Eq. (\ref{zero-params}).

The angles defining the rotation matrix, $R$ in Eq. (\ref{rotation-2hdms}), can be calculated in the general case, however the expressions are too lengthy to present here. We only show the values of the angles in the following approximation
\be
\Im\kappa_{1-5} =\Re\kappa_{1-5}=\Im\lambda_{5,9,10,12,14} =\Re\lambda_{5,9,10,12,14}=\lambda_{11,13} = \epsilon \ll 1,
\ee
which is obtained by assuming that CP-violation is small and $h_1$ is mostly CP-even and doublet-like. As a result, the angles are calculated to be as follows.
\bea
\theta_{12}&=&
\frac{v^2 \sin (2 \beta ) (\cos (2 \beta ) (\lambda_{1234}-2 \epsilon )+\lambda_1-\lambda_2)}{v^2 (\cos (4 \beta ) (\lambda_{1234})+2 \cos (2 \beta ) (\lambda_1-\lambda_2)+\lambda_1+\lambda_2+\lambda_3+\lambda_4)-2 \csc (\beta ) \sec (\beta ) \Re\mu_3^2}\nonumber
\\
\theta_{13}&=&
\frac{2 v^2 \epsilon  (\cos (2 \beta ) (\lambda_1-\lambda_2)+\lambda_{1234})}{v^2 \sin
   (2 \beta ) \left(4 \lambda_1 \lambda_2-(\lambda_3+\lambda_4)^2\right)-4 \Re\mu_3^2 (\lambda_{1234})}\nonumber
\\
\theta_{14}&=&
\frac{v \left(3 w+\sqrt{2}\right) \epsilon  \left(16 \Re\mu_3^2-2 \sin (2 \beta ) \left(v^2 \cos (4 \beta ) (\lambda_{1234})+v^2 (-\lambda_{1234})+8 \lambda_8
   w^2\right)\right)}
{\sin (2 \beta ) \left[\delta_1
+\delta_2
-32\lambda_8^2 w^4\right]
-4 \Re\mu_3^2 \left[\delta_3-8\lambda_8 w^2\right]}\nonumber
\\
\theta_{15}&=&
\frac{2 v \left(2 w+\sqrt{2}\right) \epsilon  \left(-8 \sin (2 \beta ) \Re\mu_5^2+4 \Re\mu_3^2+v^2 \sin ^3(2 \beta ) (\lambda_{1234})\right)}
{2 \Re\mu_3^2 \left[\delta_4
-16 \Re\mu_5^2\right]
+\sin (2 \beta ) \left[64 (\Re\mu_5^2)^2+\delta_5+v^4 \sin ^2(2 \beta ) \left(4 \lambda_1 \lambda_2-(\lambda_3+\lambda_4)^2\right)\right]}\nonumber
\eea
\bea
\theta_{23}&=&
\frac{\epsilon  \csc (2 \beta ) \left(8 \cot (2 \beta ) \Re\mu_3^2-v^2 (4 \cos (2 \beta ) (\lambda_1+\lambda_2)+(\cos (4 \beta )+3)
   (\lambda_1-\lambda_2))\right)}{4 \Re\mu_3^2 (\lambda_1+\lambda_2-\lambda_3-\lambda_4)+v^2 \sin (2
   \beta ) \left((\lambda_3+\lambda_4)^2-4 \lambda_1 \lambda_2\right)}\nonumber
\\
\theta_{24}&=&
\frac{16 v^3 \left(3 w+\sqrt{2}\right) \epsilon  \sin ^2(\beta ) \cos ^2(\beta ) (\cos (2 \beta ) (\lambda_{1234})+\lambda_1-\lambda_2)}
{\sin (2 \beta ) \left[\delta_6
+\delta_7-32 \lambda_8^2 w^4\right]
-4 \Re\mu_3^2 \left[\delta_8-8 \lambda_8 w^2\right]}\nonumber
\\
\theta_{25}&=&
-\frac{2 v^3 \left(2 w+\sqrt{2}\right) \epsilon  \sin ^2(2 \beta ) (\cos (2 \beta ) (\lambda_1+\lambda_2-\lambda_3-\lambda_4)+\lambda_1-\lambda_2)}
{\sin (2 \beta ) \left[\delta_9
+v^4 \sin ^2(2 \beta ) \left((\lambda_3+\lambda_4)^2-4 \lambda_1\lambda_2\right)\right]
-2 \Re\mu_3^2 \left[\delta_{10}
-16 \Re(\mu_5^2)\right]}\nonumber
\\
\theta_{34}&=&
0\nonumber
\\
\theta_{35}&=&
0\nonumber
\\
\theta_{45}&=&
\frac{\epsilon  \left(-\sqrt{2} v^2+12 w^3+4 \sqrt{2} w^2-2 \sqrt{2}\right)}{4 \lambda_8 w^3-8 w \Re\mu_5^2},
\eea
where
\bea
\delta_1&=&v^4\left[ \cos (4 \beta ) \left(4 \lambda_1 \lambda_2-(\lambda_3+\lambda_4)^2\right)-4\lambda_1 \lambda_2 +\lambda_3^2 +2 \lambda_3 \lambda_4 +\lambda_4^2 \right]\nonumber
\\
\delta_2&=&v^2 w^2 \left[16 \lambda_8 \cos (2 \beta ) (\lambda_1-\lambda_2)
+16 \lambda_1 \lambda_8
+16 \lambda_2 \lambda_8\right]\nonumber
\\
\delta_3&=&v^2\left[ \cos (4 \beta ) (\lambda_{1234})
+4 \cos (2 \beta ) (\lambda_1-\lambda_2)
+3 \lambda_1 +3 \lambda_2 +\lambda_3 +\lambda_4 \right]\nonumber
\\
\delta_4&=&v^2 \left[\cos (4 \beta ) (\lambda_{1234})
+4 \cos (2 \beta ) (\lambda_1-\lambda_2)+3 \lambda_1+3 \lambda_2+\lambda_3+\lambda_4\right]\nonumber
\\
\delta_5&=&-16 v^2 \Re\mu_5^2 [\cos (2 \beta ) (\lambda_1-\lambda_2)+\lambda_1+\lambda_2]\nonumber
\\
\delta_6&=&v^4 \left[\cos (4 \beta ) \left(4 \lambda_1 \lambda_2-(\lambda_3+\lambda_4)^2\right)
-4 \lambda_1 \lambda_2+\lambda_3^2 +2 \lambda_3 \lambda_4+\lambda_4^2 \right]\nonumber
\\
\delta_7&=&v^2 w^2\left[16 \lambda_8  \cos (2 \beta ) (\lambda_1-\lambda_2)+16 \lambda_1 \lambda_8+16 \lambda_2 \lambda_8\right]\nonumber
\\
\delta_8&=&v^2 \left[\cos (4 \beta ) (\lambda_1+\lambda_2-\lambda_3-\lambda_4)+4  \cos (2 \beta ) (\lambda_1-\lambda_2)+3 \lambda_1 +3
   \lambda_2 +\lambda_3 +\lambda_4 \right]\nonumber
\\
\delta_9&=&-64 (\Re\mu_5^2)^2+16 v^2 \Re\mu_5^2 (\cos (2 \beta ) (\lambda_1-\lambda_2)+\lambda_1+\lambda_2)\nonumber
\\
\delta_{10}&=&v^2 \left[\cos (4 \beta ) (\lambda_{1234})
+4\cos (2 \beta ) (\lambda_1-\lambda_2)+3 \lambda_1+3 \lambda_2+\lambda_3+\lambda_4\right].\nonumber
\eea



\begin{thebibliography}{99}

\bibitem{Kuzmin:1985mm} 
  V.~A.~Kuzmin, V.~A.~Rubakov and M.~E.~Shaposhnikov,
  Phys.\ Lett.\  {\bf 155B}, 36 (1985).


\bibitem{Kajantie:1996mn} 
  K.~Kajantie, M.~Laine, K.~Rummukainen and M.~E.~Shaposhnikov,
  Phys.\ Rev.\ Lett.\  {\bf 77}, 2887 (1996)
  [hep-ph/9605288].


\bibitem{Silveira:1985rk} 
  V.~Silveira and A.~Zee,
  Phys.\ Lett.\  {\bf 161B}, 136 (1985).


\bibitem{McDonald:1993ex} 
  J.~McDonald,
  Phys.\ Rev.\ D {\bf 50}, 3637 (1994)
  [hep-ph/0702143 [HEP-PH]].


\bibitem{Cline:2012hg} 
  J.~M.~Cline and K.~Kainulainen,
  JCAP {\bf 1301}, 012 (2013)
  [arXiv:1210.4196 [hep-ph]].


\bibitem{Alanne:2016wtx} 
  T.~Alanne, K.~Kainulainen, K.~Tuominen and V.~Vaskonen,
  JCAP {\bf 1608}, no. 08, 057 (2016)
  [arXiv:1607.03303 [hep-ph]].


\bibitem{Lindner:2016bgg} 
  M.~Lindner, M.~Platscher and F.~S.~Queiroz,
  Phys.\ Rept.\  {\bf 731}, 1 (2018)
  [arXiv:1610.06587 [hep-ph]].


\bibitem{Bian:2016zba} 
  L.~Bian and N.~Chen,
  Phys.\ Rev.\ D {\bf 95}, no. 11, 115029 (2017)
  [arXiv:1608.07975 [hep-ph]].


\bibitem{Kowalska:2017iqv} 
  K.~Kowalska and E.~M.~Sessolo,
  JHEP {\bf 1709}, 112 (2017)
  [arXiv:1707.00753 [hep-ph]].


\bibitem{Bennett:2006fi} 
  G.~W.~Bennett {\it et al.} [Muon g-2 Collaboration],
  Phys.\ Rev.\ D {\bf 73}, 072003 (2006)
  [hep-ex/0602035].


\bibitem{Blum:2013xva} 
  T.~Blum, A.~Denig, I.~Logashenko, E.~de Rafael, B.~Lee Roberts, T.~Teubner and G.~Venanzoni,
  arXiv:1311.2198 [hep-ph].


\bibitem{Baron:2013eja} 
  J.~Baron {\it et al.} [ACME Collaboration],
  Science {\bf 343}, 269 (2014)
  [arXiv:1310.7534 [physics.atom-ph]].


\bibitem{McDonald:1993ey} 
  J.~McDonald,
  Phys.\ Lett.\ B {\bf 323}, 339 (1994).


\bibitem{Profumo:2007wc} 
  S.~Profumo, M.~J.~Ramsey-Musolf and G.~Shaughnessy,
  JHEP {\bf 0708}, 010 (2007)
  [arXiv:0705.2425 [hep-ph]].


\bibitem{Barger:2008jx} 
  V.~Barger, P.~Langacker, M.~McCaskey, M.~Ramsey-Musolf and G.~Shaughnessy,
  Phys.\ Rev.\ D {\bf 79}, 015018 (2009)
  [arXiv:0811.0393 [hep-ph]].


\bibitem{Ahriche:2012ei} 
  A.~Ahriche and S.~Nasri,
  Phys.\ Rev.\ D {\bf 85}, 093007 (2012)
  [arXiv:1201.4614 [hep-ph]].


\bibitem{Turok:1990zg} 
  N.~Turok and J.~Zadrozny,
  Nucl.\ Phys.\ B {\bf 358}, 471 (1991).


\bibitem{Turok:1991uc} 
  N.~Turok and J.~Zadrozny,
  Nucl.\ Phys.\ B {\bf 369}, 729 (1992).


\bibitem{Funakubo:1993jg} 
  K.~Funakubo, A.~Kakuto and K.~Takenaga,
  Prog.\ Theor.\ Phys.\  {\bf 91}, 341 (1994)
  [hep-ph/9310267].


\bibitem{Davies:1994id} 
  A.~T.~Davies, C.~D.~froggatt, G.~Jenkins and R.~G.~Moorhouse,
  Phys.\ Lett.\ B {\bf 336}, 464 (1994).


\bibitem{Cline:1995dg} 
  J.~M.~Cline, K.~Kainulainen and A.~P.~Vischer,
  Phys.\ Rev.\ D {\bf 54}, 2451 (1996)
  [hep-ph/9506284].


\bibitem{Laine:2000rm} 
  M.~Laine and K.~Rummukainen,
  Nucl.\ Phys.\ B {\bf 597}, 23 (2001)
  [hep-lat/0009025].


\bibitem{Fromme:2006cm} 
  L.~Fromme, S.~J.~Huber and M.~Seniuch,
  JHEP {\bf 0611}, 038 (2006)
  [hep-ph/0605242].


\bibitem{Basler:2016obg} 
  P.~Basler, M.~Krause, M.~Muhlleitner, J.~Wittbrodt and A.~Wlotzka,
  JHEP {\bf 1702}, 121 (2017)
  [arXiv:1612.04086 [hep-ph]].


\bibitem{Basler:2017uxn} 
  P.~Basler, M.~Mühlleitner and J.~Wittbrodt,
  JHEP {\bf 1803}, 061 (2018)
  [arXiv:1711.04097 [hep-ph]].


\bibitem{Bonilla:2014xba} 
  C.~Bonilla, D.~Sokolowska, N.~Darvishi, J.~L.~Diaz-Cruz and M.~Krawczyk,
  J.\ Phys.\ G {\bf 43}, no. 6, 065001 (2016)
  [arXiv:1412.8730 [hep-ph]].


\bibitem{Kakizaki:2016dza} 
  M.~Kakizaki, A.~Santa and O.~Seto,
  Int.\ J.\ Mod.\ Phys.\ A {\bf 32}, no. 10, 1750038 (2017)
  [arXiv:1609.06555 [hep-ph]].


\bibitem{Chun:2015xfx} 
  E.~J.~Chun,
  EPJ Web Conf.\  {\bf 118}, 01006 (2016)
  [Pramana {\bf 87}, no. 3, 41 (2016)]
  [arXiv:1511.05225 [hep-ph]].


\bibitem{Broggio:2014mna} 
  A.~Broggio, E.~J.~Chun, M.~Passera, K.~M.~Patel and S.~K.~Vempati,
  JHEP {\bf 1411}, 058 (2014)
  [arXiv:1409.3199 [hep-ph]].


\bibitem{Cheung:2001hz} 
  K.~m.~Cheung, C.~H.~Chou and O.~C.~W.~Kong,
  Phys.\ Rev.\ D {\bf 64}, 111301 (2001)
  [hep-ph/0103183].


\bibitem{Harnik:2012pb} 
  R.~Harnik, J.~Kopp and J.~Zupan,
  JHEP {\bf 1303}, 026 (2013)
  [arXiv:1209.1397 [hep-ph]].


\bibitem{Burgess:2000yq} 
  C.~P.~Burgess, M.~Pospelov and T.~ter Veldhuis,
  Nucl.\ Phys.\ B {\bf 619}, 709 (2001)
  [hep-ph/0011335].


\bibitem{Davoudiasl:2004be} 
  H.~Davoudiasl, R.~Kitano, T.~Li and H.~Murayama,
  Phys.\ Lett.\ B {\bf 609}, 117 (2005)
  [hep-ph/0405097].


\bibitem{Yaguna:2008hd} 
  C.~E.~Yaguna,
  JCAP {\bf 0903}, 003 (2009)
  [arXiv:0810.4267 [hep-ph]].


\bibitem{Lerner:2009xg} 
  R.~N.~Lerner and J.~McDonald,
  Phys.\ Rev.\ D {\bf 80}, 123507 (2009)
  [arXiv:0909.0520 [hep-ph]].


\bibitem{TheATLASandCMSCollaborations:2015bln} 
  The ATLAS and CMS Collaborations,
  ATLAS-CONF-2015-044.


\bibitem{Falkowski:2015iwa} 
  A.~Falkowski, C.~Gross and O.~Lebedev,
  JHEP {\bf 1505}, 057 (2015)
  [arXiv:1502.01361 [hep-ph]].


\bibitem{Espinosa:2011eu} 
  J.~R.~Espinosa, B.~Gripaios, T.~Konstandin and F.~Riva,
  JCAP {\bf 1201}, 012 (2012)
  [arXiv:1110.2876 [hep-ph]].


\bibitem{Branco:1999fs} 
  G.~C.~Branco, L.~Lavoura and J.~P.~Silva,
  Int.\ Ser.\ Monogr.\ Phys.\  {\bf 103}, 1 (1999).


\bibitem{Lee:1973iz} 
  T.~D.~Lee,
  Phys.\ Rev.\ D {\bf 8}, 1226 (1973).


\bibitem{Gunion:1989we} 
  J.~F.~Gunion, H.~E.~Haber, G.~L.~Kane and S.~Dawson,
  Front.\ Phys.\  {\bf 80}, 1 (2000).


\bibitem{Branco:2011iw} 
  G.~C.~Branco, P.~M.~Ferreira, L.~Lavoura, M.~N.~Rebelo, M.~Sher and J.~P.~Silva,
  Phys.\ Rept.\  {\bf 516}, 1 (2012)
  [arXiv:1106.0034 [hep-ph]].

\bibitem{Keus:2015hva} 
  V.~Keus, S.~F.~King, S.~Moretti and K.~Yagyu,
  JHEP {\bf 1604}, 048 (2016)
  doi:10.1007/JHEP04(2016)048
  [arXiv:1510.04028 [hep-ph]].

\bibitem{Zarikas:1995qb}
  V.~Zarikas,
  Phys.\ Lett.\ B {\bf 384}, 180 (1996)
  [hep-ph/9509338].
 
\bibitem{Lahanas:1998wf} 
  A.~B.~Lahanas, V.~C.~Spanos and V.~Zarikas,
  Phys.\ Lett.\ B {\bf 472}, 119 (2000)
[hep-ph/9812535].

\bibitem{Aliferis:2014ofa} 
  G.~Aliferis, G.~Kofinas and V.~Zarikas,
Phys.\ Rev.\ D {\bf 91}, no. 4, 045002 (2015)
[arXiv:1406.6215 [hep-ph]].


\bibitem{Glashow:1976nt} 
  S.~L.~Glashow and S.~Weinberg,
  Phys.\ Rev.\ D {\bf 15}, 1958 (1977).


\bibitem{Paschos:1976ay} 
  E.~A.~Paschos,
  Phys.\ Rev.\ D {\bf 15}, 1966 (1977).


\bibitem{Barger:1989fj} 
  V.~D.~Barger, J.~L.~Hewett and R.~J.~N.~Phillips,
  Phys.\ Rev.\ D {\bf 41}, 3421 (1990).


\bibitem{Haber:2006ue} 
  H.~E.~Haber and D.~O'Neil,
  Phys.\ Rev.\ D {\bf 74}, 015018 (2006)
  Erratum: [Phys.\ Rev.\ D {\bf 74}, no. 5, 059905 (2006)]
  [hep-ph/0602242].


\bibitem{Davidson:2005cw} 
  S.~Davidson and H.~E.~Haber,
  Phys.\ Rev.\ D {\bf 72}, 035004 (2005)
  Erratum: [Phys.\ Rev.\ D {\bf 72}, 099902 (2005)]
  [hep-ph/0504050].


\bibitem{Mahmoudi:2009zx} 
  F.~Mahmoudi and O.~Stal,
  Phys.\ Rev.\ D {\bf 81}, 035016 (2010)
  [arXiv:0907.1791 [hep-ph]].


\bibitem{Cherchiglia:2016eui} 
  A.~Cherchiglia, P.~Kneschke, D.~Stöckinger and H.~Stöckinger-Kim,
  JHEP {\bf 1701}, 007 (2017)
  [arXiv:1607.06292 [hep-ph]].


\bibitem{Abe:2015oca} 
  T.~Abe, R.~Sato and K.~Yagyu,
  JHEP {\bf 1507}, 064 (2015)
  [arXiv:1504.07059 [hep-ph]].


\bibitem{Cherchiglia:2017uwv} 
  A.~Cherchiglia, D.~Stöckinger and H.~Stöckinger-Kim,
  arXiv:1711.11567 [hep-ph].


\bibitem{Keus:2016orl} 
  V.~Keus,
  PoS CHARGED {\bf 2016}, 017 (2016)
  [arXiv:1612.03629 [hep-ph]].


\bibitem{Grzadkowski:2009bt} 
  B.~Grzadkowski, O.~M.~Ogreid and P.~Osland,
  Phys.\ Rev.\ D {\bf 80}, 055013 (2009)
  [arXiv:0904.2173 [hep-ph]].


\bibitem{Kanemura:1993hm} 
  S.~Kanemura, T.~Kubota and E.~Takasugi,
  Phys.\ Lett.\ B {\bf 313}, 155 (1993)
  [hep-ph/9303263].


\bibitem{Akeroyd:2000wc} 
  A.~G.~Akeroyd, A.~Arhrib and E.~M.~Naimi,
  Phys.\ Lett.\ B {\bf 490}, 119 (2000)
  [hep-ph/0006035].


\bibitem{Peskin:1991sw} 
  M.~E.~Peskin and T.~Takeuchi,
  Phys.\ Rev.\ D {\bf 46}, 381 (1992).


\bibitem{Haber:2010bw} 
  H.~E.~Haber and D.~O'Neil,
  Phys.\ Rev.\ D {\bf 83}, 055017 (2011)
  [arXiv:1011.6188 [hep-ph]].


\bibitem{Baak:2014ora} 
  M.~Baak {\it et al.} [Gfitter Group],
  Eur.\ Phys.\ J.\ C {\bf 74}, 3046 (2014)
  [arXiv:1407.3792 [hep-ph]].


\bibitem{Dolle:2009fn} 
  E.~M.~Dolle and S.~Su,
  Phys.\ Rev.\ D {\bf 80}, 055012 (2009)
  [arXiv:0906.1609 [hep-ph]].


\bibitem{Misiak:2015xwa} 
  M.~Misiak {\it et al.},
  Phys.\ Rev.\ Lett.\  {\bf 114}, no. 22, 221801 (2015)
  [arXiv:1503.01789 [hep-ph]].


\bibitem{Geng:1988bq} 
  C.~Q.~Geng and J.~N.~Ng,
  Phys.\ Rev.\ D {\bf 38}, 2857 (1988)
  Erratum: [Phys.\ Rev.\ D {\bf 41}, 1715 (1990)].


\bibitem{Deschamps:2009rh} 
  O.~Deschamps, S.~Descotes-Genon, S.~Monteil, V.~Niess, S.~T'Jampens and V.~Tisserand,
  Phys.\ Rev.\ D {\bf 82}, 073012 (2010)
  [arXiv:0907.5135 [hep-ph]].


\bibitem{Amhis:2016xyh} 
  Y.~Amhis {\it et al.} [HFLAV Collaboration],
  Eur.\ Phys.\ J.\ C {\bf 77}, no. 12, 895 (2017)
  [arXiv:1612.07233 [hep-ex]].


\bibitem{Lees:2012xj} 
  J.~P.~Lees {\it et al.} [BaBar Collaboration],
  Phys.\ Rev.\ Lett.\  {\bf 109}, 101802 (2012)
  [arXiv:1205.5442 [hep-ex]].


\bibitem{Aad:2014fha} 
  G.~Aad {\it et al.} [ATLAS Collaboration],
  Phys.\ Lett.\ B {\bf 738}, 428 (2014)
  [arXiv:1407.8150 [hep-ex]].


\bibitem{Aad:2014ioa} 
  G.~Aad {\it et al.} [ATLAS Collaboration],
  Phys.\ Rev.\ Lett.\  {\bf 113}, no. 17, 171801 (2014)
  [arXiv:1407.6583 [hep-ex]].


\bibitem{Aad:2014vgg} 
  G.~Aad {\it et al.} [ATLAS Collaboration],
  JHEP {\bf 1411}, 056 (2014)
  [arXiv:1409.6064 [hep-ex]].


\bibitem{Aad:2015agg} 
  G.~Aad {\it et al.} [ATLAS Collaboration],
  JHEP {\bf 1601}, 032 (2016)
  [arXiv:1509.00389 [hep-ex]].


\bibitem{Aad:2015kna} 
  G.~Aad {\it et al.} [ATLAS Collaboration],
  Eur.\ Phys.\ J.\ C {\bf 76}, no. 1, 45 (2016)
  [arXiv:1507.05930 [hep-ex]].


\bibitem{Aad:2015wra} 
  G.~Aad {\it et al.} [ATLAS Collaboration],
  Phys.\ Lett.\ B {\bf 744}, 163 (2015)
  [arXiv:1502.04478 [hep-ex]].


\bibitem{Aad:2015xja} 
  G.~Aad {\it et al.} [ATLAS Collaboration],
  Phys.\ Rev.\ D {\bf 92}, 092004 (2015)
  [arXiv:1509.04670 [hep-ex]].


\bibitem{ATLAS-CONF-2016-056} 
  The ATLAS collaboration [ATLAS Collaboration],
  ATLAS-CONF-2016-056.


\bibitem{ATLAS-CONF-2016-062} 
  The ATLAS collaboration [ATLAS Collaboration],
  ATLAS-CONF-2016-062.


\bibitem{ATLAS-CONF-2016-059} 
  The ATLAS collaboration [ATLAS Collaboration],
  ATLAS-CONF-2016-059.


\bibitem{ATLAS-CONF-2016-085} 
  The ATLAS collaboration [ATLAS Collaboration],
  ATLAS-CONF-2016-085.


\bibitem{ATLAS-CONF-2016-074} 
  The ATLAS collaboration [ATLAS Collaboration],
  ATLAS-CONF-2016-074.


\bibitem{ATLAS-CONF-2016-044} 
  The ATLAS collaboration [ATLAS Collaboration],
  ATLAS-CONF-2016-044.


\bibitem{ATLAS-CONF-2016-082} 
  The ATLAS collaboration [ATLAS Collaboration],
  ATLAS-CONF-2016-082.


\bibitem{ATLAS-CONF-2016-079} 
  The ATLAS collaboration [ATLAS Collaboration],
  ATLAS-CONF-2016-079.


\bibitem{ATLAS-CONF-2016-071} 
  The ATLAS collaboration [ATLAS Collaboration],
  ATLAS-CONF-2016-071.


\bibitem{ATLAS-CONF-2016-004} 
  The ATLAS collaboration,
  ATLAS-CONF-2016-004.


\bibitem{ATLAS-CONF-2016-015} 
  The ATLAS collaboration,
  ATLAS-CONF-2016-015.


\bibitem{ATLAS-CONF-2016-017} 
  The ATLAS collaboration,
  ATLAS-CONF-2016-017.


\bibitem{Khachatryan:2015cwa} 
  V.~Khachatryan {\it et al.} [CMS Collaboration],
  JHEP {\bf 1510}, 144 (2015)
  [arXiv:1504.00936 [hep-ex]].


\bibitem{Khachatryan:2015lba} 
  V.~Khachatryan {\it et al.} [CMS Collaboration],
  Phys.\ Lett.\ B {\bf 748}, 221 (2015)
  [arXiv:1504.04710 [hep-ex]].


\bibitem{Khachatryan:2015tha} 
  V.~Khachatryan {\it et al.} [CMS Collaboration],
  Phys.\ Lett.\ B {\bf 755}, 217 (2016)
  [arXiv:1510.01181 [hep-ex]].


\bibitem{Khachatryan:2015tra} 
  V.~Khachatryan {\it et al.} [CMS Collaboration],
  JHEP {\bf 1511}, 071 (2015)
  [arXiv:1506.08329 [hep-ex]].


\bibitem{Khachatryan:2015yea} 
  V.~Khachatryan {\it et al.} [CMS Collaboration],
  Phys.\ Lett.\ B {\bf 749}, 560 (2015)
  [arXiv:1503.04114 [hep-ex]].


\bibitem{Khachatryan:2016sey} 
  V.~Khachatryan {\it et al.} [CMS Collaboration],
  Phys.\ Rev.\ D {\bf 94}, no. 5, 052012 (2016)
  [arXiv:1603.06896 [hep-ex]].


\bibitem{CMS-PAS-HIG-14-029} 
  CMS Collaboration [CMS Collaboration],
  CMS-PAS-HIG-14-029.


\bibitem{CMS-PAS-HIG-16-014} 
  CMS Collaboration [CMS Collaboration],
  CMS-PAS-HIG-16-014.


\bibitem{CMS-PAS-EXO-16-035} 
  CMS Collaboration [CMS Collaboration],
  CMS-PAS-EXO-16-035.


\bibitem{CMS-PAS-EXO-16-027} 
  CMS Collaboration [CMS Collaboration],
  CMS-PAS-EXO-16-027.


\bibitem{CMS-PAS-HIG-16-033} 
  CMS Collaboration [CMS Collaboration],
  CMS-PAS-HIG-16-033.


\bibitem{CMS-PAS-HIG-16-023} 
  CMS Collaboration [CMS Collaboration],
  CMS-PAS-HIG-16-023.


\bibitem{CMS-PAS-HIG-16-029} 
  CMS Collaboration [CMS Collaboration],
  CMS-PAS-HIG-16-029.


\bibitem{CMS-PAS-HIG-16-025} 
  CMS Collaboration [CMS Collaboration],
  CMS-PAS-HIG-16-025.


\bibitem{CMS-PAS-EXO-16-034} 
  CMS Collaboration [CMS Collaboration],
  CMS-PAS-EXO-16-034.


\bibitem{CMS-PAS-HIG-16-011} 
  CMS Collaboration [CMS Collaboration],
  CMS-PAS-HIG-16-011.


\bibitem{CMS-PAS-HIG-16-037} 
  CMS Collaboration [CMS Collaboration],
  CMS-PAS-HIG-16-037.


\bibitem{CMS-PAS-HIG-16-002} 
  CMS Collaboration [CMS Collaboration],
  CMS-PAS-HIG-16-002.


\bibitem{CMS-PAS-HIG-16-032} 
  CMS Collaboration [CMS Collaboration],
  CMS-PAS-HIG-16-032.


\bibitem{CMS-PAS-HIG-15-013} 
  CMS Collaboration [CMS Collaboration],
  CMS-PAS-HIG-15-013.


\bibitem{CMS-PAS-HIG-16-034} 
  CMS Collaboration [CMS Collaboration],
  CMS-PAS-HIG-16-034.


\bibitem{Agashe:2014kda} 
  K.~A.~Olive {\it et al.} [Particle Data Group],
  Chin.\ Phys.\ C {\bf 38}, 090001 (2014).


\bibitem{Pierce:2007ut} 
  A.~Pierce and J.~Thaler,
  JHEP {\bf 0708}, 026 (2007)
  [hep-ph/0703056 [HEP-PH]].


\bibitem{Aad:2014kga} 
  G.~Aad {\it et al.} [ATLAS Collaboration],
  JHEP {\bf 1503}, 088 (2015)
  [arXiv:1412.6663 [hep-ex]].


\bibitem{Khachatryan:2015qxa} 
  V.~Khachatryan {\it et al.} [CMS Collaboration],
  JHEP {\bf 1511}, 018 (2015)
  [arXiv:1508.07774 [hep-ex]].


\bibitem{Aad:2015typ} 
  G.~Aad {\it et al.} [ATLAS Collaboration],
  JHEP {\bf 1603}, 127 (2016)
  [arXiv:1512.03704 [hep-ex]].


\bibitem{ATLAS-CONF-2016-088} 
  The ATLAS collaboration [ATLAS Collaboration],
  ATLAS-CONF-2016-088.


\bibitem{CMS-PAS-HIG-16-031} 
  CMS Collaboration [CMS Collaboration],
  CMS-PAS-HIG-16-031.


\bibitem{ATLAS-CONF-2016-089} 
  The ATLAS collaboration [ATLAS Collaboration],
  ATLAS-CONF-2016-089.


\bibitem{ATLAS-CONF-2016-104} 
  The ATLAS collaboration [ATLAS Collaboration],
  ATLAS-CONF-2016-104.


\bibitem{Khachatryan:2016vau} 
  G.~Aad {\it et al.} [ATLAS and CMS Collaborations],
  JHEP {\bf 1608}, 045 (2016)
  doi:10.1007/JHEP08(2016)045
  [arXiv:1606.02266 [hep-ex]].


\bibitem{ATLAS-CONF-2016-063} 
  The ATLAS collaboration [ATLAS Collaboration],
  ATLAS-CONF-2016-063.


\bibitem{ATLAS-CONF-2016-080} 
  The ATLAS collaboration [ATLAS Collaboration],
  ATLAS-CONF-2016-080.


\bibitem{ATLAS-CONF-2016-081} 
  The ATLAS collaboration [ATLAS Collaboration],
  ATLAS-CONF-2016-081.


\bibitem{ATLAS-CONF-2016-091} 
  The ATLAS collaboration [ATLAS Collaboration],
  ATLAS-CONF-2016-091.


\bibitem{ATLAS-CONF-2016-112} 
  The ATLAS collaboration [ATLAS Collaboration],
  ATLAS-CONF-2016-112.


\bibitem{CMS-PAS-HIG-16-003} 
  CMS Collaboration [CMS Collaboration],
  CMS-PAS-HIG-16-003.


\bibitem{CMS-PAS-HIG-16-020} 
  CMS Collaboration [CMS Collaboration],
  CMS-PAS-HIG-16-020.


\bibitem{CMS-PAS-HIG-16-043} 
  CMS Collaboration [CMS Collaboration],
  CMS-PAS-HIG-16-043.


\bibitem{CMS-PAS-HIG-16-038} 
  CMS Collaboration [CMS Collaboration],
  CMS-PAS-HIG-16-038.


\bibitem{CMS-PAS-HIG-17-003} 
  CMS Collaboration [CMS Collaboration],
  CMS-PAS-HIG-17-003.














\end{thebibliography}
\end{document}